\documentclass[prd,aps,floats,floatfix,nofootinbib,preprintnumbers]{revtex4-1}
\usepackage{amsmath}
\usepackage{amsfonts}
\usepackage{graphicx}
\usepackage{slashed}
\usepackage{dcolumn}
\usepackage{bm}
\usepackage{amssymb}
\usepackage{latexsym}
\usepackage{color}
\usepackage{url}

\newcommand{\pbrac}[1]{\left( #1 \right)}
\newcommand{\tbrac}[1]{\left[ #1 \right]}
\newcommand{\cbrac}[1]{\left\{ #1 \right\}}

\begin{document}

\title{Higgs Partner Searches and Dark Matter Phenomenology\\ in a Classically Scale Invariant Higgs Boson Sector}

\author{Arsham Farzinnia}
\email[]{farzinnia@ibs.re.kr}
\affiliation{Center for Theoretical Physics of the Universe \\Institute for Basic Science (IBS), Daejeon 305-811, Republic of Korea}
\author{Jing Ren}
\email[]{jingren2004@gmail.com}
\affiliation{Institute of Modern Physics and Center for High Energy Physics\\ Tsinghua University, Beijing 100084, China}

\preprint{CTPU-14-03}

\date{\today}

\begin{abstract}
In a previous work, a classically scale invariant extension of the standard model was proposed, as a potential candidate for resolving the hierarchy problem, by minimally introducing a complex gauge singlet scalar, and generating radiative electroweak symmetry breaking by means of the Coleman-Weinberg Mechanism. Postulating the singlet sector to respect the $CP$-symmetry, the existence of a stable pseudoscalar dark matter candidate with a mass in the TeV range was demonstrated. Moreover, the model predicted the presence of another physical $CP$-even Higgs boson (with suppressed tree-level couplings), in addition to the 125~GeV scalar discovered by the LHC. The viable region of the parameter space was determined by various theoretical and experimental considerations. In this work, we continue to examine the phenomenological implications of the proposed minimal scenario by considering the constraints from the dark matter relic density, as determined by the Planck collaboration, as well as the direct detection bounds from the LUX experiment. Furthermore, we investigate the implications of the collider Higgs searches for the additional Higgs boson. Our results are comprehensively demonstrated in unified exclusion plots, which analyze the viable region of the parameter space from all relevant angles, demonstrating the testability of the proposed scenario.
\end{abstract}

\maketitle

\section{Introduction}\label{intro}

The discovery of a light and apparently fundamental Higgs-like scalar at the LHC \cite{LHCnew} brings forth the question regarding the mechanism behind its mass stabilization \cite{fine-tuning}. Classical scale invariance has been advocated \cite{Bardeen} as a potential symmetry candidate to protect the Higgs mass from large quantum corrections, as required by the naturalness criterion \cite{natural}. This is motivated by the observation that the \textit{ad hoc} mass parameter of the Higgs field in the Higgs Lagrangian constitutes the only dimensionful parameter of the SM, which explicitly (but softly) breaks the scale symmetry associated with the Lagrangian. Although the scale symmetry is anomalous and is explicitly broken by the logarithmic effects in loop integrals, such quantum scale breaking is facilitated by dimension-4 operators which cannot contribute to the dimension-2 operator of the Higgs mass. Hence, the Higgs mass in the Lagrangian might be simply viewed as a soft breaking term of scale invariance. In principle, one might then argue that the SM Higgs boson is technically natural in the absence of any other physical scale near and above the weak scale,\footnote{In principle, this assertion is valid up to the $U(1)_{Y}$ Landau pole, where an UV completion of the SM may be conjectured to remove the latter properly. Given the absence of a consistent quantum theory of gravity, we need not be concerned with the Planck scale at this point.} once its mass is calculated within a regularization scheme that respects the scale symmetry (such as dimensional regularization).

It is, nevertheless, possible to set this sole dimensionful parameter of the SM to zero in the Lagrangian, and, in principle, achieve successful spontaneous breaking of the electroweak symmetry by implementing the Coleman-Weinberg mechanism \cite{Coleman:1973jx}. Within this framework---in analogy with the QCD scale, $\lambda_{\text{QCD}}$---the dimensionful Higgs mass parameter is generated at the quantum level by means of the dimensional transmutation via the stress-tensor trace anomaly.\footnote{Intriguingly, one may entertain the possibility that \textit{all} physical scales might have a quantum origin and vanish in the $\hbar \to 0$ limit, leaving the classical world scale invariant \cite{Hill:2005wg}.} Despite its elegance, however, it is well-known that the Coleman-Weinberg mechanism does not work realistically within the pure SM, since a (loop-generated) mass for the Higgs boson consistent with the 95\%~C.L. LEP-II limit, $M_{h} >114.4$~GeV \cite{Barate:2003sz}, renders the one-loop potential unbounded from below. Therefore, additional fields beyond the SM content are required to stabilize the potential. This notion, among others, has motivated many (recent) efforts in the community to formulate a classically scale invariant extension of the SM \cite{SI-other}.

Recently, we proposed such an extension of the SM, by minimally introducing a complex electroweak and color singlet scalar into the classically scale invariant potential \cite{Farzinnia:2013pga}. In analogy to the ordinary SM scalar sector, the singlet sector was postulated to be $CP$-symmetric, rendering the pseudoscalar singlet stable, and providing a dark matter candidate. The $CP$-even singlet and the SM Higgs boson, both accruing non-zero vacuum expectation values (VEV), mixed with one another and produced two physical Higgs bosons, one of which was successfully identified with the 125~GeV scalar discovered by the LHC \cite{LHCnew}. Furthermore, by introducing singlet right-handed Majorana neutrinos, mass terms for the SM neutrinos were generated by means of the see-saw mechanism \cite{seesaw}, and it was demonstrated that a Yukawa interaction between the right-handed Majorana neutrinos and the singlet scalar successfully led to the generation of weak-scale masses for the former. We systematically analyzed the theoretical constraints arising from vacuum stability, perturbative unitarity, and triviality, as well as the experimental bounds from electroweak precision tests and LHC direct measurements of the 125~GeV state, and determined the viable region of the parameter space.\footnote{In principle, the proposed effective scenario may contain Landau poles associated with the scalar (self-)couplings, which would determine the upper range of validity of the theory. Although these poles generally develop below the $U(1)_Y$ Landau pole, they may lie close to the latter for suitable numerical choices of the couplings \cite{Gabrielli:2013hma}. It is possible to avoid the ``little" hierarchy problem for the potential poles positioned below the Planck scale with an UV completion of the current effective theory, which has a ``small" coupling with the SM and retains properly classical scale invariance \cite{smallcouple} (see e.g. \cite{UVcompl} for implementation of scale invariance at quantum level).}

The present study is devoted to further investigating the phenomenological implications of the proposed scenario by considering the dark matter relic density and direct detection constraints, as well as the application of the collider Higgs search data to the additional Higgs boson predicted by the model. A similar analysis of the dark matter relic abundance and direct detection experiments was previously performed by other authors \cite{Gabrielli:2013hma}, without inclusion of the right-handed Majorana neutrinos. In the current treatment, however, we continue to systematically include the latter and investigate its effects. Our results are comprehensively demonstrated in unified exclusion plots, which examine the viable region of the parameter space from all relevant angles. We determine that the collider search data as applied to the additional Higgs boson further restrict the parameter space of the model in a manner complementary to the previously deduced experimental bounds. Moreover, identifying the pseudoscalar as the sole or dominant component of dark matter in the universe, the relic density and direct detection considerations tightly constrain a mixing between the SM Higgs boson and the singlet scalar to small values and favor a heavy TeV mass cold dark matter, rendering the scenario highly predictive.

We start by reviewing the formalism of the proposed minimal scenario \cite{Farzinnia:2013pga} in Section~\ref{review}, highlighting the important aspects of physics and explicitly exhibiting the relevant quantities. Section~\ref{sigma} is devoted to examining the collider Higgs search constraints and its application to the additional Higgs boson of the current scenario, utilizing the data from LEP \cite{Barate:2003sz} as well as LHC \cite{LHCHeavyH} Higgs searches. In Section~\ref{chi}, we systematically analyze the dark matter constraints arising from the relic abundance, as determined by the Planck collaboration \cite{Ade:2013zuv}, and the data from the LUX direct detection experiments \cite{LUX2013}. For illustration, we also display the projected constraints by the future Xenon1T experiment \cite{Aprile:2012zx}, which in the absence of any dark matter signal discovery is expected to reduce the upper bound of the interaction cross section additionally by two orders of magnitude. We discuss the combined results of our analysis in Section~\ref{disc}, and present a comprehensive view of the model's parameter space by unified exclusion plots. Finally, we conclude the work in Section~\ref{concl}.

\section{Review of the Minimal Classically Scale Invariant Higgs Sector}\label{review}

In this section, we provide a brief review of the minimal classically scale invariant extension of the standard model (SM), proposed in \cite{Farzinnia:2013pga}. In this framework, the electroweak Higgs doublet, $H$, is augmented by a complex scalar, $S$, singlet under the SM gauge interactions. Specifically, the scalar Lagrangian reads
\begin{equation}\label{Lscalar}
\mathcal{L}_{\text{scalar}} = (D^\mu H)^\dagger D_\mu H + \partial^\mu S^* \partial_\mu S - V^{(0)}(H,S) \ ,
\end{equation}
where the electroweak doublet and singlet are, respectively, defined as
\begin{equation}\label{HS}
H= \frac{1}{\sqrt{2}}
\begin{pmatrix} \sqrt{2}\,\pi^+ \\ v_\phi+\phi+i\pi^0 \end{pmatrix} \ , \qquad S =\frac{1}{\sqrt 2} \pbrac{v_\eta + \eta + i\chi} \ .
\end{equation}
In \eqref{HS}, $\phi$ represents the SM Higgs boson with a corresponding vacuum expectation value (VEV) $v_{\phi}=246$~GeV, $\pi^{0, \pm}$ are the usual electroweak Nambu-Goldstone bosons, $\eta$ denotes a $CP$-even singlet scalar degree of freedom acquiring a VEV $v_{\eta}$, and $\chi$ represents the $CP$-odd component of the complex singlet scalar. One should keep in mind that, within the current framework, the non-zero VEVs are generated dynamically at the quantum level via the Coleman-Weinberg mechanism \cite{Coleman:1973jx} (see below).

Requiring a $CP$-symmetric scalar sector, the most general classically scale invariant potential of this model contains the following six entities
\begin{equation}\label{V0}
V^{(0)}(H,S) = \frac{\lambda_1}{6} \pbrac{H^\dagger H}^2 + \frac{\lambda_2}{6} |S|^4 + \lambda_3 \pbrac{H^\dagger H}|S|^2 + \frac{\lambda_4}{2} \pbrac{H^\dagger H}\pbrac{S^2 + S^{*2}} + \frac{\lambda_5}{12} \pbrac{S^2 + S^{*2}} |S|^2 + \frac{\lambda_6}{12} \pbrac{S^4 + S^{*4}} \ ,
\end{equation}
with all six couplings, $\lambda_i$, real and dimensionless. The potential \eqref{V0} formally accommodates a mixing between the electroweak doublet and singlet by means of the parameters $\lambda_3$ and $\lambda_4$. Employing the following definitions
\begin{equation}\label{couprel}
\lambda_\phi \equiv \lambda_{1} \ , \quad \lambda_\eta \equiv \lambda_{2} + \lambda_{5} + \lambda_{6} \ , \quad \lambda_\chi \equiv \lambda_{2} - \lambda_{5} + \lambda_{6} \ , \quad \lambda_{\eta \chi} \equiv \frac{1}{3}\lambda_{2} - \lambda_{6} \ , \quad \lambda_m^+ \equiv \lambda_{3} + \lambda_{4} \ , \quad \lambda_m^- \equiv \lambda_{3} - \lambda_{4} \ ,
\end{equation}
the quartic part of the potential \eqref{V0} in terms of the field components may be conveniently expressed as
\begin{equation}\label{V0quart}
\begin{split}
V^{(0)}_{\text{quartic}} =&\, \frac{1}{24} \tbrac{ \lambda_\phi \phi^4 + \lambda_\eta \eta^4 + \lambda_\chi \chi^4 + \lambda_\phi\pbrac{\pi^0\pi^0 + 2 \pi^+ \pi^-}^2}+ \frac{1}{4}\tbrac{\lambda_m^+ \phi^2 \eta^2 + \lambda_m^- \phi^2 \chi^2 +\lambda_{\eta \chi} \eta^2 \chi^2}
\\
&+ \frac{1}{12} \tbrac{\lambda_{\phi} \phi^2 + 3\lambda_m^+ \eta^2 + 3\lambda_m^- \chi^2}\pbrac{\pi^0\pi^0 + 2 \pi^+ \pi^-}\ .
\end{split}
\end{equation}
One can then show that the tree-level potential is bounded from below \cite{Kannike:2012pe}, once the following conditions are satisfied\footnote{As explained in \cite{Kannike:2012pe}, the relations \eqref{stabtree1} and \eqref{stabtree2} represent the sufficient and necessary conditions for vacuum stability. The current conditions are \textit{less restrictive} than those derived previously in \cite{Farzinnia:2013pga}.}
\begin{subequations}
\begin{align}
&
\lambda_\phi^{} > 0 \ , \qquad \lambda_\eta^{} > 0 \ , \qquad
\lambda_\chi^{} > 0 \,, \qquad
\lambda_{\eta\chi}^{} > -\frac{1}{3}\!\sqrt{\lambda_{\eta}^{}\lambda_{\chi}^{}} \ , \qquad
\lambda_m^+ > -\frac{1}{3} \sqrt{\lambda_\phi \lambda_\eta} \ , \qquad
\lambda_m^- > -\frac{1}{3} \sqrt{\lambda_\phi \lambda_\chi} \ ,
\label{stabtree1}
\\[2mm]
&
\lambda_{\eta\chi} \sqrt {\lambda_\phi} + \lambda_m^+ \sqrt{\lambda_{\chi}} + \lambda_m^- \sqrt{\lambda_{\eta}} > -\frac{1}{3} \tbrac{\sqrt{\lambda_{\phi} \lambda_{\eta} \lambda_{\chi}} + \sqrt{2\pbrac{3\lambda_{\eta\chi}+\sqrt{\lambda_{\eta} \lambda_{\chi}}}\pbrac{3\lambda_{m}^{+}+\sqrt{\lambda_{\phi} \lambda_{\eta}}}\pbrac{3\lambda_{m}^{-}+\sqrt{\lambda_{\phi} \lambda_{\chi}}}}} \ .
\label{stabtree2}
\end{align}
\end{subequations}

The non-zero VEVs, $v_{\phi}$ and $v_{\eta}$, induce formal mass terms for the scalar fields of the Lagrangian. In addition, due to the mixing parameters, $\lambda_3$ and $\lambda_4$, they give rise to a mixing between the $CP$-even scalars, $\phi$ and $\eta$. The physical masses of the latter scalars may, subsequently, be determined by means of an orthogonal rotation matrix \cite{Farzinnia:2013pga}
\begin{equation}\label{hs}
\begin{pmatrix} \phi\\ \eta \end{pmatrix}
= \begin{pmatrix} \cos\omega & \sin\omega \\ -\sin\omega & \cos\omega \end{pmatrix} \begin{pmatrix} h \\ \sigma \end{pmatrix} \ , \qquad \cot(2\omega) \equiv \frac{1}{4\lambda_m^+} \tbrac{ (\lambda_\eta-\lambda_m^+) \frac{v_\eta}{v_\phi} - (\lambda_\phi-\lambda_m^+) \frac{v_\phi}{v_\eta} }\ ,
\end{equation}
where $h$ and $\sigma$ represent the mass eigenstates of the $CP$-even scalars. With all the other scalar masses remaining automatically diagonal, one obtains at tree-level \cite{Farzinnia:2013pga}
\begin{equation}\label{masstree}
\begin{split}
& M_h^2 = \frac{1}{2} \tbrac{\lambda_\phi v_\phi^2+\lambda_m^+ v_\eta \pbrac{v_\eta - 2 v_\phi \tan \omega}} \ , \qquad
M_\chi^2 =  \frac{1}{2} \tbrac{\lambda_m^- v_\phi^2 + \lambda_{\eta \chi} v_\eta^2} \ , \\
& M_\sigma^2 = \frac{1}{2} \tbrac{\lambda_\phi v_\phi^2+\lambda_m^+ v_\eta \pbrac{v_\eta + 2v_\phi \cot \omega}} \ ,  \qquad M_{\pi^0}^2 = M_{\pi^\pm}^2 = \frac{1}{6}\tbrac{\lambda_{\phi} v_\phi^2 + 3\lambda_m^+ v_\eta^2} \ .
\end{split}
\end{equation}
The $h$ scalar degree of freedom is identified with the 125~GeV state discovered at the LHC \cite{LHCnew}; i.e. $M_{h} = 125$~GeV.\footnote{As discussed in \cite{Farzinnia:2013pga}, identifying the $\sigma$ scalar with the discovered 125~GeV state in this minimal scenario is ruled out by the obtained theoretical and experimental bounds.}

Furthermore, invoking the see-saw mechanism \cite{seesaw}, we account for the non-zero neutrino masses (deduced from the experimentally observed neutrino oscillations) by including three heavy right-handed Majorana neutrino flavors, $\mathcal{N}^{i}$. The masses of the latter are generated via their Yukawa interactions with the complex singlet scalar, $S$ in \eqref{HS}. For simplicity, these Yukawa couplings---and hence the right-handed neutrino masses---are chosen to be flavor-universal. Demanding the pure gauge-singlet sector, in addition, to be $CP$-invariant \cite{Farzinnia:2013pga}, we may write
\begin{equation}\label{LRHN}
\mathcal{L}_{\mathcal N} = - \tbrac{Y^\nu_{ij}\, \bar{L}_{\ell}^{i} \tilde{H} \mathcal{N}^{j} + \text{h.c.}} -\frac{1}{2}y^N \mathcal{I}_{3\times3} \pbrac{S + S^*} \bar{\mathcal{N}}^{i}\mathcal{N}^{i} \ ,
\end{equation}
where $Y^\nu_{ij}$ is the (complex) Dirac neutrino Yukawa matrix, coupling the SM Higgs doublet $H$ to the left-handed lepton doublet $L_{\ell}^{i}$ and the right-handed neutrino $\mathcal N^{j}$, $y^N$ represents the (real) flavor-universal right-handed Majorana neutrino Yukawa coupling, $\mathcal{N}_{i} = \mathcal{N}_{i}^{c}$ is the 4-component gauge-singlet Majorana spinor, and $\tilde{H} \equiv i \sigma^2 H^*$. A flavor-universal mass scale for the right-handed Majorana neutrinos is induced once the $CP$-even component of $S$ acquires a non-zero VEV, $v_{\eta}$,\footnote{The Dirac Yukawa couplings $Y^{\nu}$ are of the same order as the SM electron Yukawa coupling \cite{Farzinnia:2013pga}; hence, we ignore them altogether in the rest of this analysis.}
\begin{equation}\label{mN}
M_{N} = \sqrt 2 \, y^{N} v_{\eta} \ .
\end{equation}

The classical scale invariance is explicitly broken by the logarithmic quantum effects; hence, a one-loop study of the scalar potential is necessary, in order to determine the true vacuum of the system. To this end, we express the full one-loop scalar potential as 
\begin{equation}\label{V01}
V(H,S) = V^{(0)}(H,S) + V^{(1)}(H,S) \ ,
\end{equation}
containing the tree-level potential $V^{(0)}(H,S)$, given by \eqref{V0}, and the one-loop contribution $V^{(1)}(H,S)$ from all relevant degrees of freedom in the loop.

The minimization of the one-loop potential \eqref{V01}---although analytically difficult in general---may be performed perturbatively using the Gildener-Weinberg prescription \cite{Gildener:1976ih}, where initially only the tree-level potential $V^{(0)}(H,S)$ \eqref{V0} is minimized with respect to its constituent fields, $H$ and $S$. In this approach, the tree-level minimization, nevertheless, occurs at a definite mass scale $\Lambda$. This is due to the fact that the couplings of the tree-level potential run with the renormalization scale $\mu$ at the quantum level. At the energy scale $\mu = \Lambda$, a flat direction among the non-zero VEVs in the potential may be identified by the tree-level minimization. The one-loop corrections will, then, become dominant along this particular direction, where they lift the flatness of the potential and determine the physical vacuum---thereby, breaking the classical scale symmetry.

Performing the described tree-level minimization \cite{Farzinnia:2013pga}, one deduces the following relations, valid at the scale $\mu=\Lambda$
\begin{equation}\label{mincond}
\frac{v_\phi^2}{v_\eta^2} = \frac{-3\lambda_m^+(\Lambda)}{\lambda_\phi(\Lambda)}=\frac{\lambda_\eta(\Lambda)}{-3\lambda_m^+(\Lambda)} \ ,
\end{equation}
which defines the flat direction of the potential, and eliminates one of the couplings in favor of the dimensional transmutation scale $\Lambda$. Along this direction, one obtains for the mixing angle: $\cot \omega = v_{\eta} / v_{\phi}$ (c.f. \eqref{hs}), and the tree-level expressions for the masses \eqref{masstree} and \eqref{mN} now yield
\begin{equation}\label{masstreemincond}
\begin{split}
& M_h^2 = \frac{v_\phi^2}{3} \tbrac{\lambda_\phi(\Lambda) -3\lambda_m^+(\Lambda) } \ , \quad M_\chi^2 =  \frac{v_\phi^2}{6\lambda_m^+(\Lambda)} \tbrac{3\lambda_m^+(\Lambda)\lambda_m^-(\Lambda) - \lambda_{\phi}(\Lambda)\lambda_{\eta\chi}(\Lambda)} \ ,\\
& M_\sigma^2 =M_{\pi^0}^2 = M_{\pi^\pm}^2 = 0 \ , \qquad\;\quad M_N = y^N v_\phi \sqrt{\frac{2 \lambda_\phi(\Lambda)}{-3\lambda_m^+(\Lambda)}} \ .
\end{split}
\end{equation}
The electroweak Nambu-Goldstone bosons are massless, as expected, which remains true to all orders in perturbation theory. Furthermore, it is worth noting that the $\sigma$ scalar has a vanishing mass at tree-level. This is attributed to the fact that the $\sigma$ field serves as the (pseudo) Nambu-Goldstone boson of the classical scale symmetry, and becomes massive at one-loop due to the Coleman-Weinberg mechanism \cite{Coleman:1973jx}.

As explained in \cite{Farzinnia:2013pga}, the one-loop effective potential of the $\phi$ field (c.f. \eqref{HS}) with the massive $h$ scalar, $\chi$ pseudoscalar, $W^{\pm}$ and $Z$ vector bosons, top quark, and the heavy right-handed neutrinos in the loop may be expressed, at the scale $\mu=\Lambda$, according to
\begin{equation}\label{V1h0}
V(\phi) = \alpha\, \phi^4 + \beta\, \phi^4 \log \frac{\phi^2}{\Lambda^2} \ ,
\end{equation}
where the coefficients in the $\overline{\text{MS}}$ scheme are defined as
\begin{subequations}
\begin{align}
&\alpha = \frac{1}{64\pi^2 v_\phi^4} \Bigg\{M_h^4\pbrac{-\frac{3}{2}+\log\frac{M_h^2}{v_\phi^2}} + M_\chi^4\pbrac{-\frac{3}{2}+\log\frac{M_\chi^2}{v_\phi^2}} + 6M_W^4\pbrac{-\frac{5}{6}+\log\frac{M_W^2}{v_\phi^2}} \notag \\
&\qquad \qquad\qquad+ 3M_Z^4\pbrac{-\frac{5}{6}+\log\frac{M_Z^2}{v_\phi^2}} - 12M_t^4\pbrac{-1+\log\frac{M_t^2}{v_\phi^2}} - 6M_N^4\pbrac{-1+\log\frac{M_N^2}{v_\phi^2}}\Bigg\} \ , \label{Ap} \\
&\beta = \frac{1}{64\pi^2 v_\phi^4} \pbrac{M_h^4 + M_\chi^4 + 6M_W^4+3M_Z^4 -12 M_t^4 - 6 M_N^4} \label{Bp} \ .
\end{align}
\end{subequations}

The energy scale $\Lambda$ may be determined explicitly by minimizing \eqref{V1h0} with respect to $\phi$ at $\phi = v_{\phi}$, yielding
\begin{equation}\label{Lambdafin} 
\Lambda = v_\phi \exp\tbrac{\frac{\alpha}{2\beta}+\frac{1}{4}} \ . 
\end{equation}
Hence, inserting \eqref{Lambdafin}, the one-loop effective potential \eqref{V1h0} reduces to
\begin{equation}\label{V1fin}
V(\phi) =\beta\, \phi^4 \tbrac{\log \frac{\phi^2}{v_{\phi}^2}-\frac{1}{2}} \ ,
\end{equation}
which is guaranteed to be bounded from below for large $\phi$ values if $\beta > 0$. This corresponds to demanding the following relation between the masses to be satisfied (c.f. \eqref{Bp})\footnote{Notice that the stability relation \eqref{staboneloop} cannot be satisfied within the SM alone.}
\begin{equation}\label{staboneloop}
M_\chi^4 - 6 M_N^4 > 12 M_t^4 - 6M_W^4 - 3M_Z^4 - M_h^4 \ .
\end{equation}
Furthermore, it is easy to show \cite{Farzinnia:2013pga} that the one-loop effective potential generates a radiative mass for the $\sigma$~scalar, via the Coleman-Weinberg mechanism \cite{Coleman:1973jx}, as a function of the other parameters of the theory
\begin{equation}\label{ms}
m_\sigma^2 (\omega, M_{\chi}, M_{N}) = 8 \beta\, v_{\phi}^2 \sin^{2} \omega \ .
\end{equation}
The positivity of $m_\sigma^2$ is automatically ensured by the stability condition $\beta > 0$ \eqref{staboneloop}.

The model contains five free parameters \cite{Farzinnia:2013pga}, which, without loss of generality, may be taken as the set
\begin{equation}\label{inputs}
\cbrac{\omega, M_\chi, M_N, \lambda_\chi, \lambda_m^-}  \ .
\end{equation}
Note that, in principle, either of the first three parameters in \eqref{inputs} may be traded for the mass of the $\sigma$~boson $m_{\sigma}$, by virtue of \eqref{ms} and \eqref{Bp}. The remaining Lagrangian parameters are expressed in terms of the set \eqref{inputs} according to
\begin{equation}\label{eq:paraIV2}
\begin{split}
\lambda_\phi&=3\frac{M_h^2}{v_{\phi }^2} \cos ^2\omega\ ,\quad
\lambda_{m}^+=-\frac{M_h^2}{v_{\phi }^2} \sin ^2\omega\ ,\quad
\lambda_{\eta}=3\frac{M_h^2}{v_{\phi }^2} \sin^2\omega\tan^2\omega\ , \\
\lambda_{\eta\chi}&=\left(2\frac{M_\chi^2}{v_{\phi }^2}-\lambda _{m}^-\right)\tan^2\omega\ ,\quad
y_N=\frac{M_N}{\sqrt{2}v_\phi}\tan\omega \ .
\end{split}
\end{equation}
One observes from \eqref{eq:paraIV2} that the sign of the mixing angle, $\omega$, does not affect any of the parameters in the scalar potential; therefore, in the following, we shall confine the analysis to $0\leq\sin \omega \leq 1$, without loss of generality. This observation is, however, not true for the $\lambda^{-}_{m}$ parameter. As we shall discuss in the forthcoming sections, the sign of the latter leads to interesting phenomenological consequences. Moreover, we note that, using \eqref{eq:paraIV2}, the last stability expression of the tree-level potential in \eqref{stabtree1} dictates a formal relation among the input parameters $\lambda_\chi$, $\lambda_m^-$, and the mixing angle $\omega$ in \eqref{inputs},\footnote{One can verify that the remaining conditions in \eqref{stabtree1} and \eqref{stabtree2} are automatically satisfied along the flat direction \eqref{mincond}.}
\begin{equation}\label{treeineq}
\frac{\lambda_m^-}{\sqrt{\lambda_\chi}\cos\omega} > - \frac{M_h}{\sqrt3 \, v_{\phi }} \ .
\end{equation}
As a consequence, fixed values of $\lambda_\chi$ and $\lambda_m^-$ impose formal restrictions on the viable range of $\sin \omega$; we shall further elaborate on this observation and its implications in Section~\ref{disc}.

This concludes our brief review of the formal aspects of the proposed scenario. The viable region of the free parameter space has been previously explored in \cite{Farzinnia:2013pga}, by imposing theoretical constraints from stability of the potential, unitarity, and triviality, as well as experimental bounds from electroweak precision tests and LHC direct measurements of the 125~GeV scalar. In the following sections, we investigate additional constraints on the parameter space arising from LEP and LHC Higgs searches, as well as dark matter relic density and direct detection data.

\section{Collider Search Constraints on the {\large $\sigma$} Scalar}\label{sigma}

It was demonstrated in \cite{Farzinnia:2013pga}, and briefly reviewed in Section~\ref{review}, that a mixing between the electroweak doublet and singlet \eqref{HS} in the scalar potential \eqref{V0} necessarily leads to a mixing between their $CP$-even components which acquire non-zero VEVs. As a consequence, the model predicts the existence of two physical Higgs bosons; namely, the $h$ and $\sigma$ scalars (c.f. \eqref{hs}). Both of these scalars are capable of interacting with the particle content of the SM electroweak sector; although, their tree-level coupling strengths are suppressed with respect to a pure SM Higgs boson by $\cos \omega$ and $\sin \omega$, respectively, due to the mixing. 

The $h$~boson is, as mentioned, identified with the 125~GeV state discovered at the LHC \cite{LHCnew}, $M_{h} = 125$~GeV (c.f. \eqref{masstreemincond}), and the constraints on the model's free parameters, arising from the direct measurements of its properties, were analyzed in \cite{Farzinnia:2013pga}. The latter study favored $\sin \omega \leq 0.44$.\footnote{Upon reexamining our previous analysis in \cite{Farzinnia:2013pga}, we have discovered a minor unfortunate error in the fitting code. Correcting this error leads to a slightly weaker constraint on the mixing angle; namely, $\sin \omega \leq 0.44$, as opposed to the previously reported incorrect value $\sin \omega \leq 0.37$. We will employ the correctly derived value throughout the current analysis.} Therefore, the $h$~Higgs is expected to be mostly SM-like, whereas the $\sigma$~boson is mostly singlet-like in nature.

The (radiatively generated) mass of the $\sigma$~scalar is given by \eqref{ms}, which exhibits a dependence on the masses of the pseudoscalar and the right-handed neutrinos, in addition to the mixing angle. The $\sigma$~boson may, thus, be lighter or heavier than the 125~GeV $h$~Higgs, while maintaining perturbative unitarity of the theory  \cite{Farzinnia:2013pga}. In spite of its mixing-angle suppressed tree-level coupling, the $\sigma$~boson can interact with the SM degrees of freedom; consequently, the data from the (heavy) Higgs collider searches may be used to constrain its properties.

In particular, we employ the available data from the LEP Higgs searches \cite{Barate:2003sz}, probing the mass range 10--120~GeV at 95\%~C.L., as well as the LHC (heavy) Higgs searches at $\sqrt s = 7,8$~TeV \cite{LHCHeavyH}, extending the mass reach at 95\%~C.L. to 1~TeV. In order to analyze these experimental data within the current framework, we construct an effective Lagrangian, which describes the tree-level interactions of the $\sigma$~boson with the heavy vector bosons and heavy fermions, its one-loop effective couplings to gluons and photons, as well as its additional tree-level non-SM couplings to a pair of $h$~Higgses and right-handed Majorana neutrinos\footnote{A decay of the $\sigma$~boson into a pair of pseudoscalars $\chi$, although formally present, is kinematically forbidden for a $\sigma$~boson lighter than 1~TeV, as $m_{\sigma} < 2 M_{\chi}$ for all choices of the free parameters within this mass range.}
\begin{subequations}
\begin{align}
\mathcal{L}^\sigma_{\text{effective}} =&\,
   c^\sigma_V \frac{2M_W^2}{v_\phi}\,\sigma\,W_{\mu}^{+}W^{-\mu}
  +c^\sigma_V \frac{M_Z^2}{v_\phi}\,\sigma\,Z_{\mu}Z^{\mu}
  -c^\sigma_t \frac{M_t}{v_\phi}\,\sigma\,\bar{t}t
  -c^\sigma_b \frac{M_b}{v_\phi}\,\sigma\,\bar{b}b
  -c^\sigma_c \frac{M_c}{v_\phi}\,\sigma\,\bar{c}c
  -c^\sigma_\tau \frac{M_\tau}{v_\phi}\,\sigma\,\bar{\tau}\tau \label{LefftreeSM}\\
&
  +c^\sigma_g\frac{\alpha_s}{12\pi{v_\phi}}\,\sigma\,G^a_{\mu\nu}G^{a\, \mu\nu}
  +c^\sigma_\gamma\frac{\alpha}{\pi{v_\phi}}\,\sigma\,A_{\mu\nu}A^{\mu\nu}
  +c^\sigma_h \,\sigma\,h h
  +c^\sigma_\mathcal{N} \,\sigma\,\bar{\mathcal{N}}^{i}\mathcal{N}^{i} \label{Leffrest} \ .
\end{align}
\end{subequations}

In this effective Lagrangian, the (dimensionless) tree-level coefficients in \eqref{LefftreeSM} parametrize the deviation of their couplings from the corresponding SM values. Accordingly, their values are determined by the mixing-angle suppression factor
\begin{equation}\label{cstreeSM}
c^\sigma_V=c^\sigma_t=c^\sigma_b=c^\sigma_c=c^\sigma_\tau=\sin\omega \ .
\end{equation}
The current scenario does not introduce any new degrees of freedom carrying either color or electric charges; therefore, involving only the usual SM states in the loop, a similar situation arises for the one-loop interaction of the $\sigma$~boson with pairs of gluons and photons in \eqref{Leffrest}. Their corresponding (dimensionless) coefficients are just those calculated within the SM (at the $\sigma$ mass) multiplied by the suppression factor
\begin{equation}\label{csloopSM}
c^\sigma_g= \sin \omega\times c_g^{\phi}(m_{\phi} = m_{\sigma}) \ , \qquad c^\sigma_\gamma= \sin \omega\times c_\gamma^{\phi}(m_{\phi} = m_{\sigma}) \ .
\end{equation}
The remaining non-SM tree-level (dimensionful) $\sigma hh$ and (dimensionless) $\sigma \bar{\mathcal{N}}^{i}\mathcal{N}^{i}$ couplings in \eqref{Leffrest} may be determined from their corresponding Lagrangians (see Appendix~\ref{FR} for the relevant Feynman rules), which yield
\begin{equation}\label{csnonSM}
c^\sigma_h = - \frac{M_{h}^{2}}{v_{\phi}}\sin \omega \ , \qquad c^\sigma_\mathcal{N} = - \frac{M_{N}}{2v_{\phi}}\sin \omega \ .
\end{equation}

Given the described effective formalism, we may now proceed to determine the total decay width of the $\sigma$~boson within our model. Expressing the $\sigma$~boson's width according to that of a corresponding SM Higgs with the same mass ($\Gamma^{\phi}_\text{total}(m_{\phi}=m_\sigma)$), one may easily deduce from the effective Lagrangian,
\begin{equation}\label{Ctots}
\begin{split}
\Gamma_\text{total}^{\sigma} =&\, \sin^2\omega\tbrac{\textrm{BR}^{\textrm{SM}}_{WW} + \textrm{BR}^{\textrm{SM}}_{ZZ} + \textrm{BR}^{\textrm{SM}}_{gg} + \textrm{BR}^{\textrm{SM}}_{\gamma \gamma}+\textrm{BR}^{\textrm{SM}}_{\bar{t}t}+\textrm{BR}^{\textrm{SM}}_{\bar{b}b}+\textrm{BR}^{\textrm{SM}}_{\bar{c}c}+\textrm{BR}^{\textrm{SM}}_{\bar{\tau}\tau}} \Gamma^{\phi}_\text{total}(m_{\phi}=m_\sigma) \\
&+\Gamma(\sigma \to hh)+\Gamma(\sigma \to\bar{\mathcal{N}}^{i}\mathcal{N}^{i}) \ ,
\end{split}
\end{equation}
where, $\textrm{BR}^{\textrm{SM}}_{ij}$ denotes the SM branching ratio of the Higgs decay into the $ij$ final states. Using \eqref{csnonSM}, the non-SM partial decay widths may be computed
\begin{equation}\label{nonSMwidth}
\Gamma(\sigma \to hh)=\sin^{2}\omega\,\frac{M_{h}^4}{8\pi v_{\phi}^{2} \,m_\sigma}\sqrt{1-\pbrac{\frac{2M_{h}}{m_\sigma}}^{2}} \ , \qquad \Gamma(\sigma \to \bar{\mathcal{N}}^{i}\mathcal{N}^{i})=\sin^{2}\omega\,\frac{m_\sigma M_{N}^2}{16\pi v_{\phi}^{2}}\tbrac{1-\pbrac{\frac{2M_{N}}{m_\sigma}}^{2}}^{3/2} \ .
\end{equation}

\begin{figure}
\includegraphics[width=.45\textwidth]{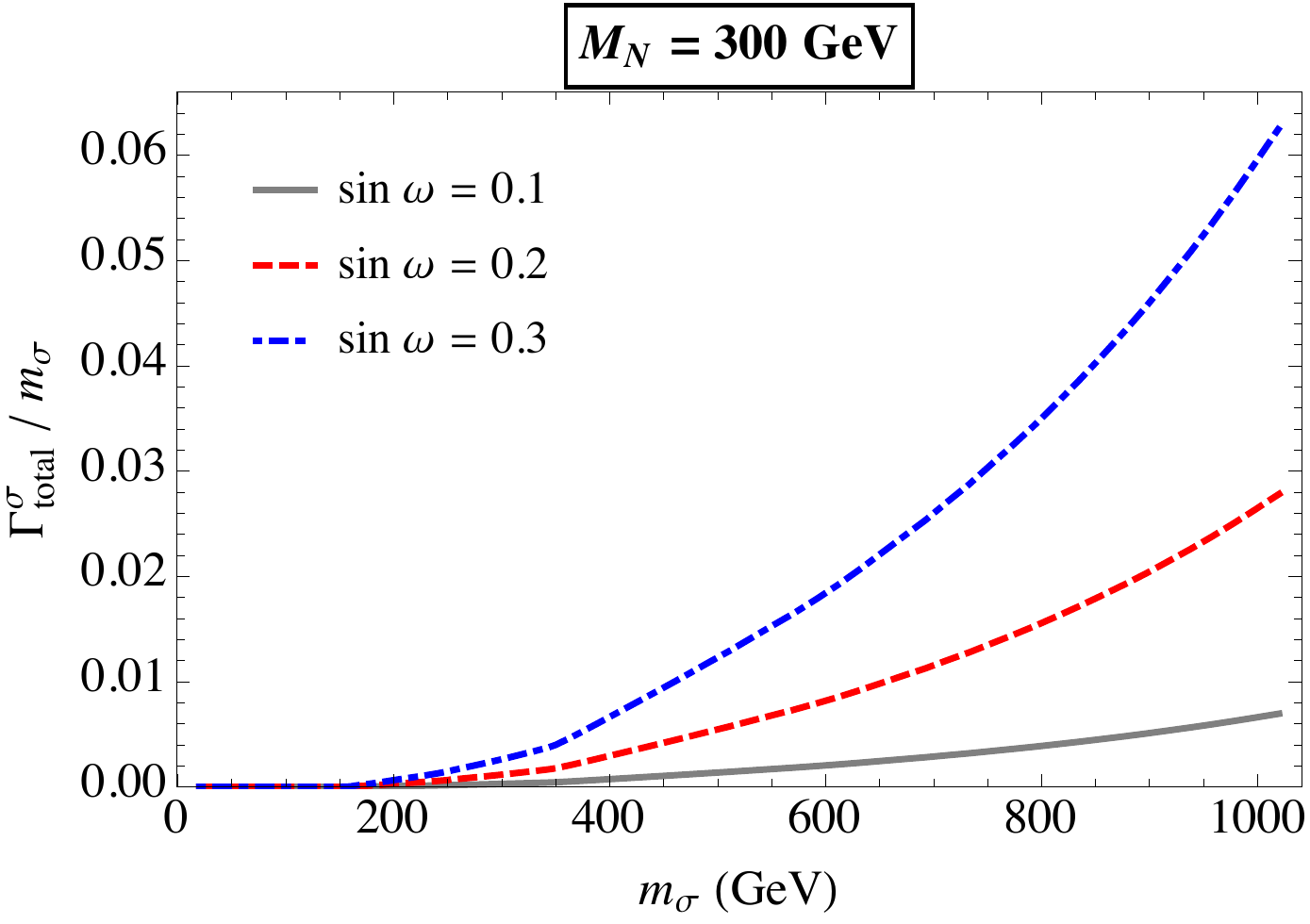} \qquad
\includegraphics[width=.45\textwidth]{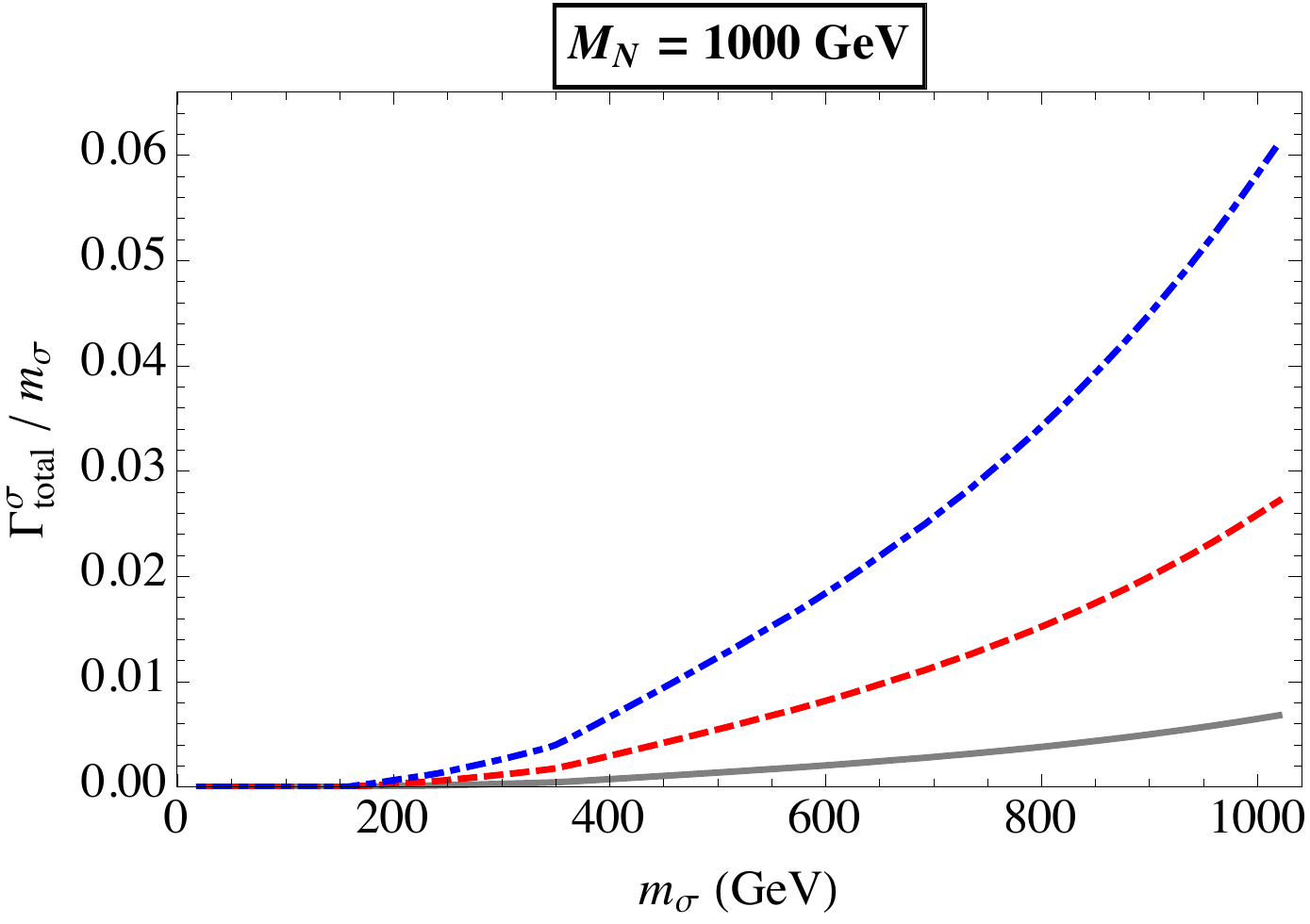}
\caption{Validity of the narrow-width approximation for the $\sigma$~boson. The panels display the ratio of the boson's total width to its mass as a function of its mass for $m_{\sigma}\leq 1$~TeV. Three values of the mixing angle---motivated by the experimental constraints \cite{Farzinnia:2013pga}---are selected for illustration in each panel. In the left panel ($M_{N}=300$~GeV), the decay channel of the $\sigma$~scalar to a pair of right-handed neutrinos is kinematically open, whereas in the right panel ($M_{N}=1000$~GeV), such a decay is not permitted. The effect of this non-SM decay mode is, thus, negligible in the entire mass range of interest.}
\label{NWA}
\end{figure}

The analysis of the experimental data from the Higgs searches \cite{Barate:2003sz,LHCHeavyH} depends crucially on the validity of the narrow-width approximation, in which the ratio of the Higgs total decay width to its mass is assumed to remain small within the entire mass range of the searches. In this spirit, let us examine the validity of the narrow-width approximation for the $\sigma$~scalar of the current scenario. As evident from \eqref{Ctots}, the total decay width of the $\sigma$~boson formally depends on the mixing angle and the right-handed neutrino mass scale, in addition to its own mass. Fig.~\ref{NWA} depicts the ratio of the $\sigma$~boson's calculated total decay width \eqref{Ctots} to its mass for $m_{\sigma}\leq 1$~TeV. In the left panel of this figure, representing a right-handed neutrino mass $M_{N}=300$~GeV, an on-shell decay of the $\sigma$~boson to a pair of right-handed neutrinos is kinematically allowed within the displayed range, whereas such a decay is kinematically forbidden in the right panel, where $M_{N}=1000$~GeV. Furthermore, each panel depicts three different values of the mixing angle, $\sin \omega$, for illustration purposes; the latter are motivated by the experimental bounds, as studied in \cite{Farzinnia:2013pga}. It is evident from Fig.~\ref{NWA} that the narrow-width approximation remains valid for the $\sigma$~boson in the entire mass range of interest for the collider searches, and for all (allowed) choices of the model's free parameters. Moreover, the effect of the $\sigma$~boson's potential decay into right-handed Majorana neutrinos may be safely ignored, and we do not consider this effect in the remainder of this section. It is interesting to note that the width of the $\sigma$~boson is, in fact, much narrower than that of a corresponding SM Higgs with the same mass. This is attributable to the suppression by the mixing-angle factor (c.f. \eqref{Ctots}), while the additional non-SM contributions to the width are not sufficiently large to compensate for this tree-level suppression within the mass range of interest.

Next, we analyze the experimental data from LEP \cite{Barate:2003sz} and LHC \cite{LHCHeavyH} Higgs searches as applied to the $\sigma$~scalar of the current scenario, taking into account the most stringent bounds in the search data. For the LHC, the strongest constraints arise from Higgs decays to $W^{+}W^{-}$ and $ZZ$ final states; whereas, for the LEP, they are given by the $b\bar b$ decay mode. The production of the $\sigma$~boson, on the other hand, may equally proceed via the vector-boson-fusion (VBF), vector-Higgs (VH) associated production, or the gluon-fusion channels---all suppressed by the mixing angle factor, with respect to a pure SM Higgs. In the narrow-width approximation---assumed in the aforementioned searches and appropriate for our $\sigma$~boson (c.f. Fig.~\ref{NWA})---the signal rate may be constructed by multiplying the production cross section by the branching ratio of the appropriate decay mode. In order to make a comparison between the current model and the ordinary SM predicted signal rates, we construct the $\mu$~parameter, defined by
\begin{equation}\label{iisVV}
\mu(ii \to \sigma \to jj)\equiv \frac{\sigma(ii \to \sigma) \times \textrm{BR}(\sigma \to jj)}{\sigma(ii \to \phi) \times \textrm{BR}(\phi \to jj)} = \sin^4\omega \, \frac{\Gamma^{\phi}_\text{total}(m_{\phi}=m_\sigma)}{\Gamma_\text{total}^{\sigma}} \ ,
\end{equation}
where $ii$ denotes the VBF, VH associated production, or the gluon-fusion production channels, $jj$ stands for the $W^{+}W^{-}$, $ZZ$, or $b\bar b$ final states, and we have used the coefficients of the effective Lagrangian \eqref{LefftreeSM} and \eqref{Leffrest}. Inserting the $\sigma$~boson's total width \eqref{Ctots} in \eqref{iisVV}, and neglecting a decay into the right-handed neutrinos (c.f. Fig.~\ref{NWA}), one notes that the $\mu$~parameter is explicitly a function of $m_{\sigma}$ and $\omega$. One may, then, compare the constructed $\mu$~parameter of the model with the one quoted by the experimental searches \cite{Barate:2003sz,LHCHeavyH}, and derive bounds on the input parameters \eqref{inputs}.

\begin{figure}
\includegraphics[width=.59\textwidth]{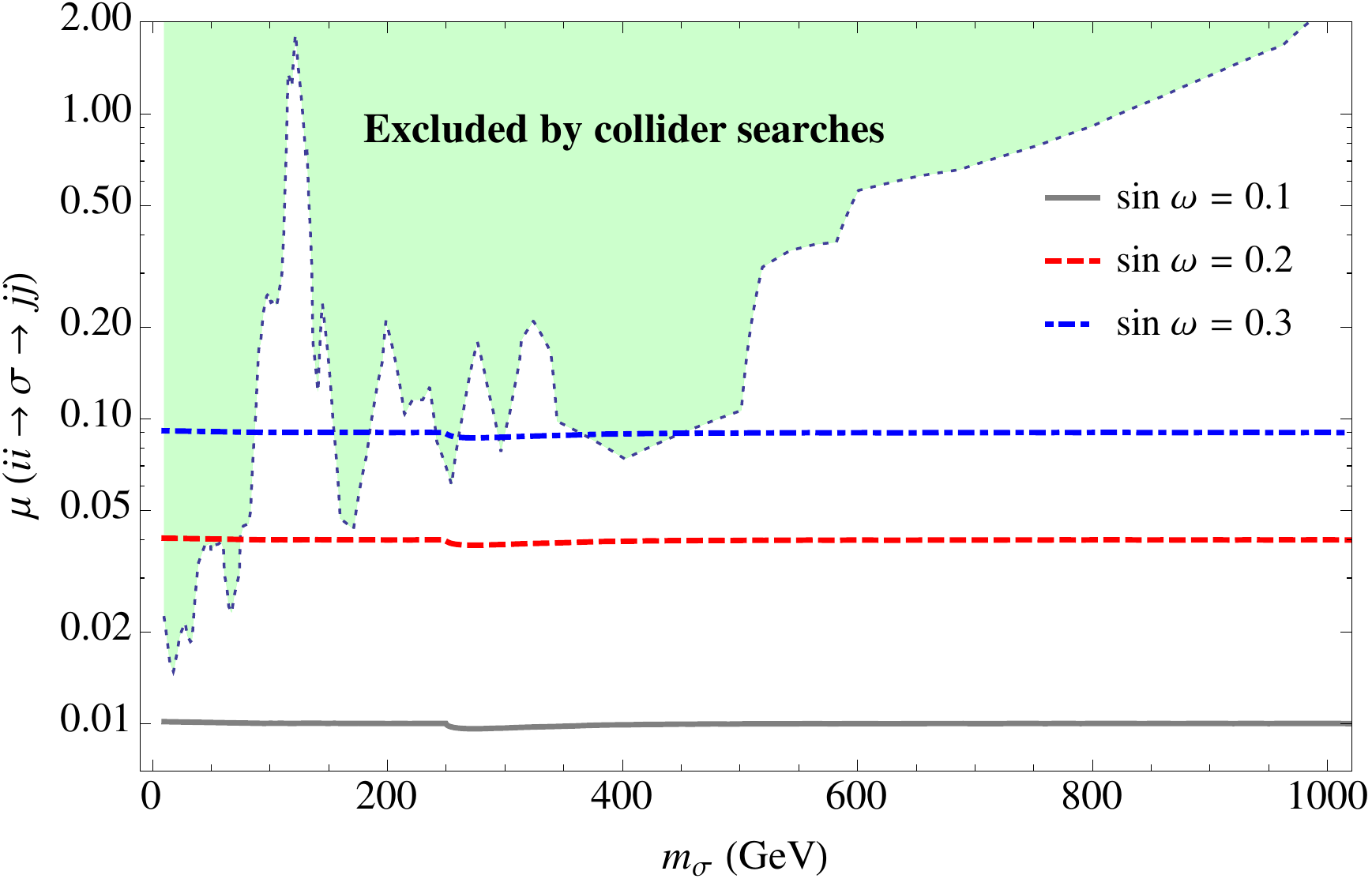}
\includegraphics[width=.4\textwidth]{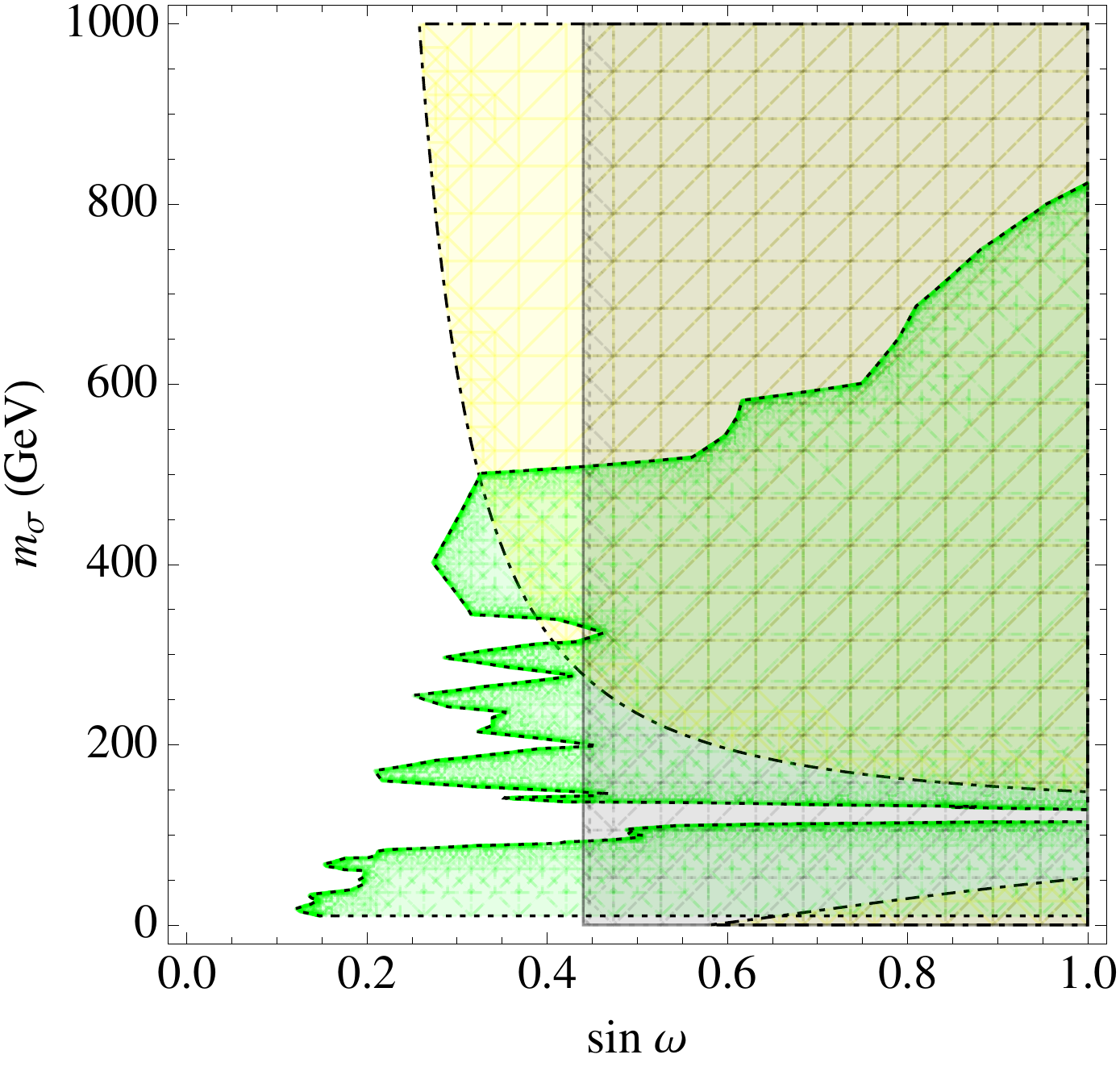}
\caption{\textit{Left}: Theoretical curves of the $\mu$~parameter \eqref{iisVV} as a function of the $\sigma$~boson mass, for three representative values of the mixing angle, along with the most stringent experimental upper limit from LEP \cite{Barate:2003sz} and LHC \cite{LHCHeavyH} Higgs searches at 95\%~C.L. \textit{Right}: The experimental exclusion limits in the $\sin\omega-m_{\sigma}$~plane. All colored regions are excluded at 95\%~C.L., taking into account the electroweak precision tests (dot-dashed), direct measurements of the LHC 125~GeV $h$~Higgs' properties (solid), and the LEP and LHC Higgs searches (dotted).}
\label{mu}
\end{figure}

In the left panel of Fig.~\ref{mu}, the theoretical values of the $\mu$~parameter \eqref{iisVV} are depicted as a function of the $\sigma$~boson mass, for three selected values of the mixing angle. In addition, the most stringent upper bound arising from LEP \cite{Barate:2003sz} and LHC \cite{LHCHeavyH} Higgs searches, together covering a mass range 10--1000~GeV at 95\%~C.L., is displayed in the same figure. One concludes that the collider Higgs searches generally exclude a light $\sigma$~boson with a large mixing angle. The exclusion limits from these searches are complementary to the experimental bounds derived in \cite{Farzinnia:2013pga}, considering the electroweak precision tests and the direct measurements of the 125~GeV $h$~Higgs at the LHC. This fact is illustrated in the right panel of Fig.~\ref{mu}, where the mass of the $\sigma$~boson is plotted as a function of the mixing angle, $\sin \omega$. These experimental considerations are insensitive to the values of the remaining input parameters, $M_{N}$, $\lambda_{\chi}$, and $\lambda_{m}^{-}$, and exclude together the mixing region $\sin\omega \gtrsim 0.3$ for most values of the $\sigma$~boson masses.

\section{Dark Matter Constraints on the {\large $\chi$} Pseudoscalar}\label{chi}

Demanding the pure singlet sectors (i.e., the scalar potential \eqref{V0}, and the Yukawa interaction between the singlet and the right-handed neutrinos in \eqref{LRHN}) to be $CP$-invariant results in the pseudoscalar state, $\chi$, always appearing in pairs \cite{Farzinnia:2013pga}, rendering it stable and a potential weakly-interacting massive particle (WIMP) dark matter candidate. Furthermore, it was demonstrated that this degree of freedom may be heavy, with a mass potentially in the TeV region; hence, appropriate within the cold dark matter framework. This section is devoted to study the constraints on the parameter space which arise from the WIMP relic abundance considerations as determined by the Planck collaboration \cite{Ade:2013zuv}, as well as the limits set by the LUX direct detection experiment \cite{LUX2013},\footnote{As mentioned in Section~\ref{intro}, a similar analysis of $\chi$ as a dark matter candidate, for small values of the mixing angle, has previously been presented in \cite{Gabrielli:2013hma} without considering the right-handed Majorana neutrinos. In the current treatment, we include the latter, which formally influences the lower bound on $M_{\chi}$ (c.f. \eqref{staboneloop}), as well as providing an additional channel for the dark matter annihilation.} assuming $\chi$ constitutes an $\mathcal O (1)$ fraction of the dark matter in the universe.

\subsection{Thermal Relic Density}\label{relden}

As a WIMP dark matter candidate, the heavy $\chi$ pseudoscalars are initially thermalized in the early universe, where $T \gg M_{\chi}$. As the temperature continues to drop due to the expansion of the universe, the rate of the dark matter pair-annihilation decreases accordingly; therefore, maintaining the thermal equilibrium becomes progressively difficult. Once the scattering rate is approximately below the Hubble expansion rate, the dark matter density essentially freezes out, decoupling from the remaining relevant species---all of which are assumed to be lighter, and hence, remain thermalized at that epoch. In this fashion, the frozen abundance of the dark matter survives until the present time. We may, thus, derive constraints on the model's free parameters, by estimating its prediction for the relic density of the $\chi$~pseudoscalar and comparing the latter with the latest observational data from the Planck satellite \cite{Ade:2013zuv}. To this end, we follow the standard approach presented in \cite{Srednicki:1988ce,Gondolo:1990dk,Kolb:book}.

Defining the number of dark matter particles per comoving volume as $Y\equiv n/s$, with $n$ the number density and $s$ the entropy density, one may express the rate of change of $Y$ by the Boltzmann equation \cite{Kolb:book}
\begin{equation}\label{Yrate}
\frac{d Y}{d x}=-\frac{x\, s\, \langle\sigma v \rangle_{\text{ann}}}{H(M_\chi)}\pbrac{ Y^2-Y_{\text{eq}}^2} \qquad \qquad \pbrac{x\equiv M_\chi/T} \ ,
\end{equation}
with $Y_{\text{eq}}\equiv n_{\text{eq}}/s$ the equilibrium value. In the non-relativistic limit, $x\gg 3$ \cite{Kolb:book}, the equilibrium value of the comoving number density is given by
\begin{equation}\label{Yeq}
Y_{\text{eq}}(x) \equiv a\, x^{3/2} e^{-x} \ , \qquad a\equiv \frac{45}{2^{5/2}\, \pi^{7/2}}\frac{1}{g^{\ast}_{s}} \ ,
\end{equation}
with $g^{\ast}_{s}$ the effective entropy degrees of freedom. Similarly, the Hubble rate at the time of freeze out, $T\sim M_{\chi}$, reads
\begin{equation}\label{HMX}
H(M_{\chi}) = \frac{2\pi^{3/2}}{3}\sqrt{\frac{g_{\text{rad}}}{5}} \, \frac{M_{\chi}^{2}}{M_{P}} \ ,
\end{equation}
where, $M_{P} = G_{N}^{-1/2} = 1.22\times10^{19}$~GeV is the Planck mass. The number of effective relativistic degrees of freedom at the time of freeze out is $g_{\text{rad}} = 106.75 + N_{s}$, and approximately equals $g^{\ast}_{s}$ for our purposes.\footnote{This is exact if all the particle species in the universe have the same temperature.} $N_{s}$ represents the number of non-SM scalar contributions, which in the current model constitutes only the $\sigma$ state. Hence, we have $g^{\ast}_{s} \simeq g_{\text{rad}} = 107.75$.

\begin{figure}
\includegraphics[width=.8\textwidth]{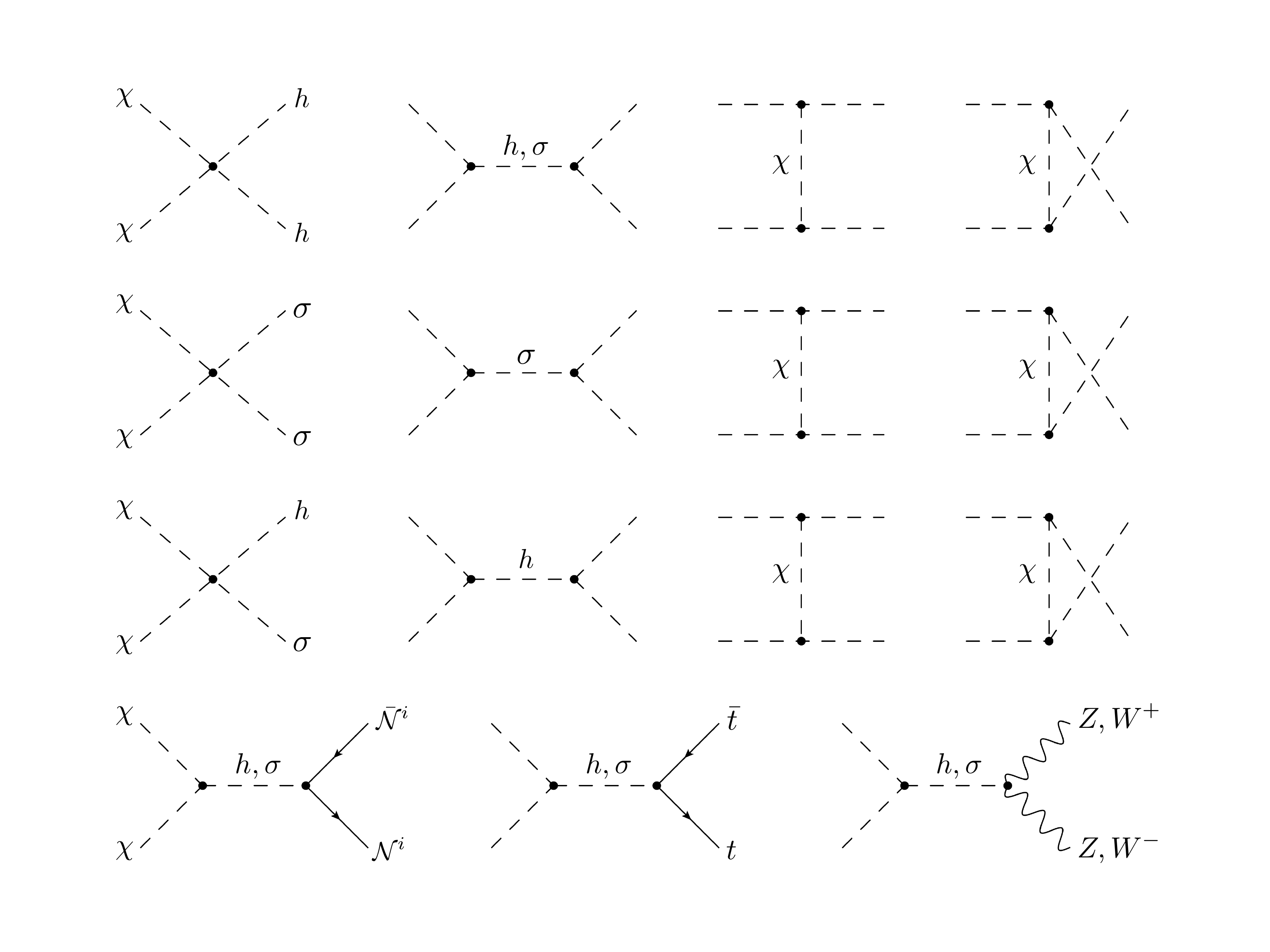}
\caption{Pair-annihilation of the $\chi$ WIMP dark matter into the (dominant) pairs of scalar, fermion, and vector final states. Diagrams in the top three rows illustrate their scattering process in all possible channels with the corresponding mediators.}
\label{DMann}
\end{figure}

The thermally averaged cross section of the dark matter pair-annihilation into $jj'$ final states, $\chi \chi \to jj'$, is determined according to \cite{Srednicki:1988ce,Kolb:book}
\begin{equation}\label{tacs}
\langle\sigma v \rangle_{\text{ann}} \equiv \frac{1}{n_{\text{eq}}^2}\int d^3p_1d^3p_2f(E_1)f(E_2)\, v_{12}\, \sigma_\text{ann} \simeq \left.\frac{E_1E_2\, v_{12}\, \sigma_\text{ann}}{M_\chi^2}\right|_{s_{\text{cm}}=4M_\chi^2}+\mathcal{O}\pbrac{\frac{1}{x}}\ ,
\end{equation}
where, $f(E_{i})$ is the Boltzmann distribution of particle $i$ with energy $E_{i}$, $v_{12}$ is the relative velocity of the dark matter pair, and $\sigma_\text{ann}$ is the total $2\to2$ scattering cross section. The leading-order expression on the right-hand side of \eqref{tacs} corresponds to the zero-temperature limit, and should be evaluated at the center of mass energy, $s_{\text{cm}}=4M_\chi^2$; for the non-relativistic $\chi$ pair-annihilation it is sufficient to use this approximating term. The dominant $2\to2$ processes, contributing to the total annihilation cross section of the dark matter pair, are depicted in Fig.~\ref{DMann}, where an annihilation into the right-handed neutrinos and the $\sigma$~scalars is taken into account, in addition to the $h$~Higgs pair, $t\bar t$, $W^{+}W^{-}$, and $ZZ$ final states. The corresponding expressions for the thermally averaged cross section are provided in Appendix~\ref{CSexp}.

Accordingly, one may derive an approximate analytical solution for the freeze out temperature \cite{Kolb:book}
\begin{equation}\label{xfo}
x_{\text{fo}}\equiv \frac{M_{\chi}}{T_{\text{fo}}} \simeq \log \lambda a - \frac{1}{2} \log \log \lambda a \ ,
\end{equation}
with $a$ as defined in \eqref{Yeq}, and the parameter $\lambda$ given by
\begin{equation}\label{lambda}
\lambda \equiv \frac{2\pi^{2}}{45}\, \frac{g^{\ast}_{s}M_{\chi}^{3} \langle\sigma v \rangle_{\text{ann}}}{H(M_\chi)} \ .
\end{equation}
The comoving number density of dark matter at the present time, $Y_{\infty}$, may then be found by integrating the Boltzmann equation \eqref{Yrate} from $x_{\text{fo}}$ to $\infty$, which approximately yields \cite{Kolb:book}
\begin{equation}\label{Yinf}
Y_{\infty} \simeq \frac{x_{\text{fo}}}{\lambda} \ .
\end{equation}
The WIMP thermal relic density---defined as the ratio of its mass density to the critical density at the present time, $\rho_{\text{crit}}/h^{2}= 1.878\times 10^{-29}$~g cm$^{-3}$ \cite{Beringer:1900zz} with $h=0.673$ the Hubble scale factor---is determined as \cite{Kolb:book}
\begin{equation}\label{relic}
\Omega_{\chi} h^{2} = \frac{M_{\chi}\,Y_{\infty}s_{\infty}}{\rho_{\text{crit}}/h^{2}} \ ,
\end{equation}
where, $s_{\infty} = 2891$~cm$^{-3}$ \cite{Beringer:1900zz} is the present value of the entropy density. Expression \eqref{relic} constitutes the prediction of our model for the present time thermal relic density of the cold dark matter pseudoscalar, $\chi$, and depends on the input parameters $\omega$, $\lambda_{m}^{-}$, $M_{\chi}$, and $M_{N}$. Comparing this expression with the observed value, $\Omega_{\chi} h^{2} = 0.1199 \pm 0.0027$, from the Planck collaboration \cite{Ade:2013zuv} imposes further bounds on these free parameters.

\begin{figure}
\includegraphics[width=.329\textwidth]{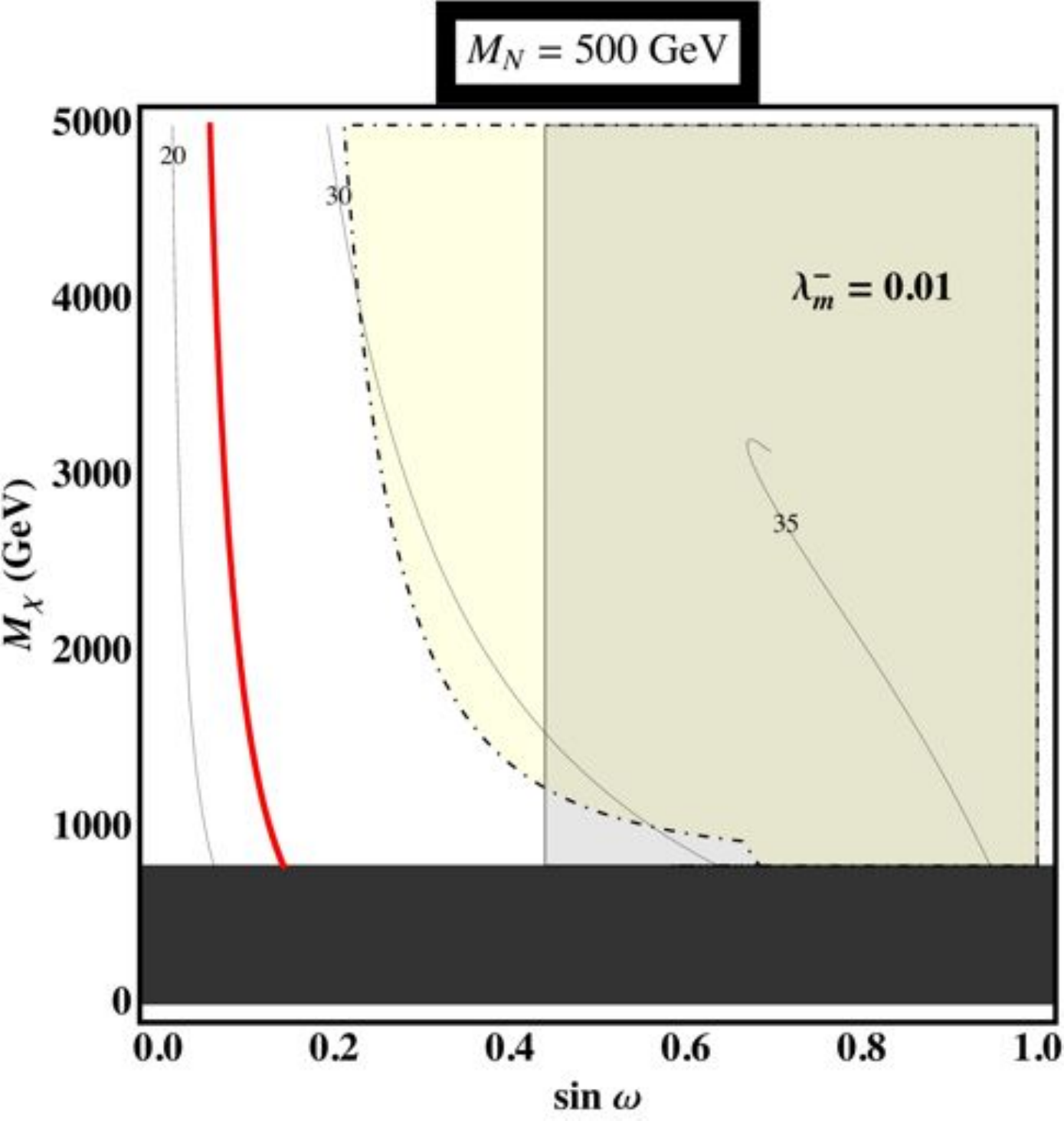}
\includegraphics[width=.329\textwidth]{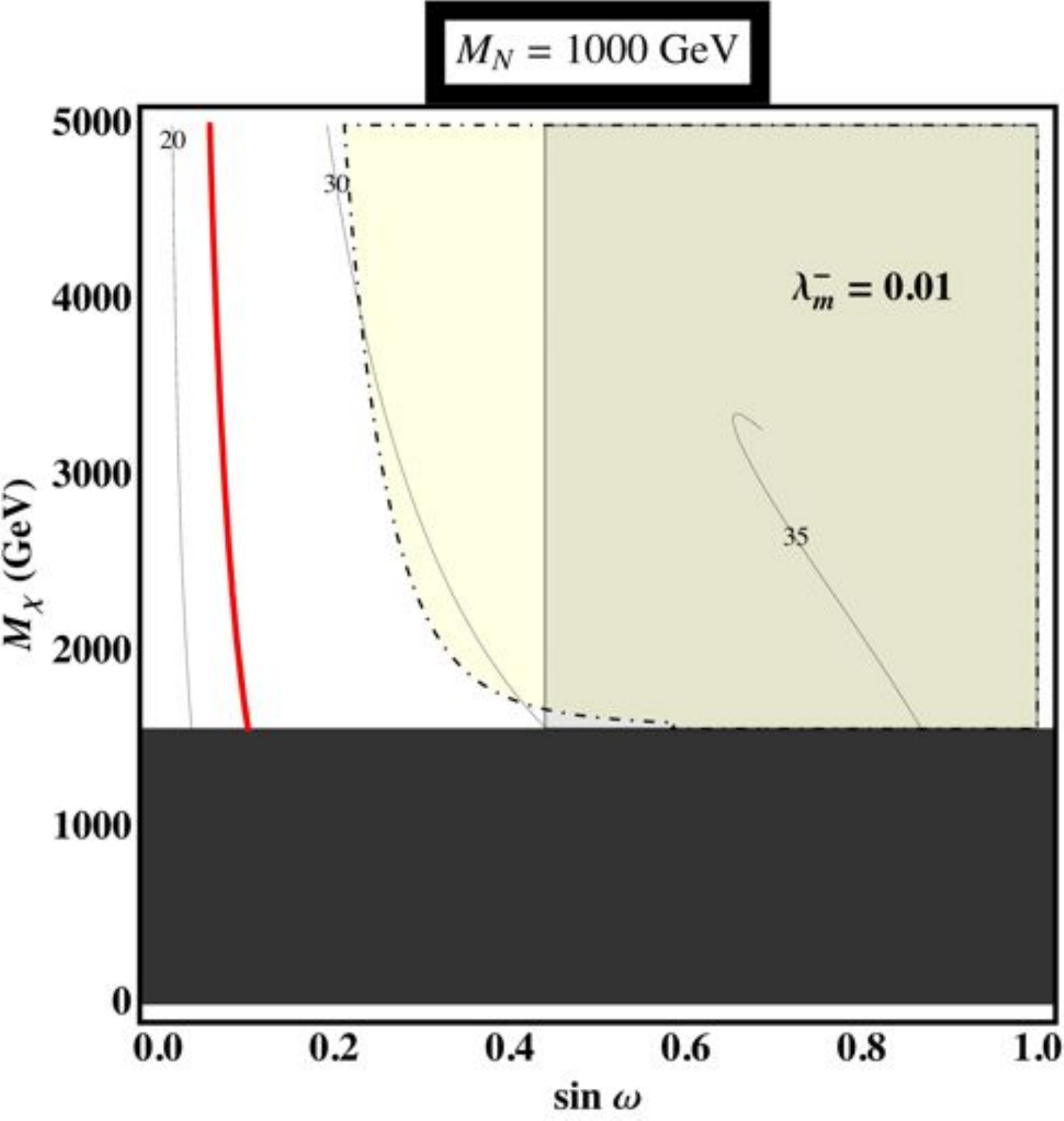}
\includegraphics[width=.329\textwidth]{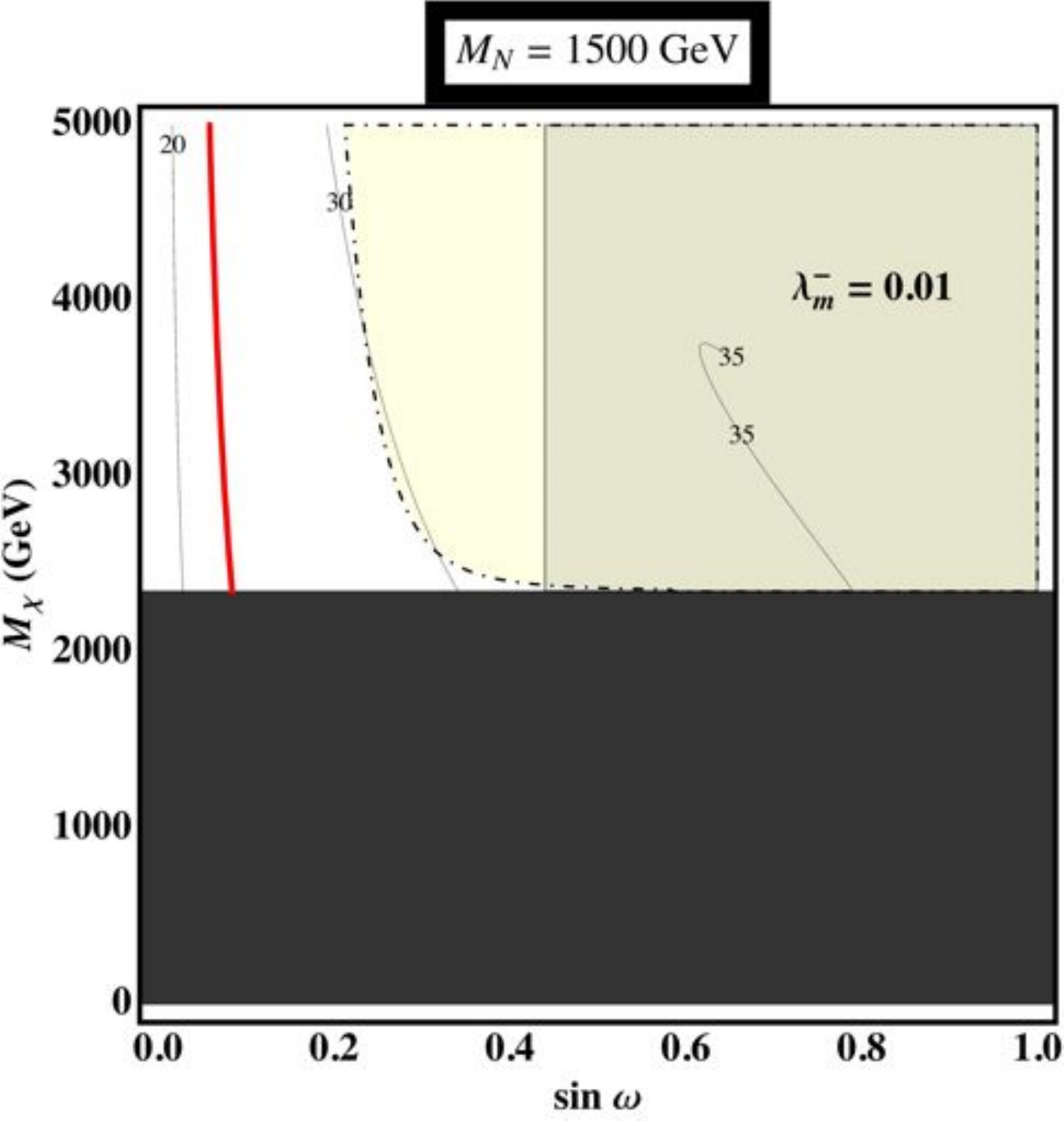}
\includegraphics[width=.329\textwidth]{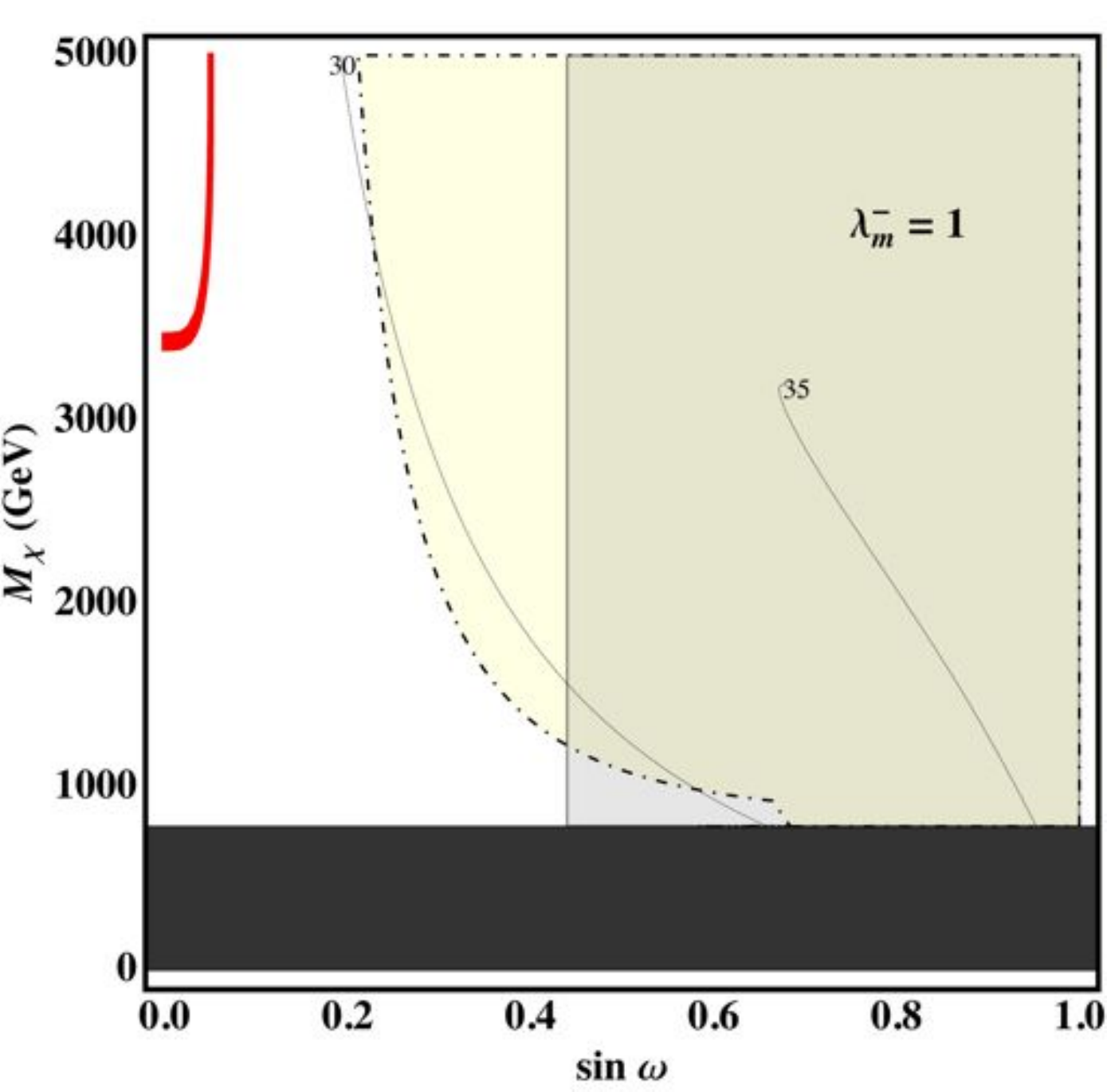}
\includegraphics[width=.329\textwidth]{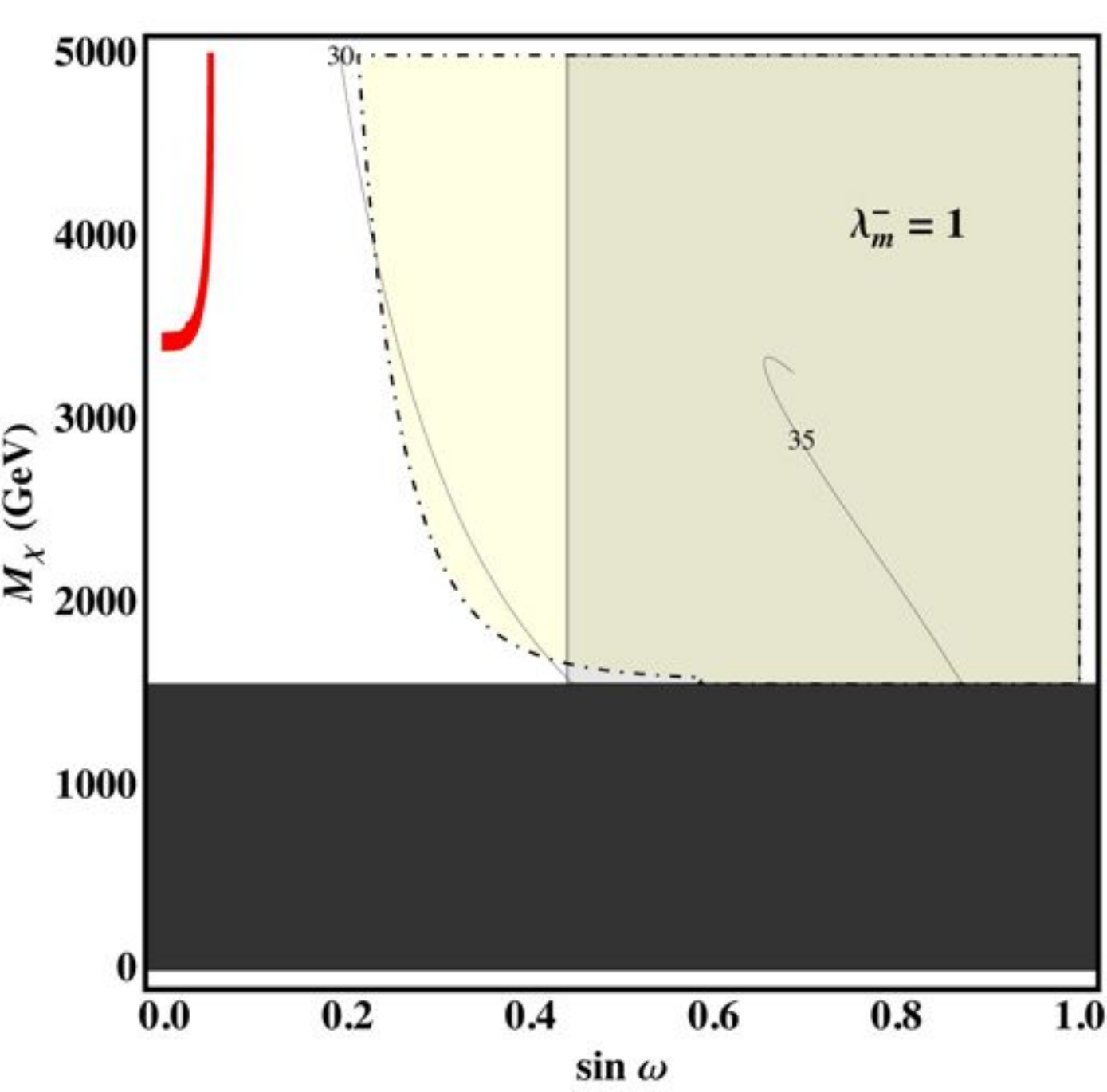}
\includegraphics[width=.329\textwidth]{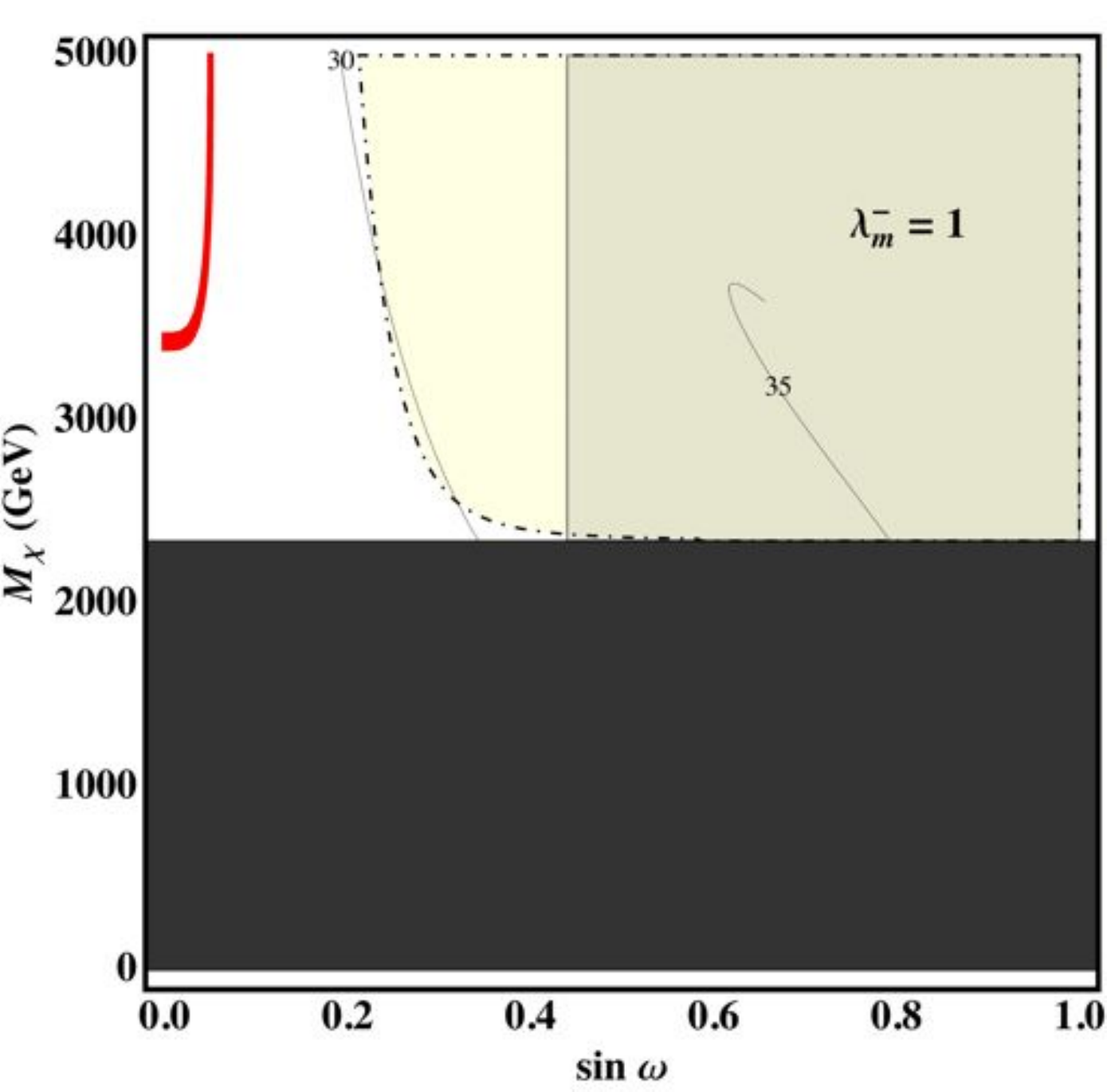}
\caption{Constraint from the dark matter relic abundance in the $\sin\omega-M_{\chi}$ plane, for benchmark values of the right-handed neutrino mass, $M_{N}$ (columns), and the input parameter $\lambda_{m}^{-}$ (rows). The thick (red) band represents the thermal relic density of the cold WIMP pseudoscalar as constrained by the data from the Planck collaboration \cite{Ade:2013zuv}. The thickness of the line corresponds to the 1$\sigma$ uncertainty quoted by the collaboration. A dependence on the sign of the $\lambda_{m}^{-}$ parameter is negligible. In addition, the experimental exclusion bounds from the electroweak precision tests (dot-dashed) and the direct measurements of the LHC 125~GeV Higgs' properties (solid) at 95\%~C.L. are displayed, which set the upper limit on the mixing angle. The solid black region, inferred from the stability condition of the one-loop potential \eqref{staboneloop}, determines the formal lower bound on the WIMP mass, $M_{\chi}$, for each selected value of $M_{N}$. The enumerated thin contours represent the values of the $x_{\text{fo}}$ parameter \eqref{xfo}, illustrating the validity of the non-relativistic treatment ($x_{\text{fo}}\gg3$), and hence the cold dark matter nature of the pseudoscalar~$\chi$. The observed relic density is comfortably accommodated within the allowed region of the model's parameter space.}
\label{relpix}
\end{figure}

This fact has been illustrated in Fig.~\ref{relpix}, which displays the constraint from the dark matter relic abundance in the $\sin\omega-M_{\chi}$ plane, along with the discussed \cite{Farzinnia:2013pga} experimental 95\%~C.L. bounds arising from the electroweak precision tests and the direct measurements of the LHC 125~GeV Higgs' properties. The panels correspond to representative values of the remaining parameters; namely, the right-handed neutrino mass, $M_{N}$, and the input parameter $\lambda_{m}^{-}$. It is evident that the observed value of the relic density (thick red band) is comfortably accommodated within the viable range of the model's parameters, and that a larger value of $\lambda_{m}^{-}$ necessitates a heavier dark matter to comply with the observational data. We have verified that a dependence on the sign of the $\lambda_{m}^{-}$ parameter is negligible in this analysis. In addition, in order to investigate the validity of the non-relativistic treatment of $\chi$ during the freeze out epoch---as indicated by the condition $x_{\text{fo}}\gg3$ \cite{Kolb:book}---various contours of the $x_{\text{fo}}$ parameter \eqref{xfo} are shown in the same figure. One concludes that the relic density constraint generally lies within the range $20<x_{\text{fo}}<30$; whence, the assumption of $\chi$ as a cold dark matter candidate is justified.

\subsection{Direct Detection}\label{dirdet}

Having discussed the implications of the Planck observations for the thermal relic density of our scenario's dark matter candidate, let us further investigate the limits inferred from the experiments for its potential direct detection. Given the heavy TeV nature of the $\chi$~WIMP, we shall focus on the results obtained by the LUX experiment \cite{LUX2013}, which currently define the most stringent constraints on potential direct detection of dark matter particles heavier than $\sim 100$~GeV.

\begin{figure}
\includegraphics[width=.25\textwidth]{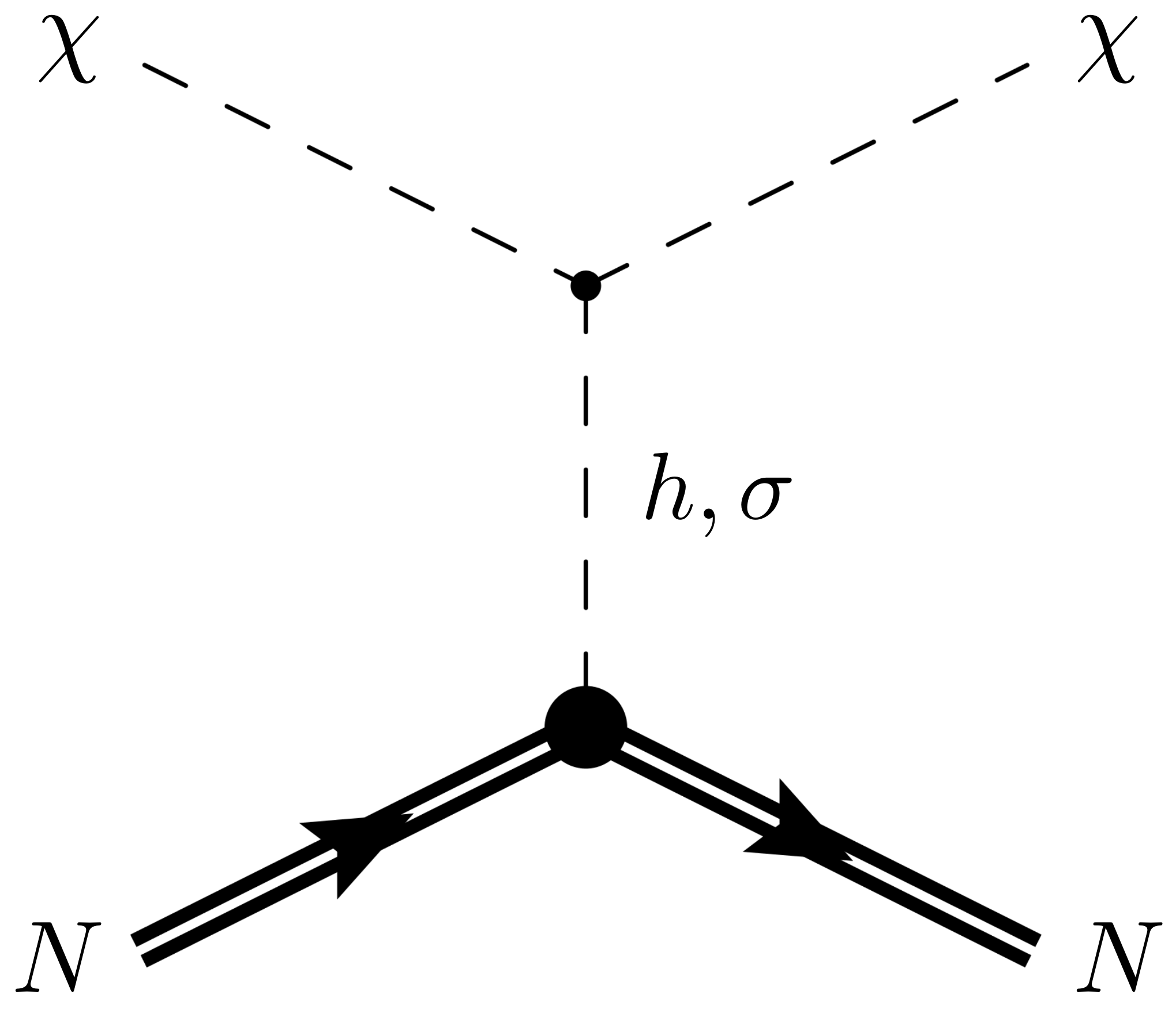}
\caption{Elastic scattering of a dark matter WIMP, $\chi$, off a nucleon, $N$. The process is mediated by the exchange of the $h$ and $\sigma$ scalars.}
\label{DMdir}
\end{figure}

Within the current framework, the $\chi$~WIMP interacts with the nucleons by exchanging the $h$ and $\sigma$ bosons in the $t$-channel, as illustrated in Fig.~\ref{DMdir}. The amplitude for the elastic scattering is given by the expression
\begin{equation}\label{DMdiramp}
i \mathcal{M}_{N\chi \to N\chi} = \frac{g_{W}}{2M_{W}}\, m_{N}f_{N} \tbrac{\frac{\lambda_{\chi\chi h}}{t-M_{h}^{2}}\cos \omega + \frac{\lambda_{\chi\chi\sigma}}{t-m_{\sigma}^{2}}\sin \omega} \bar{u}(p_{f}) u(p_{i}) \ ,
\end{equation}
where, $g_{W}$ is the weak coupling, $i\lambda_{\chi\chi h}$ and $i\lambda_{\chi\chi\sigma}$ are the couplings of their corresponding scalar mediators to the dark matter (see Appendix~\ref{FR}), $p_{i}$ and $p_{f}$ are the initial and final momenta of the nucleon, respectively, and $t$ is the Mandelstam variable representing the square of the exchanged momentum. For the nucleon mass, we take the average value of the proton and neutron masses, $m_{N} = 0.939$~GeV. The nucleon form factor, $f_{N}$, parametrizes the coupling of the SM Higgs, $\phi$ (c.f. \eqref{hs}), to the nucleon, and takes the approximate value $f_{N} \simeq 0.345$ \cite{Cline:2013gha,Agrawal:2010fh,Ellis:2000ds} (see also \cite{Crivellin:2013ipa} for further discussions).

Since the dark matter particle is, within the model, much heavier than the nucleon (i.e. $M_{\chi} \gg m_{N}$), the center of mass frame virtually coincides with the $\chi$ rest frame. Given that the typical momentum exchange for direct detection is below $\mathcal{O}$(GeV), the spin-independent cross section (appropriate for scalar dark matter) of the scattering process is easily obtained as
\begin{equation}\label{DMdirCS}
\sigma^{\text{SI}}_{N\chi \to N\chi} =\frac{g_{W}^{2}}{16\pi} \, \frac{m_{N}^{4}f_{N}^{2}}{M_{W}^{2} M_{\chi}^{2}} \tbrac{\frac{\lambda_{\chi\chi h}}{M_{h}^{2}}\cos \omega + \frac{\lambda_{\chi\chi\sigma}}{m_{\sigma}^{2}}\sin \omega}^{2} \ .
\end{equation}
As with the study of the thermal relic abundance \eqref{relic}, the spin-independent cross section \eqref{DMdirCS} is a function of the input parameters $\omega$, $\lambda_{m}^{-}$, $M_{\chi}$, and $M_{N}$ (c.f. \eqref{ms} and Appendix~\ref{FR}), and the experimentally determined results for this quantity, quoted by the LUX experiment at 90\%~C.L. \cite{LUX2013}, may be utilized to constrain, once again, the parameter space.

\begin{figure}
\includegraphics[width=.329\textwidth]{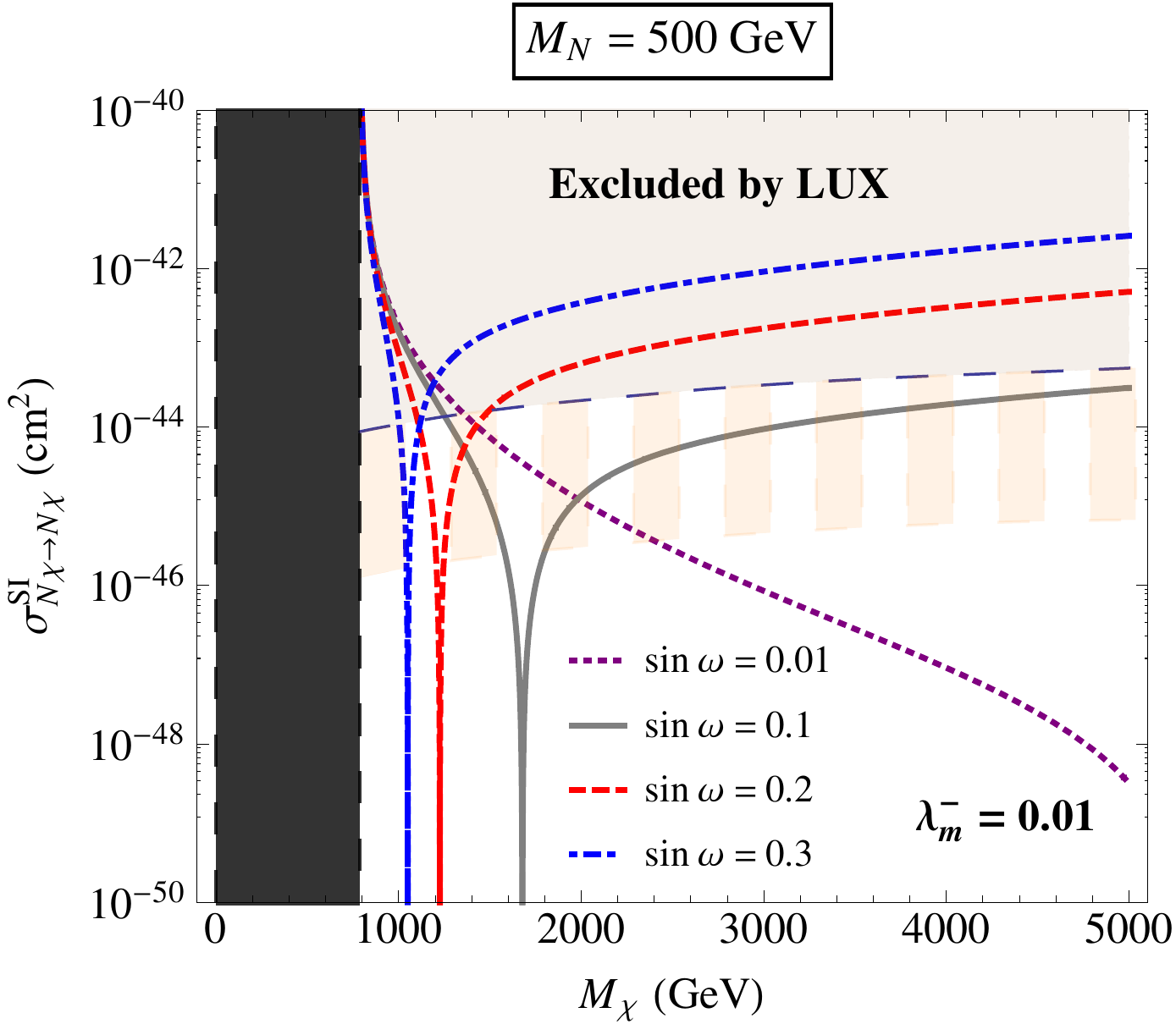}
\includegraphics[width=.329\textwidth]{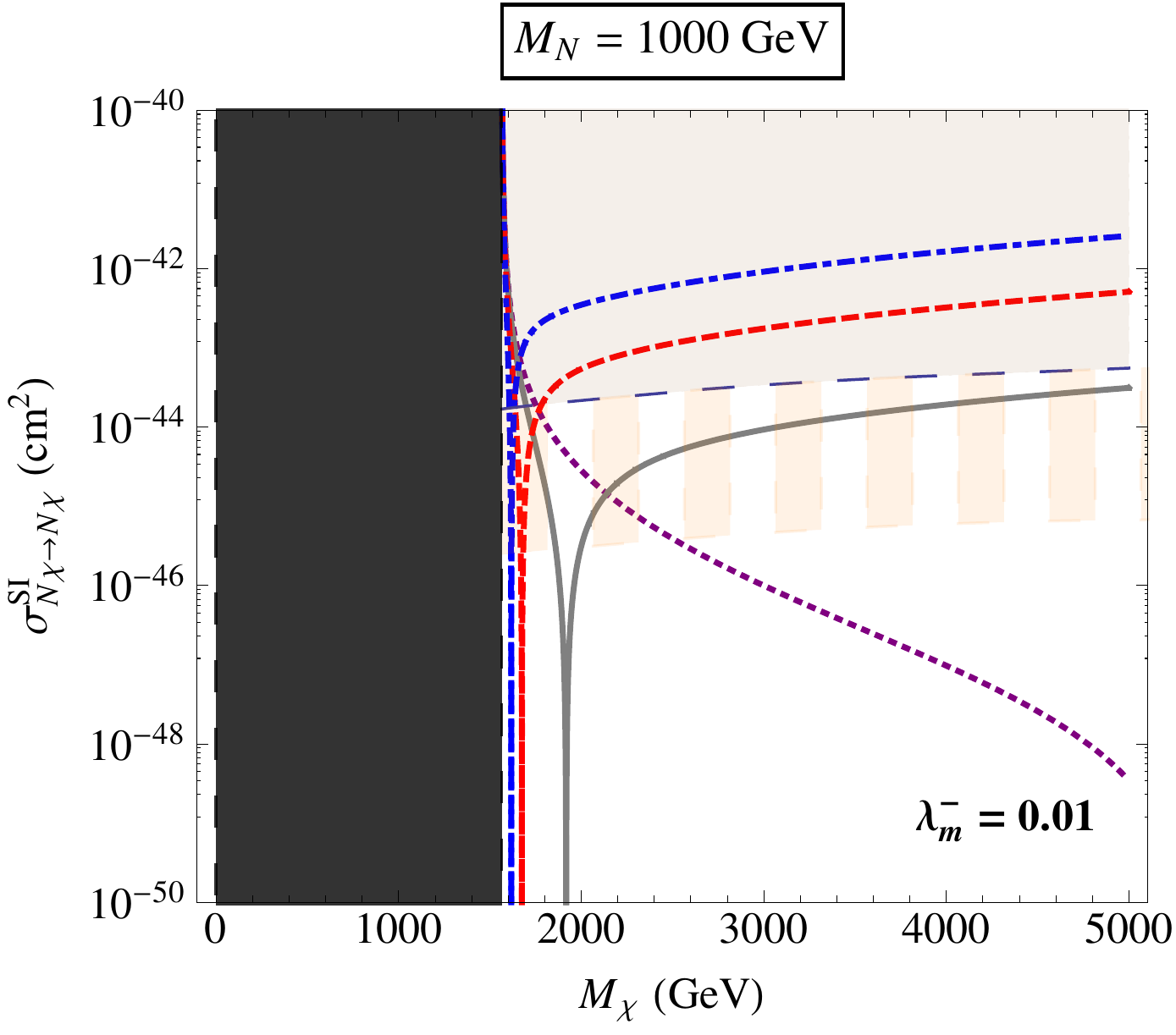}
\includegraphics[width=.329\textwidth]{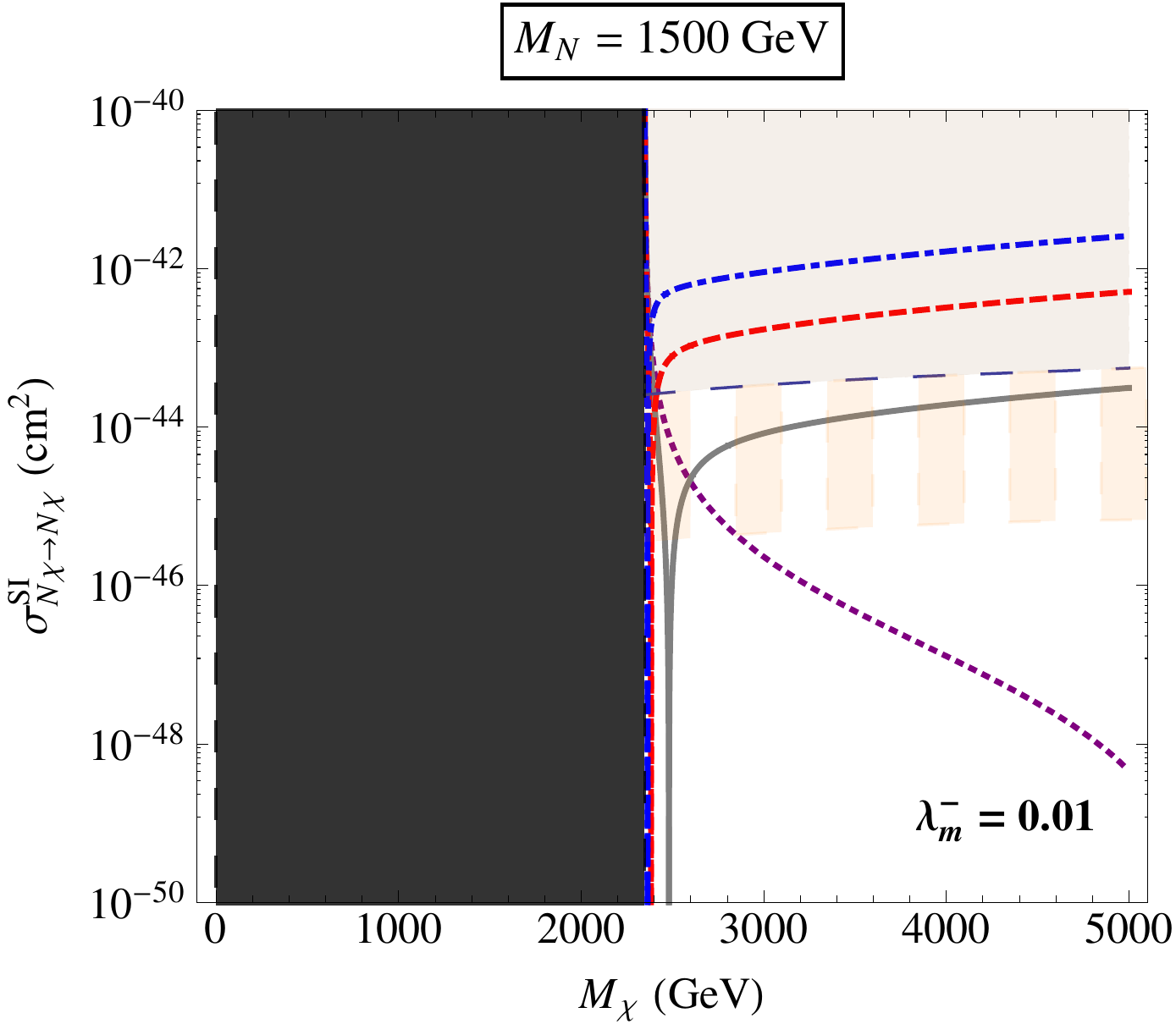}
\includegraphics[width=.329\textwidth]{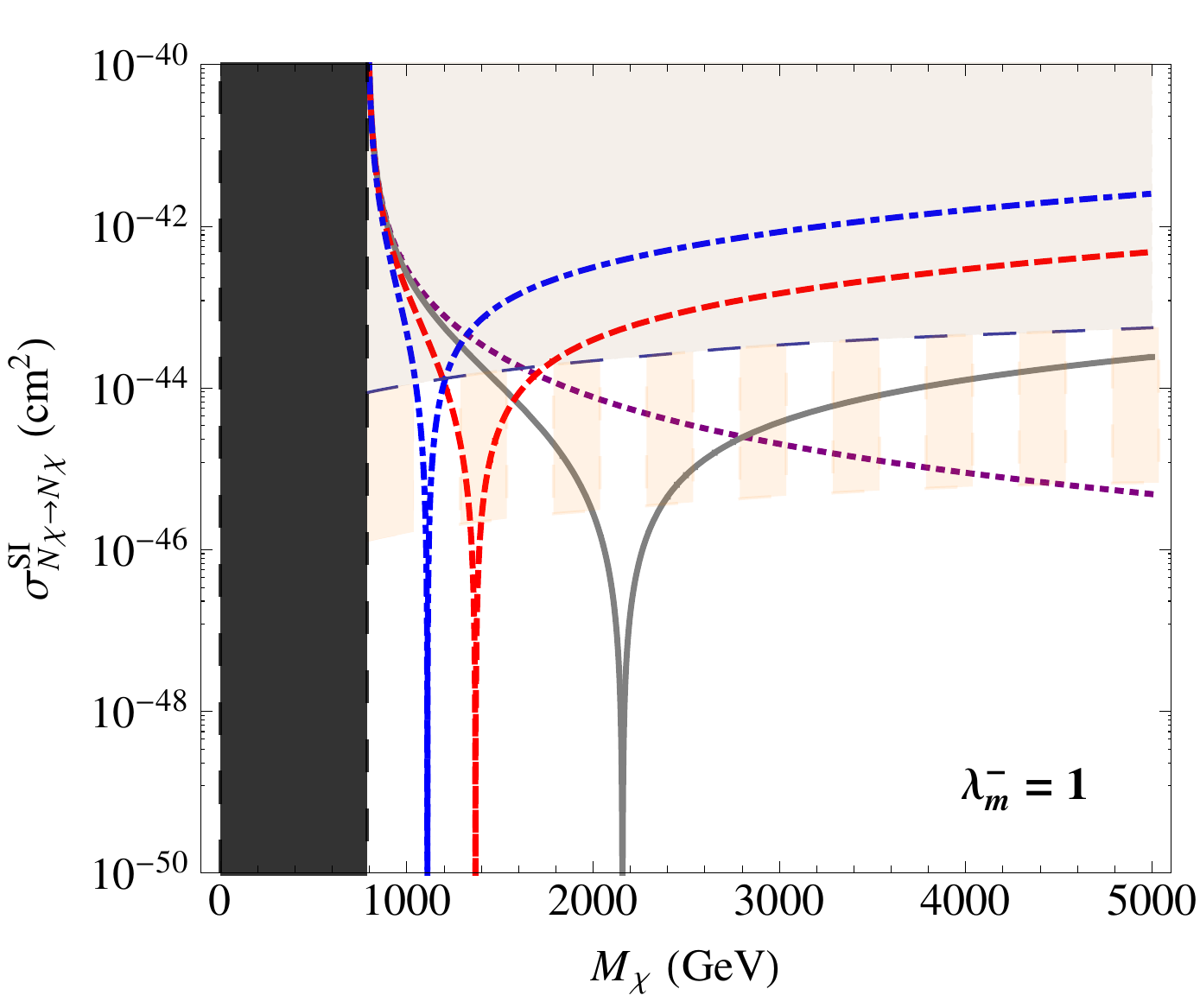}
\includegraphics[width=.329\textwidth]{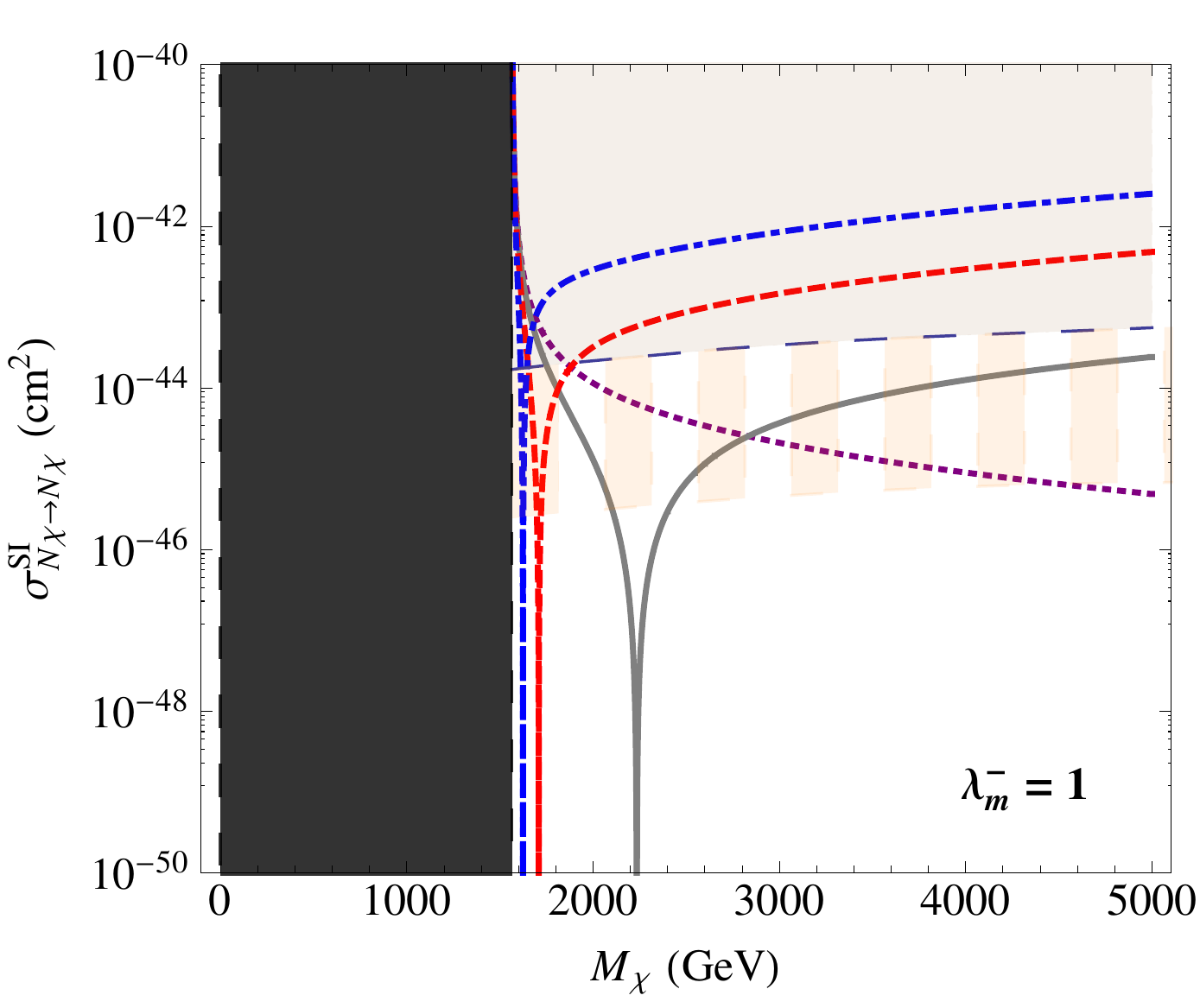}
\includegraphics[width=.329\textwidth]{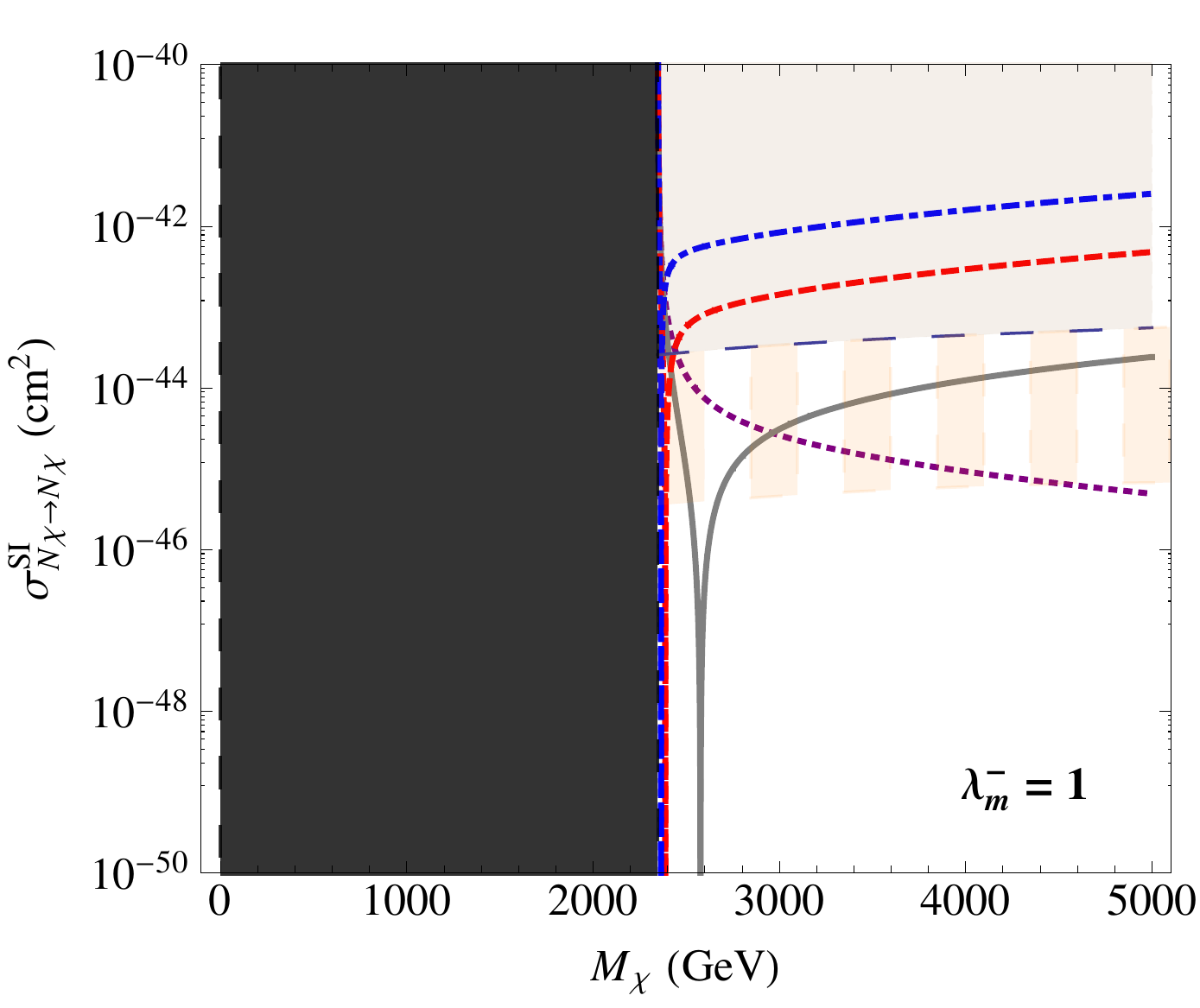}
\includegraphics[width=.329\textwidth]{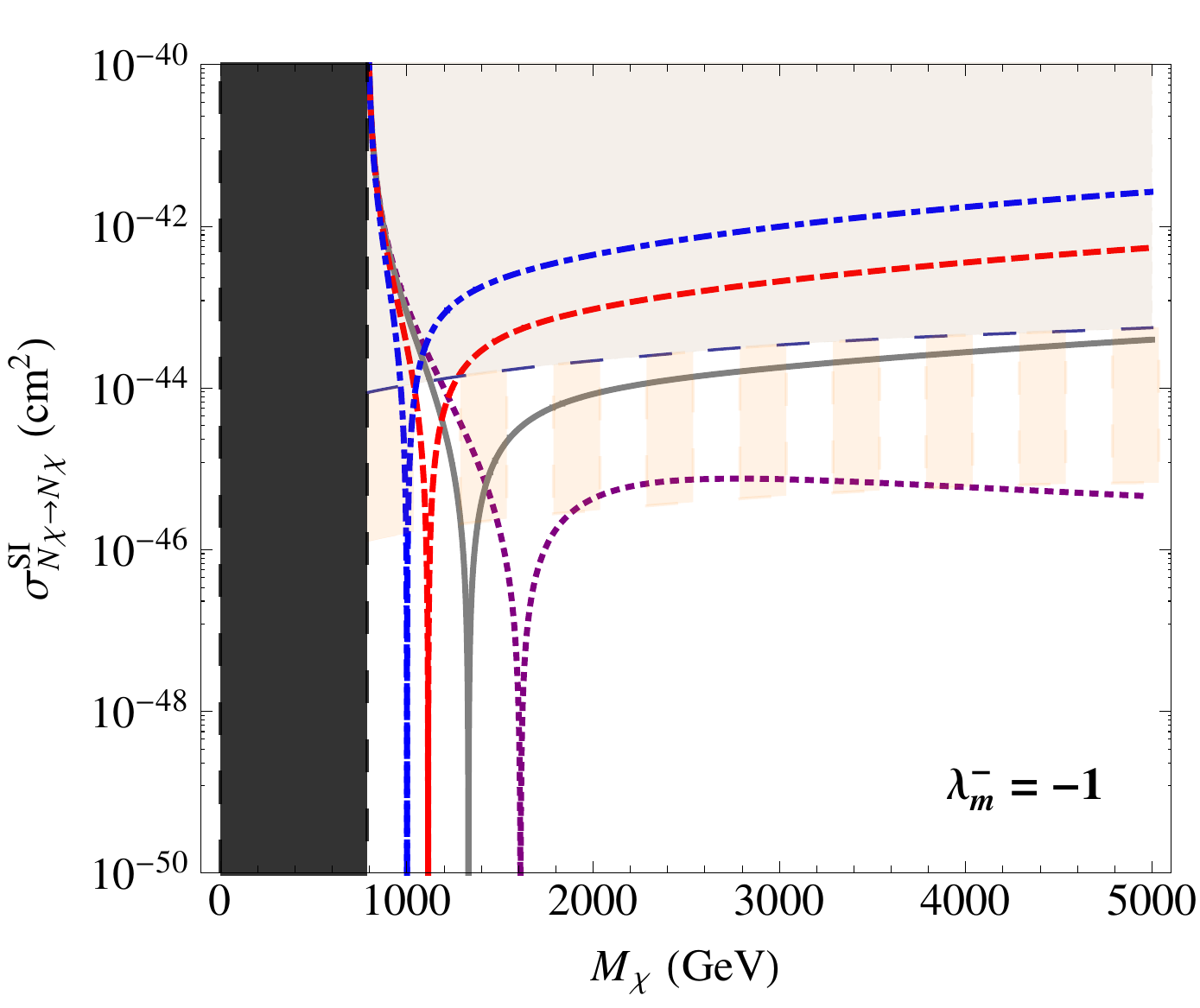}
\includegraphics[width=.329\textwidth]{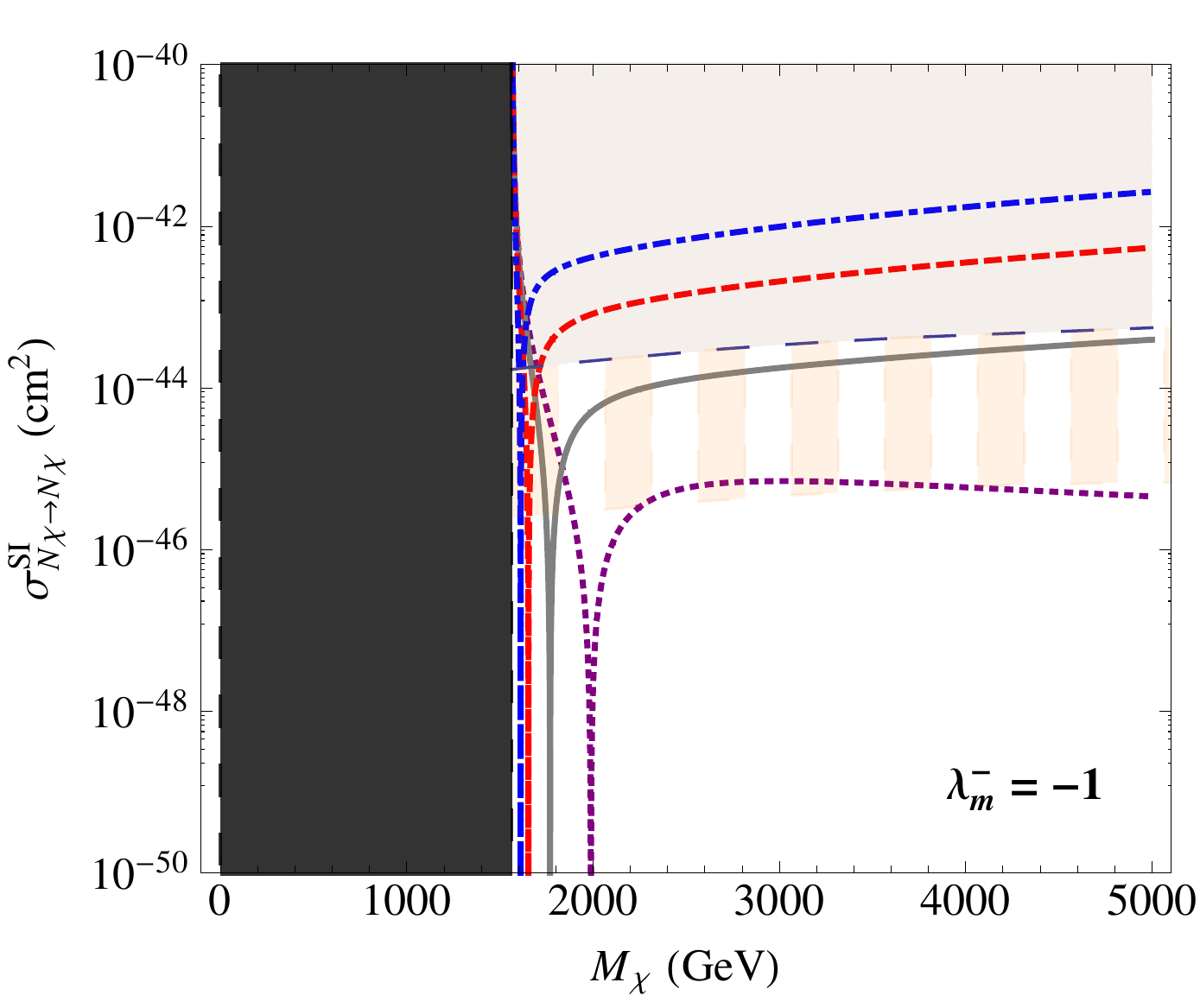}
\includegraphics[width=.329\textwidth]{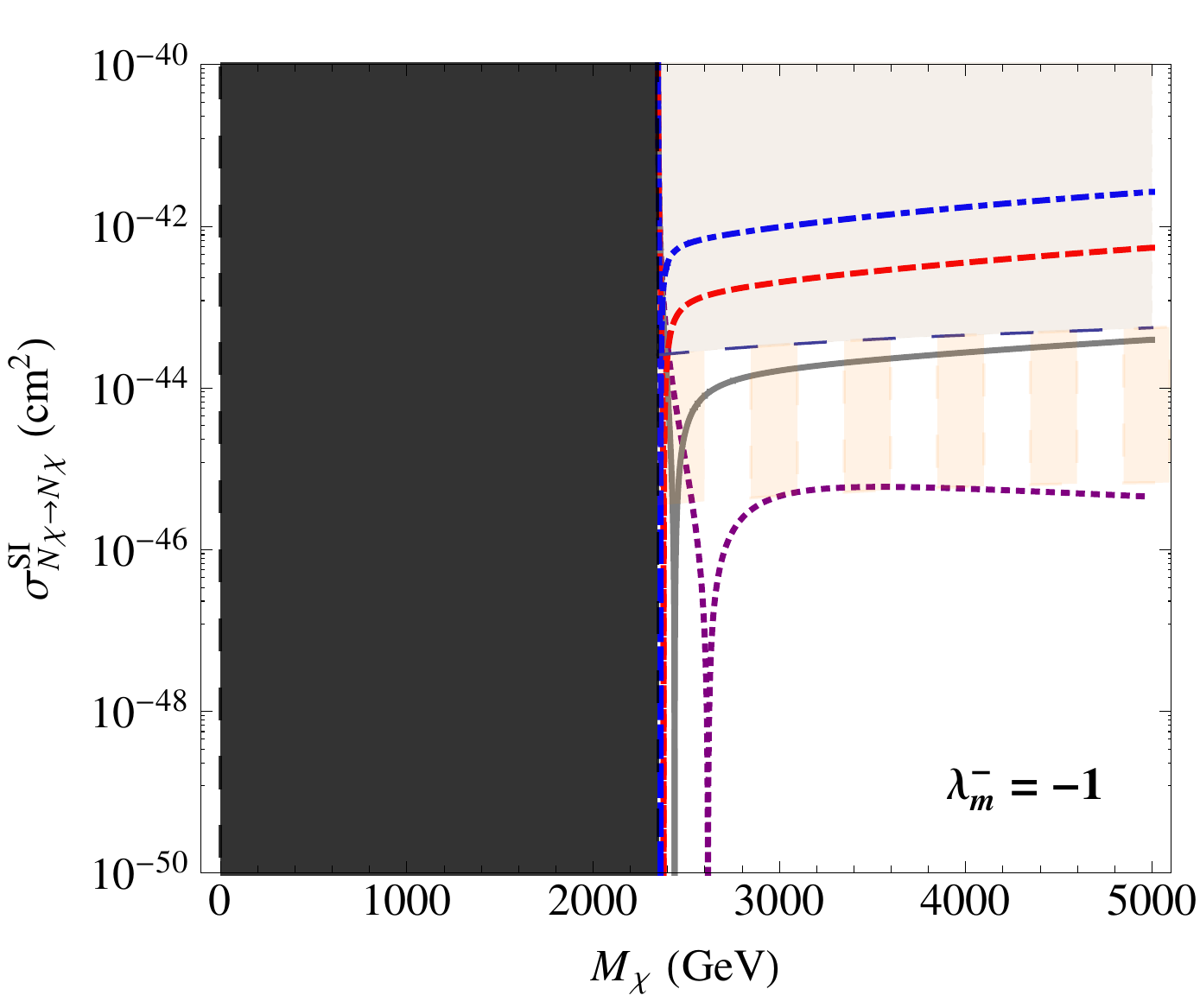}
\caption{Theoretically calculated curves of the WIMP-nucleon elastic scattering spin-independent cross section \eqref{DMdirCS} as a function of the WIMP mass, for four representative values of the mixing angle motivated by the experimental constraints. The panels correspond to benchmark values of the right-handed neutrino mass, $M_{N}$ (columns), and the input parameter $\lambda_{m}^{-}$ (rows). In addition, the exclusion bound from the LUX direct detection experiment \cite{LUX2013} is displayed (dashed), setting the current upper limit on the scattering cross section. The solid black region, inferred from the stability condition of the one-loop potential \eqref{staboneloop}, determines the formal lower bound on the WIMP mass, $M_{\chi}$, for each selected value of $M_{N}$. Furthermore, the shaded region indicates the projected exclusion bound from the future Xenon1T experiment \cite{Aprile:2012zx}.}
\label{dirpix}
\end{figure}

The results are depicted in Fig.~\ref{dirpix}, where the theoretical curve of the elastic scattering spin-independent cross section \eqref{DMdirCS} is shown as a function of the WIMP mass for $M_{\chi} \leq 5$~TeV, together with the cross section upper bound at 90\%~C.L. as reported by the LUX direct detection experiment \cite{LUX2013}. For comparison, the projected upper limits of the scattering cross section from the future Xenon1T experiment \cite{Aprile:2012zx} are also indicated, which, in the absence of a positive detection signal is expected to further reduce this upper bound by about two orders of magnitude at 90\%~C.L.\footnote{For the purpose of the current illustration, we extrapolate the Xenon1T projections up to 5~TeV (see also \url{http://dendera.berkeley.edu/plotter/entryform.html}).} The curves within each panel correspond to several experimentally motivated values of the mixing angle, and the panels represent various choices of the remaining input parameters, $M_{N}$ and $\lambda_{m}^{-}$. Once more, larger mixings are disfavored by the experimental data for heavier WIMP masses. Nevertheless, in the lower mass region, the cross section drops once there is a cancellation between the two competing scalar mediator channels in \eqref{DMdirCS}. Using the explicit form of the couplings (Appendix~\ref{FR}), one can show that this cancellation occurs once the following relation between the free parameters is satisfied
\begin{equation}\label{dirrel}
\lambda_{m}^{-} \sim \frac{2 M_{\chi}^{2}}{v_{\phi}^{2}} \tbrac{1-\frac{M_{h}^{2}}{m_{\sigma}^{2}}} \sin^{2} \omega \ ,
\end{equation}
opening up a window of compatibility for the larger values of the mixing angle with the observations. Interestingly, one notes that the sign of $\lambda_{m}^{-}$ is highly relevant in this analysis, since it plays a crucial role in the mentioned cancellation.

\section{Discussion}\label{disc}

In this section, we provide a concise summary of our study, by combining the results from collider Higgs searches and dark matter analyses into unified exclusion plots of the model's parameter space. In addition, we incorporate the previously analyzed \cite{Farzinnia:2013pga} experimental results from the electroweak precision tests and direct measurements of the 125~GeV $h$~Higgs at the LHC, as well as formal bounds obtained from stability of the potential and perturbative unitarity considerations.

\begin{figure}
\includegraphics[width=.329\textwidth]{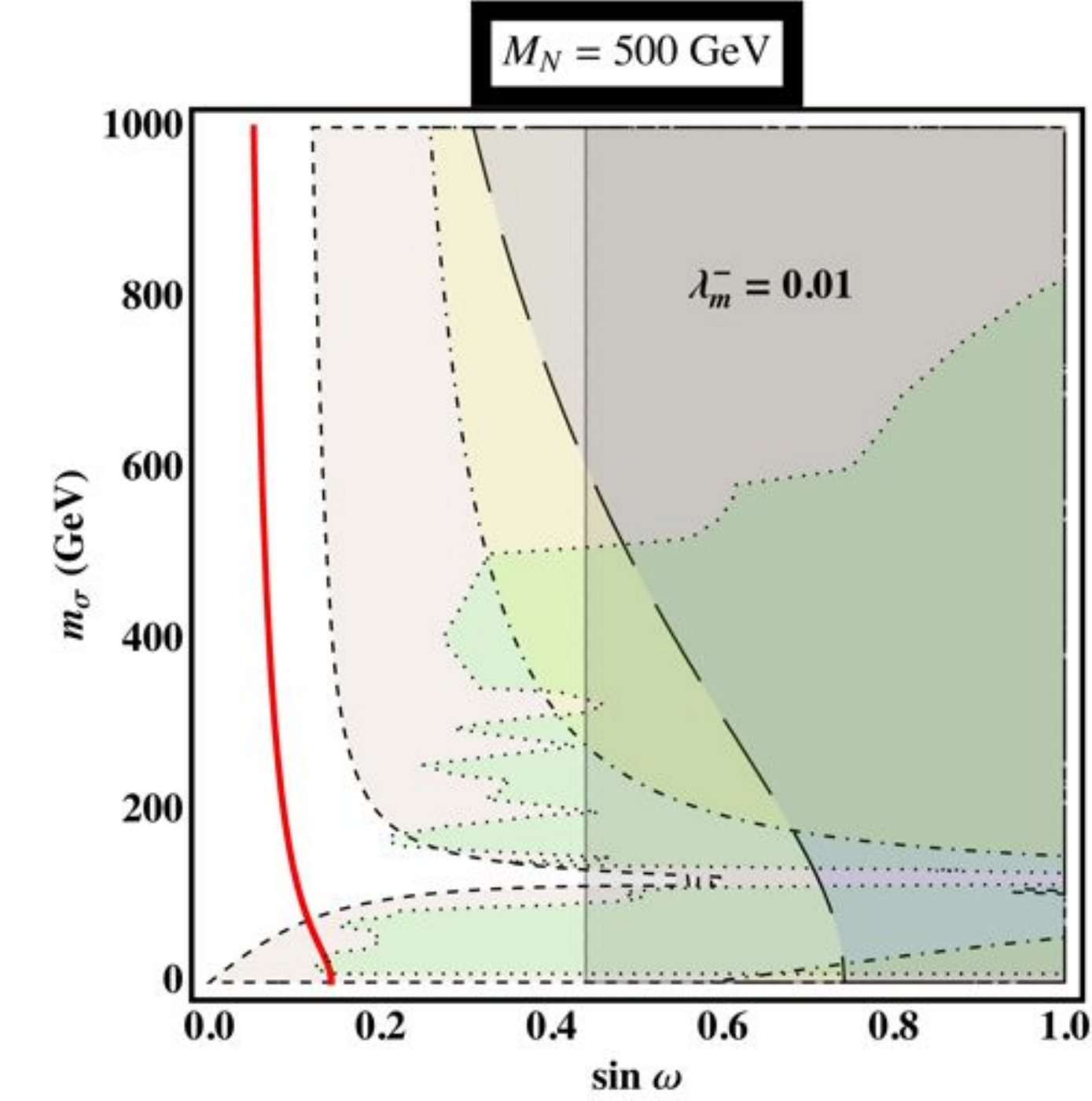}
\includegraphics[width=.329\textwidth]{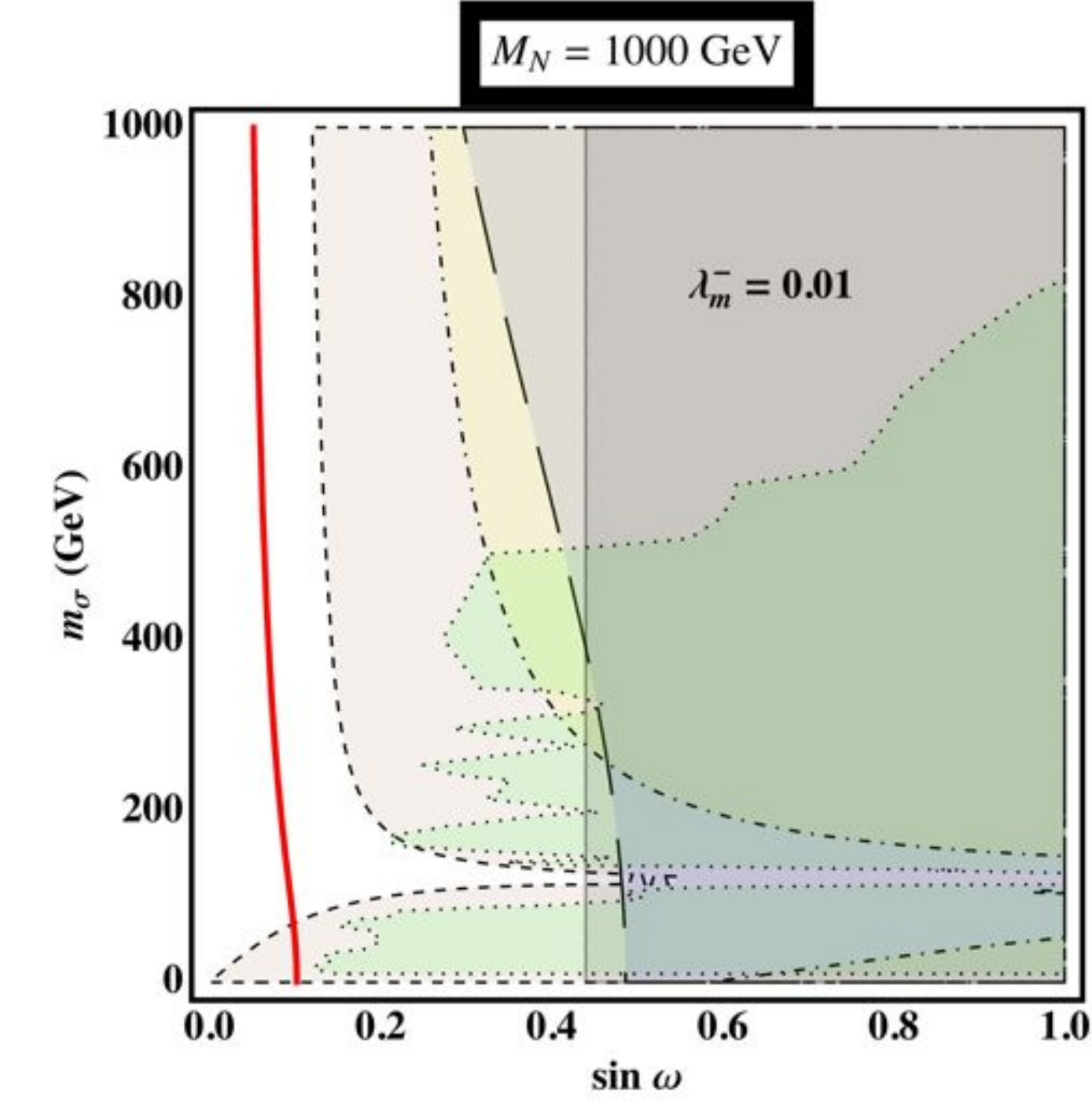}
\includegraphics[width=.329\textwidth]{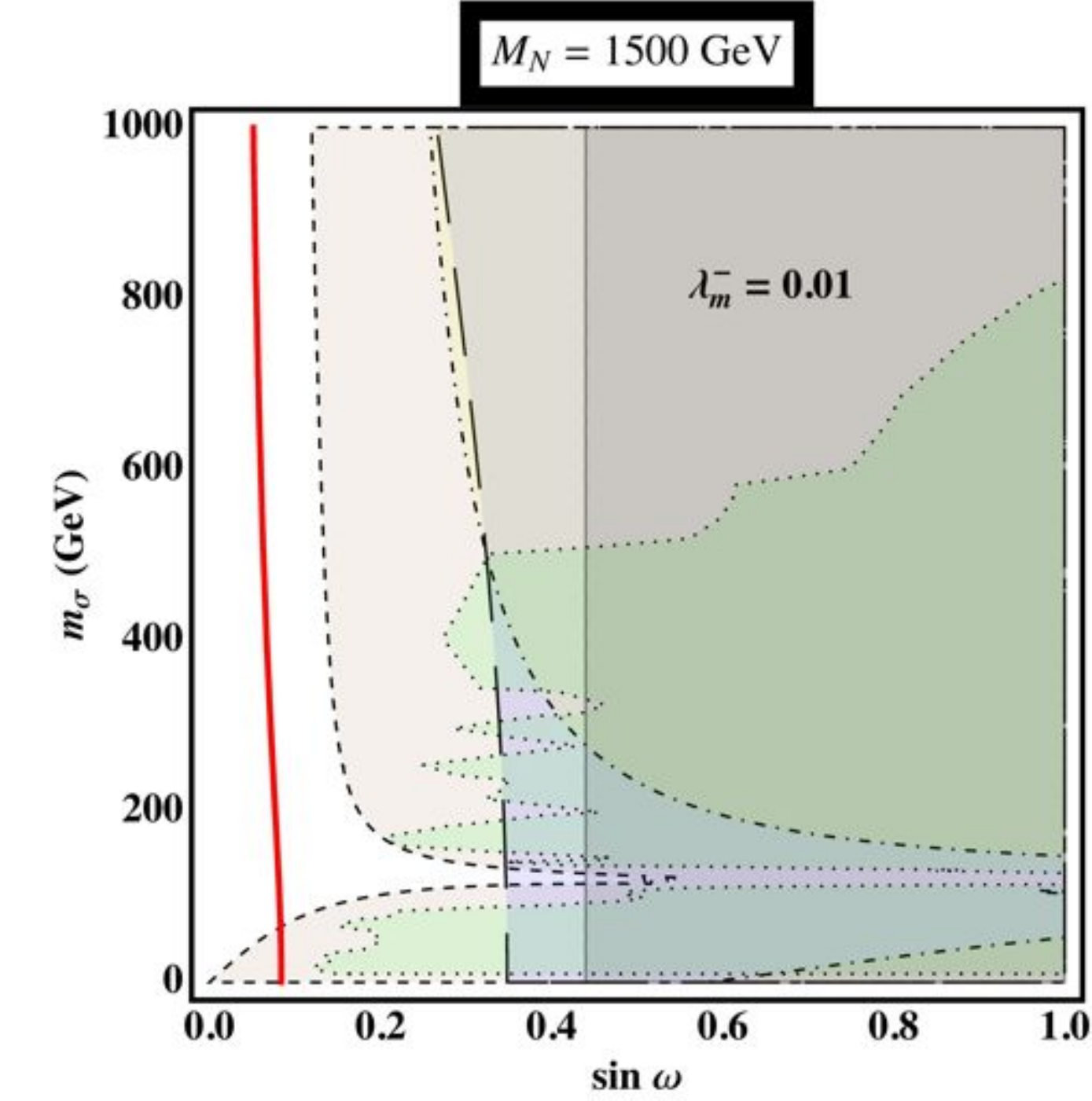}
\includegraphics[width=.329\textwidth]{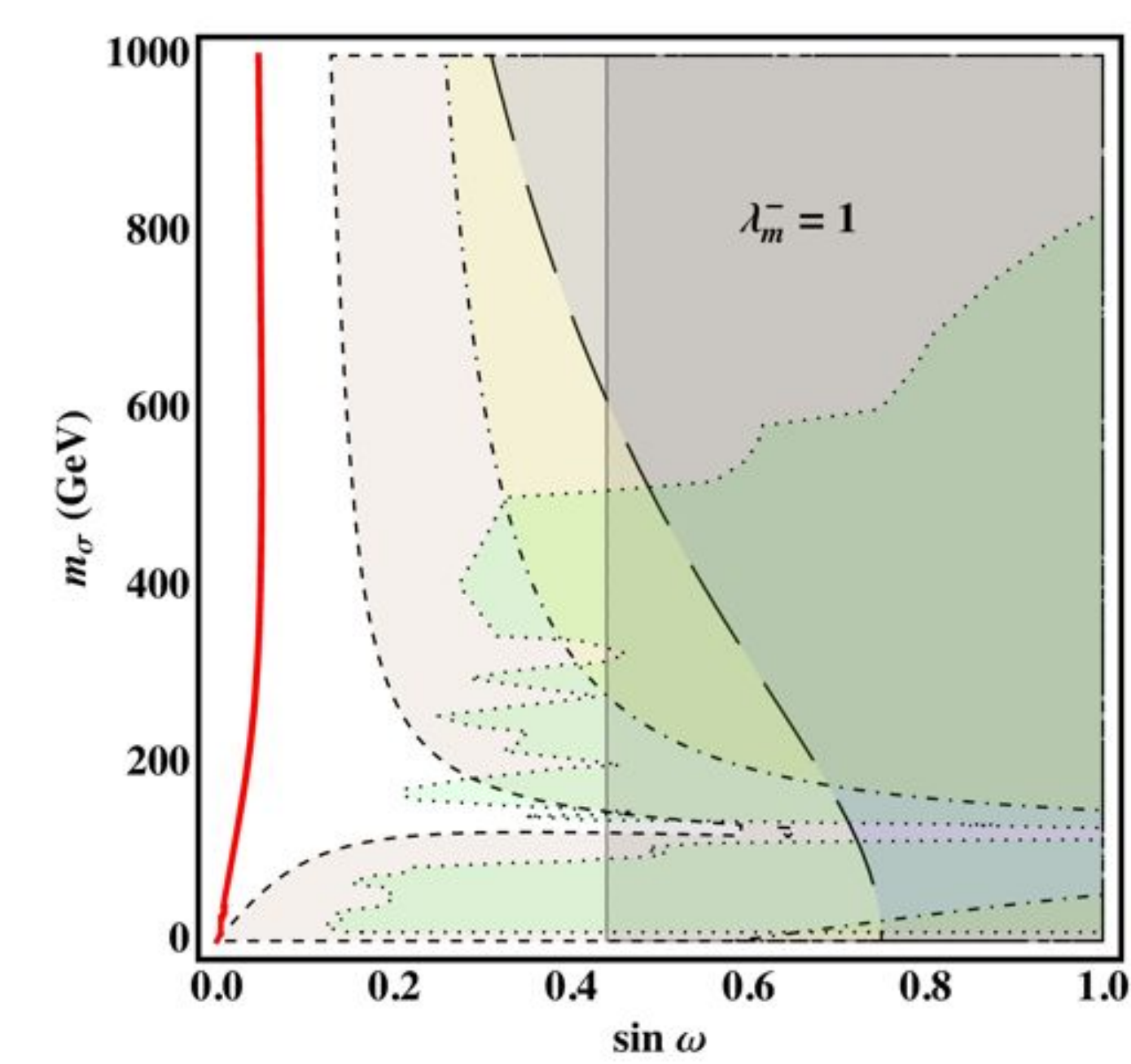}
\includegraphics[width=.329\textwidth]{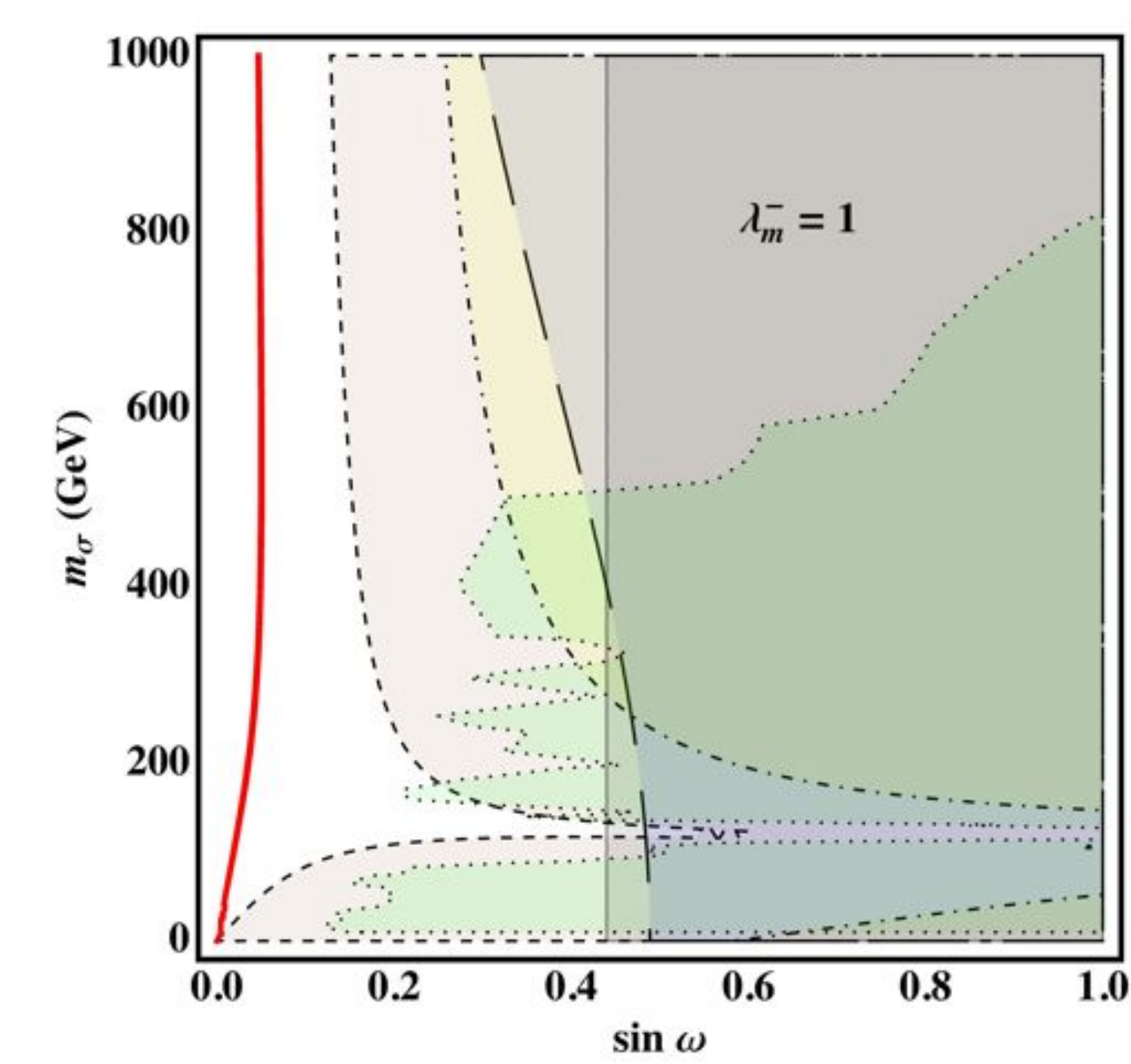}
\includegraphics[width=.329\textwidth]{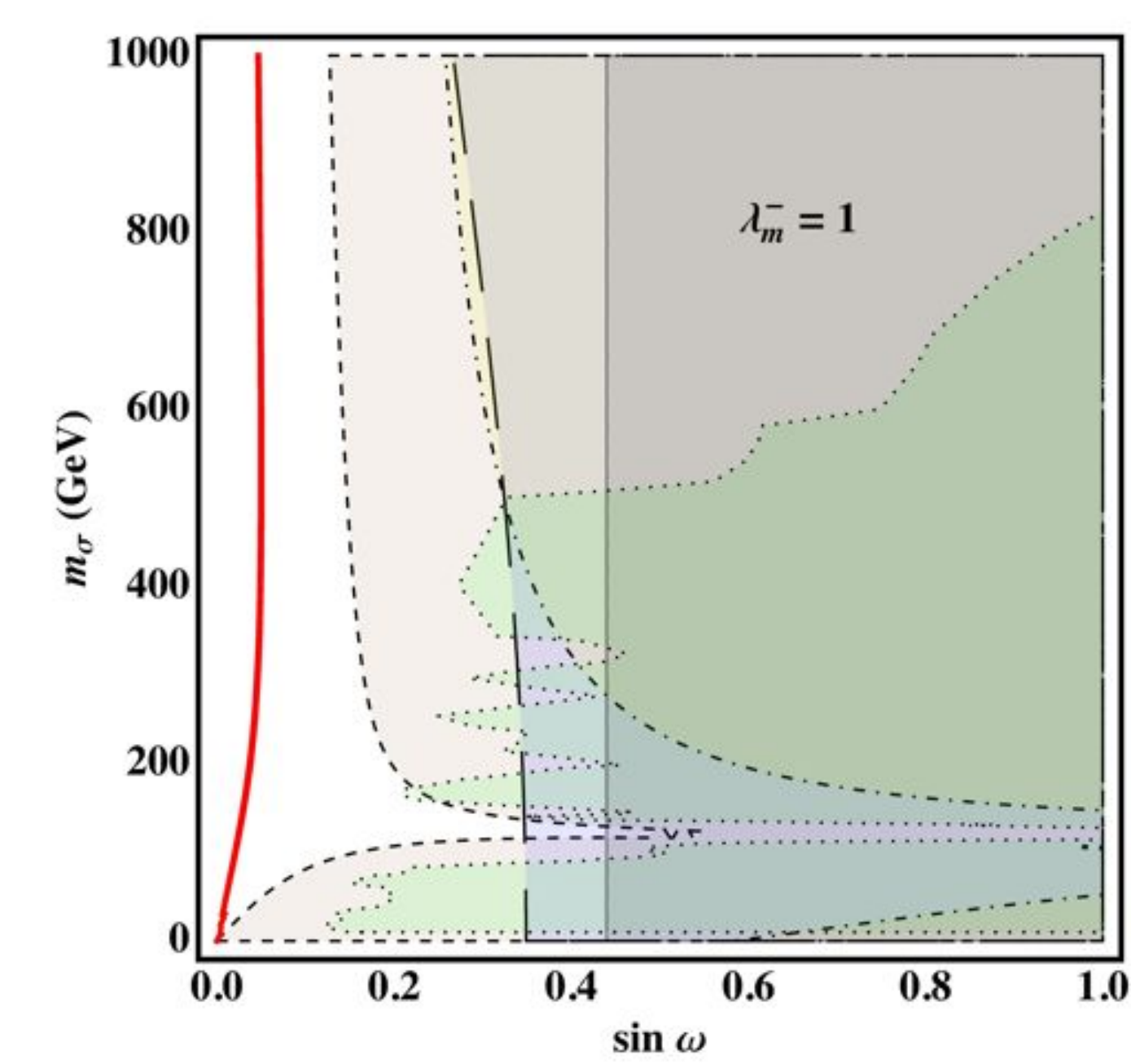}
\includegraphics[width=.329\textwidth]{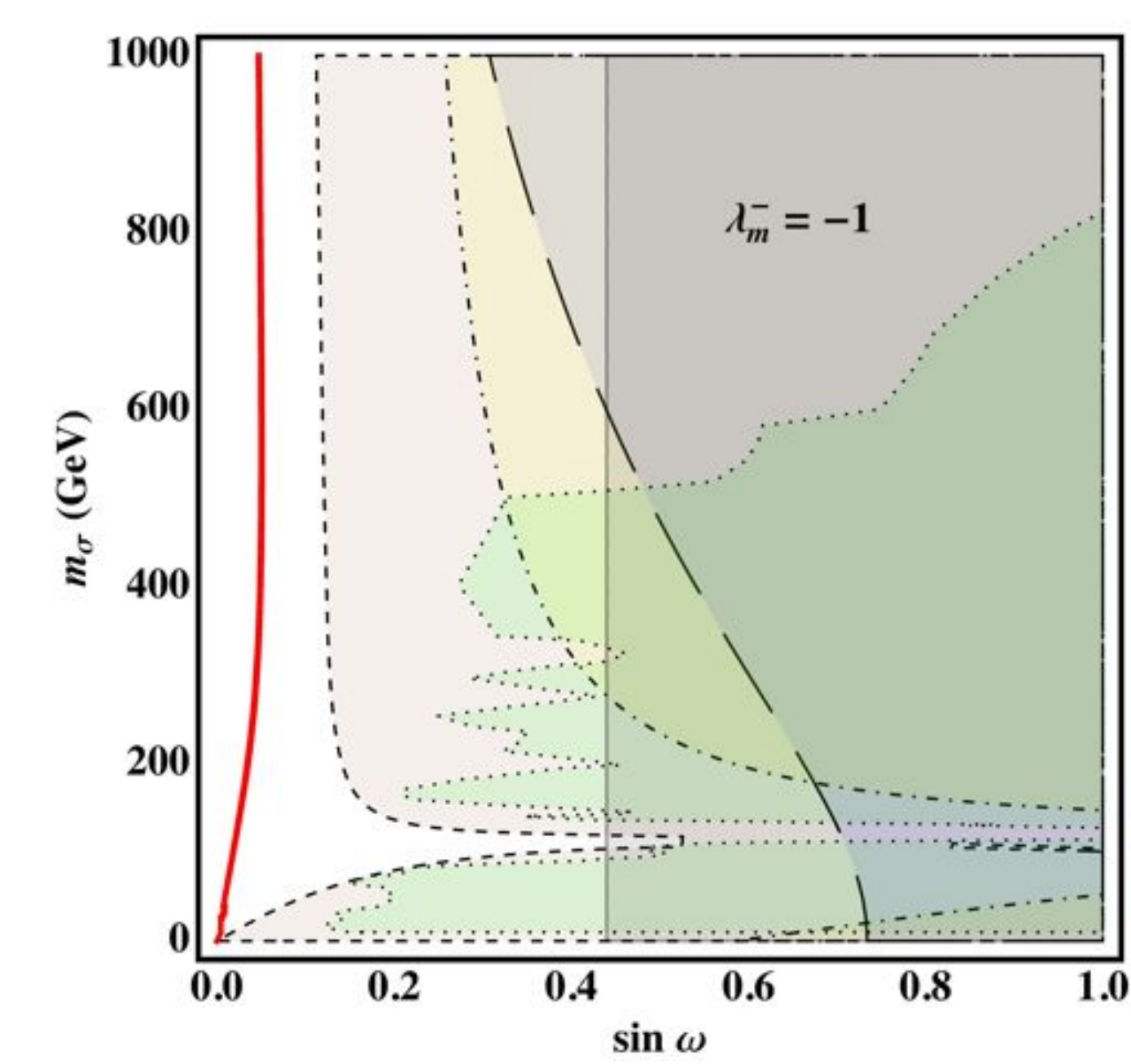}
\includegraphics[width=.329\textwidth]{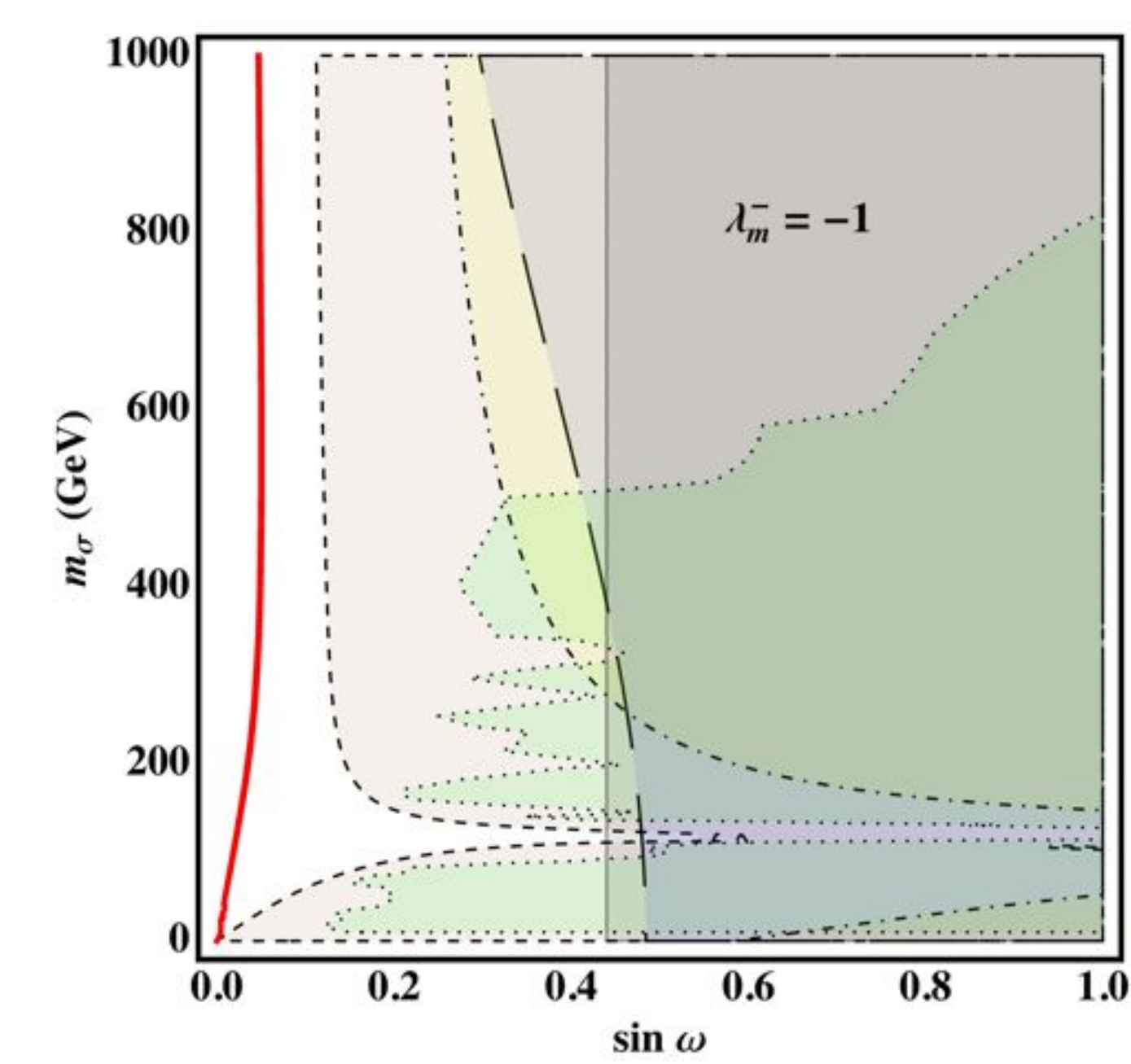}
\includegraphics[width=.329\textwidth]{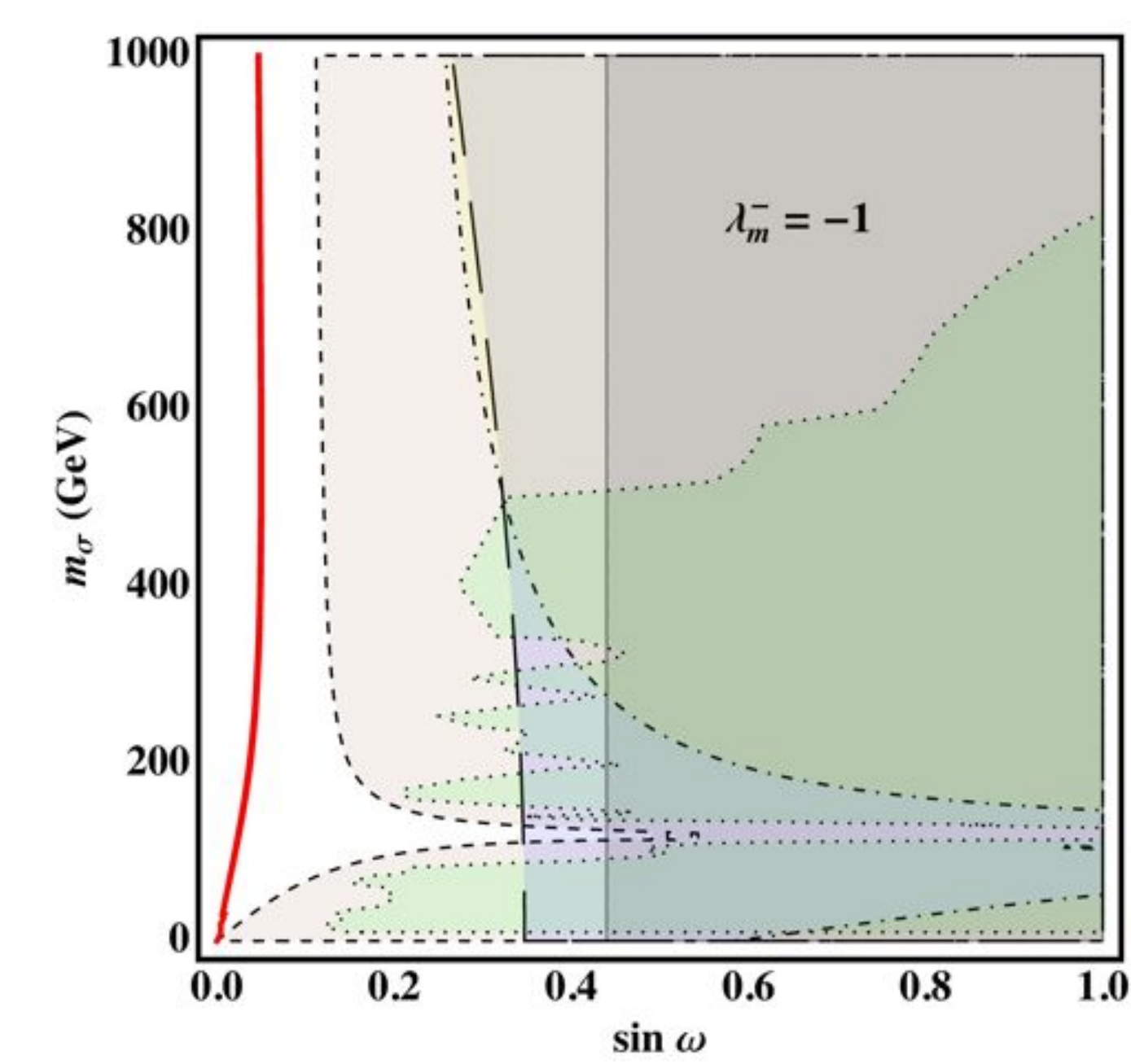}
\caption{Constraining the $\sin\omega - m_{\sigma}$ plane for various choices of the parameters $\lambda_{m}^{-}$ and $M_{N}$. All colored regions are excluded. The experimental constraints are, at 95\%~C.L., derived from the electroweak precision tests (dot-dashed), direct measurements of the LHC 125~GeV Higgs' properties (solid), and the LEP and LHC Higgs searches (dotted). The formal perturbative unitarity bound (long-dashed) is also depicted. In all panels, the parameter space is, nonetheless, most severely restricted by the LUX direct detection data (short-dashed) at 90\%~C.L., permitting only small mixings, and setting a lower limit on the $\sigma$~mass. The thick (red) band, within the allowed region, corresponds to satisfying the observational value of the WIMP relic abundance within the $1\sigma$ uncertainty quoted by the Planck collaboration.}
\label{msom}
\end{figure}

\begin{figure}
\includegraphics[width=.329\textwidth]{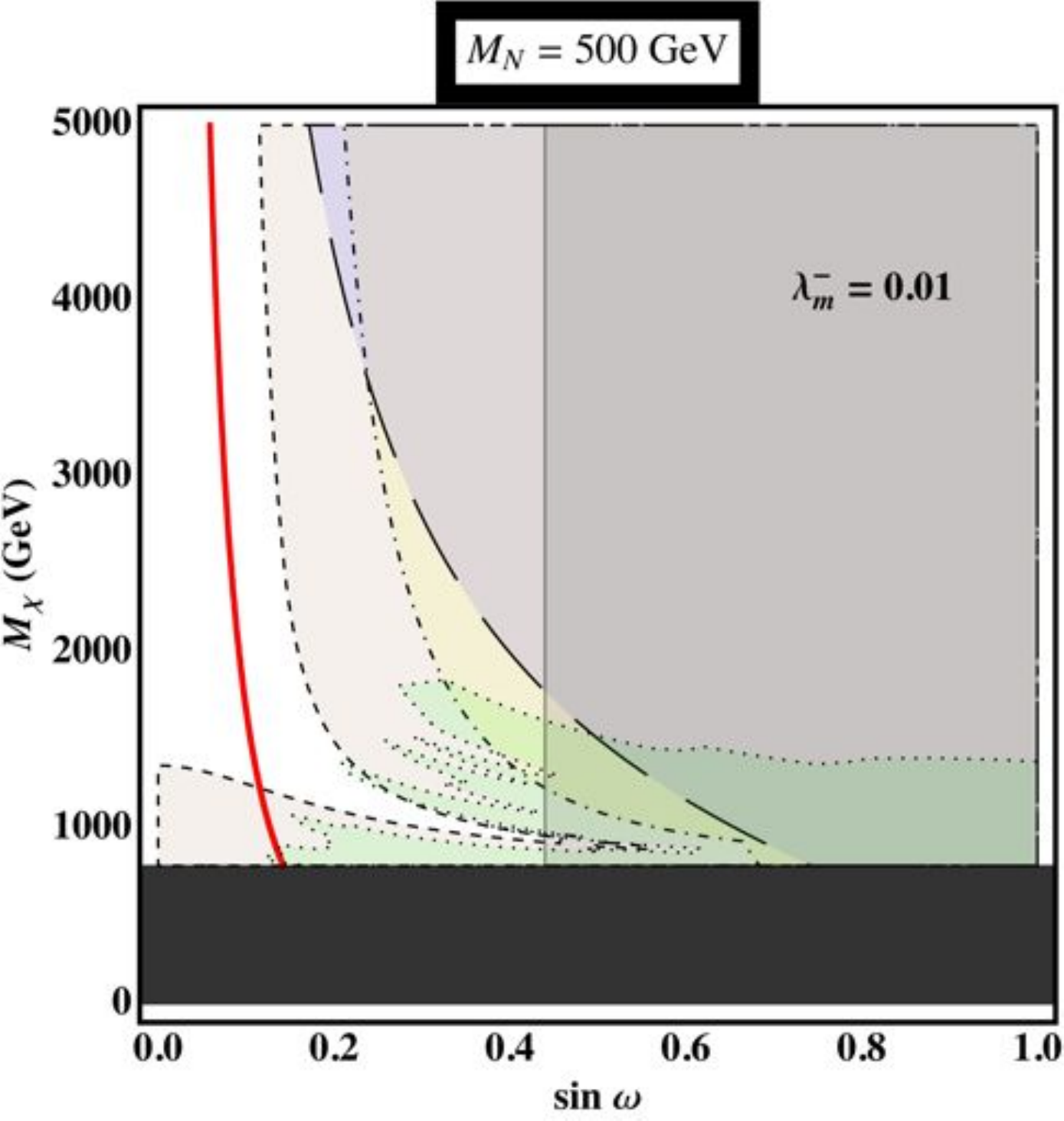}
\includegraphics[width=.329\textwidth]{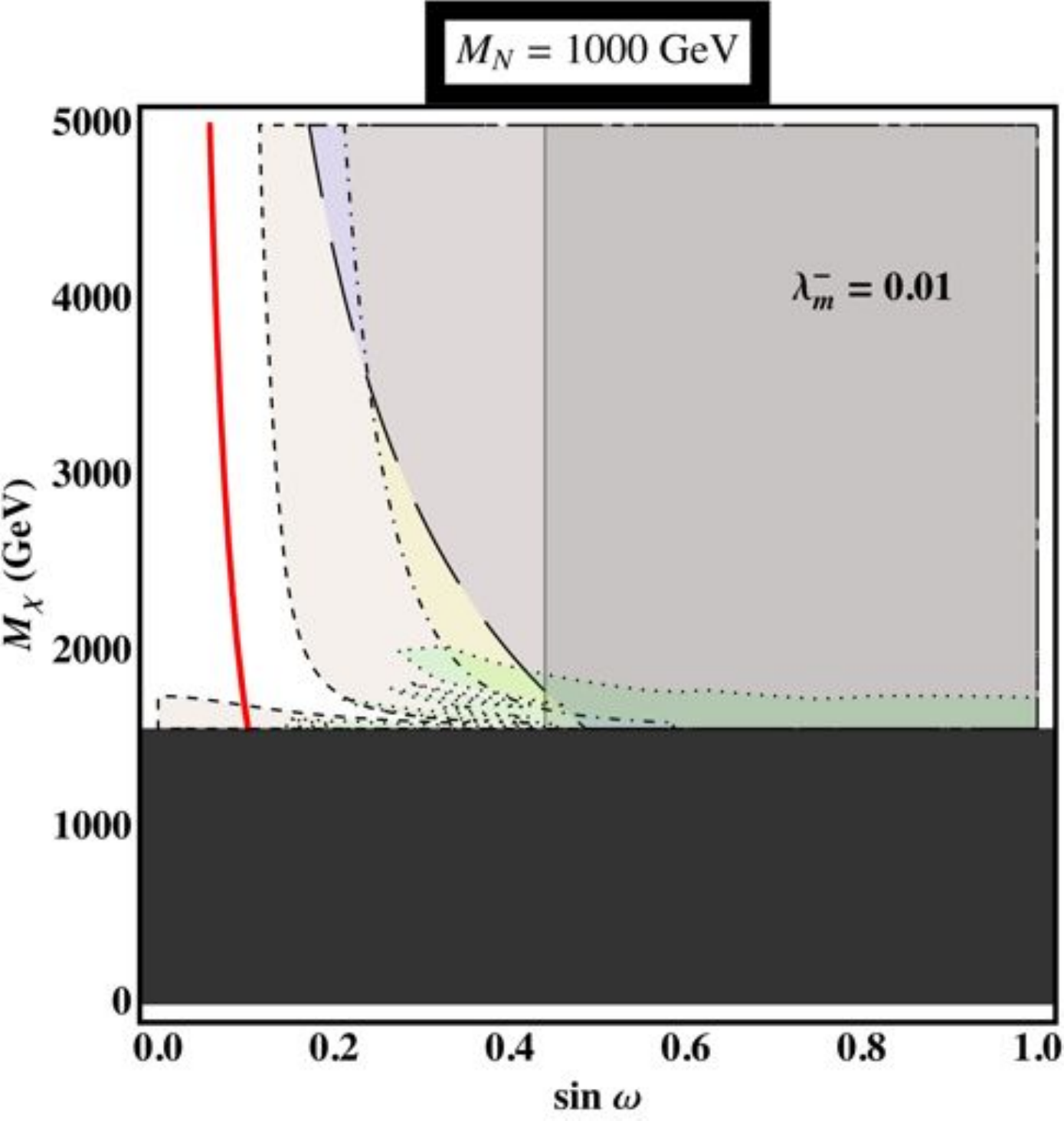}
\includegraphics[width=.329\textwidth]{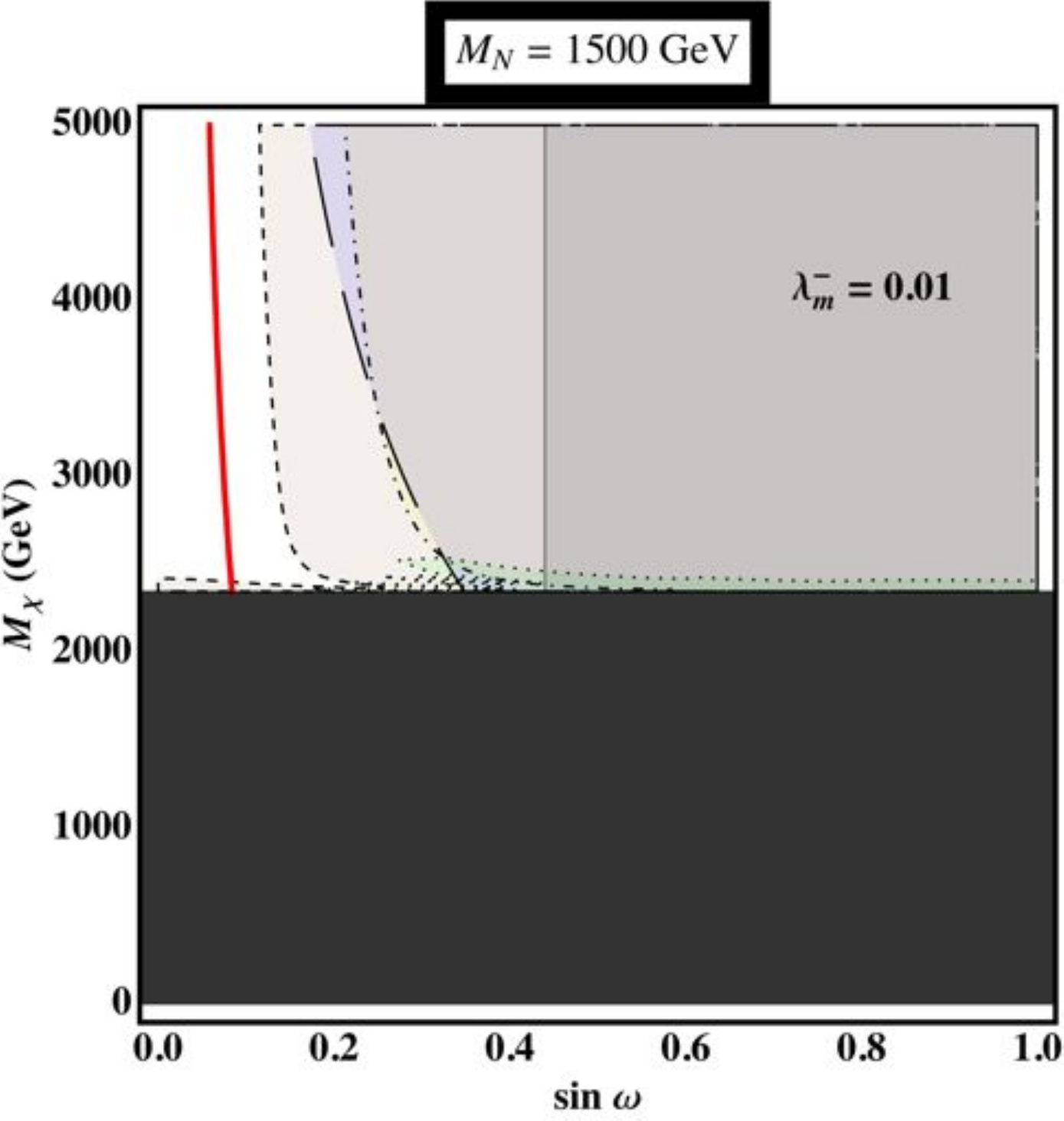}
\includegraphics[width=.329\textwidth]{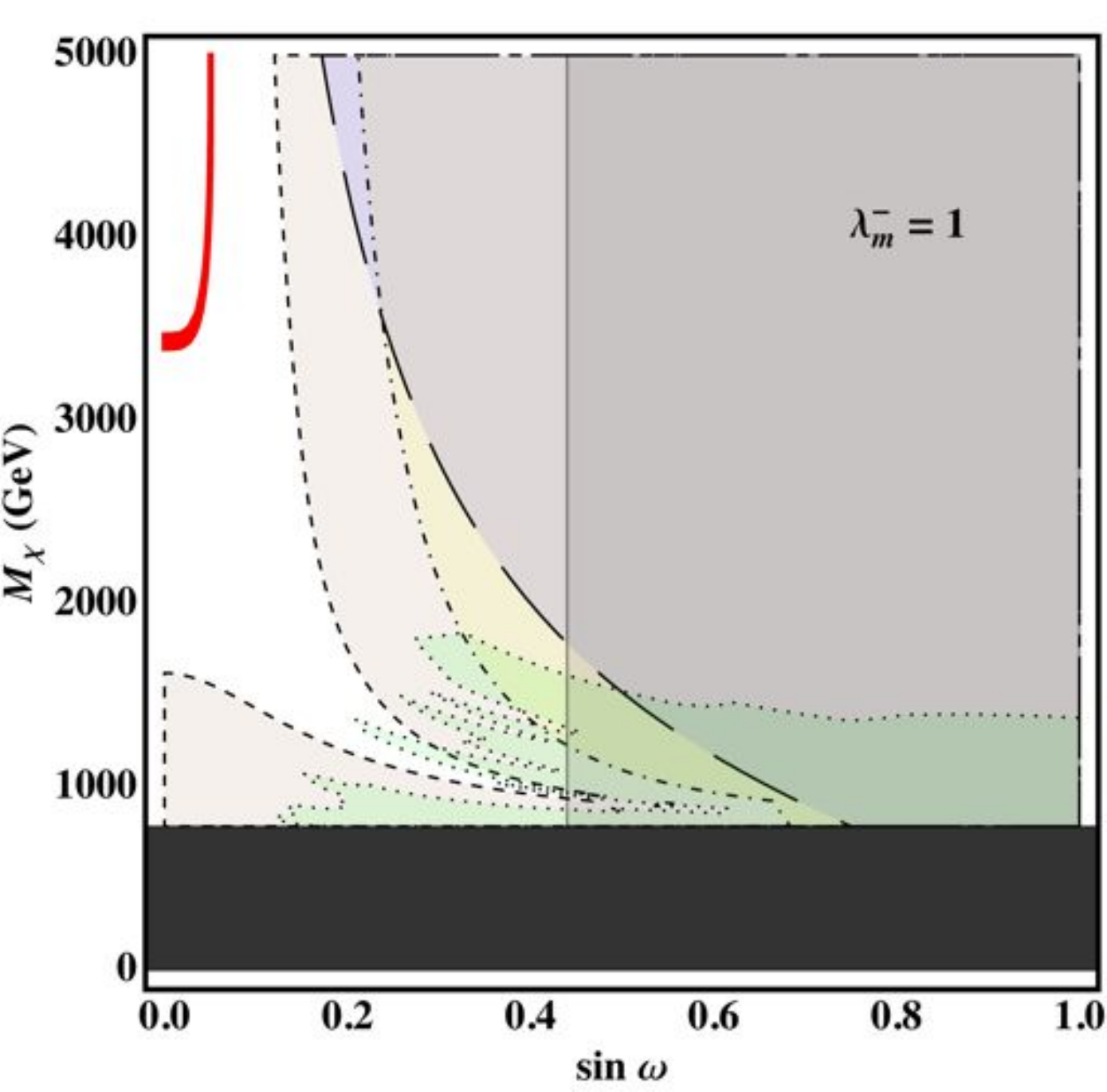}
\includegraphics[width=.329\textwidth]{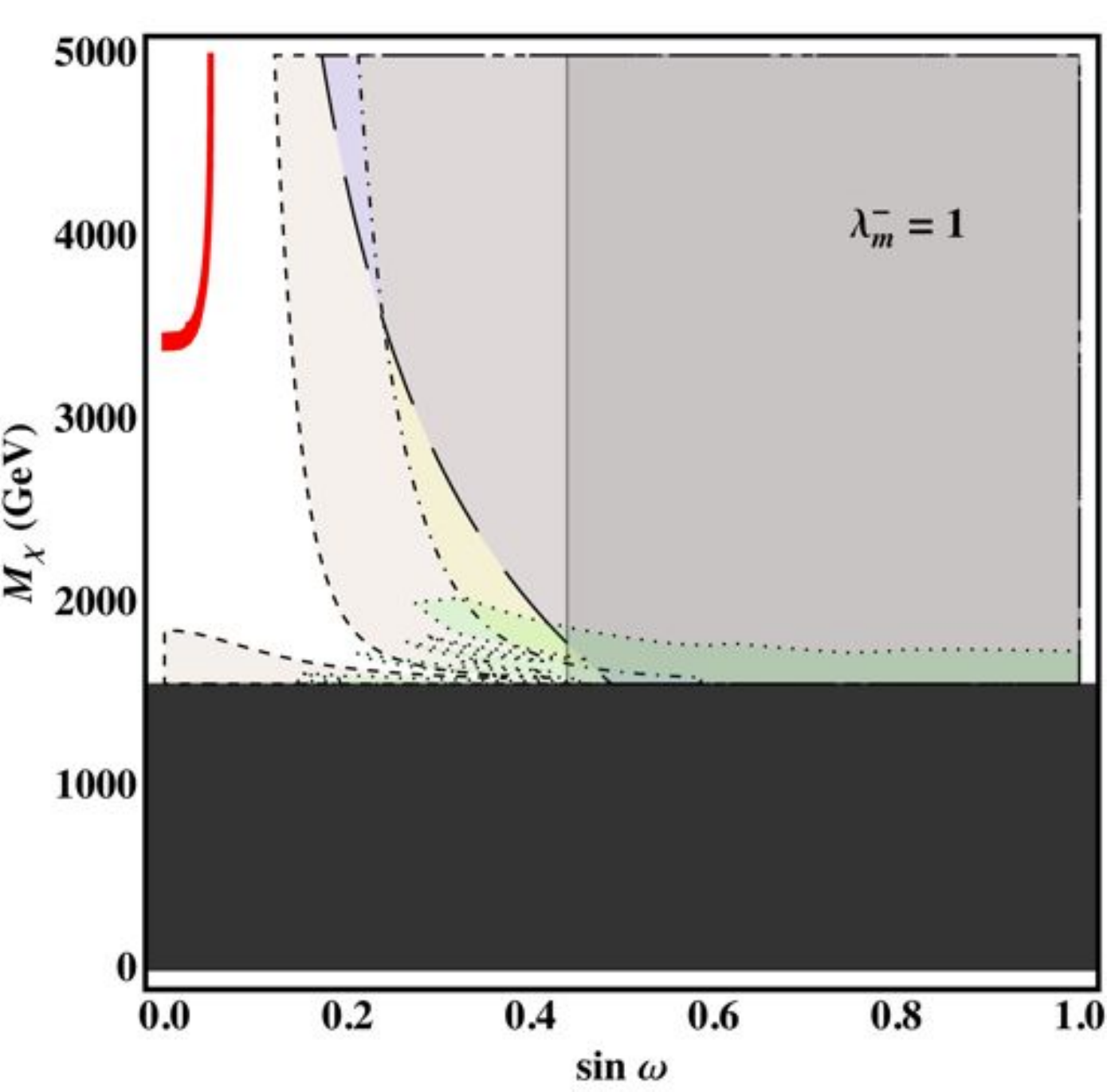}
\includegraphics[width=.329\textwidth]{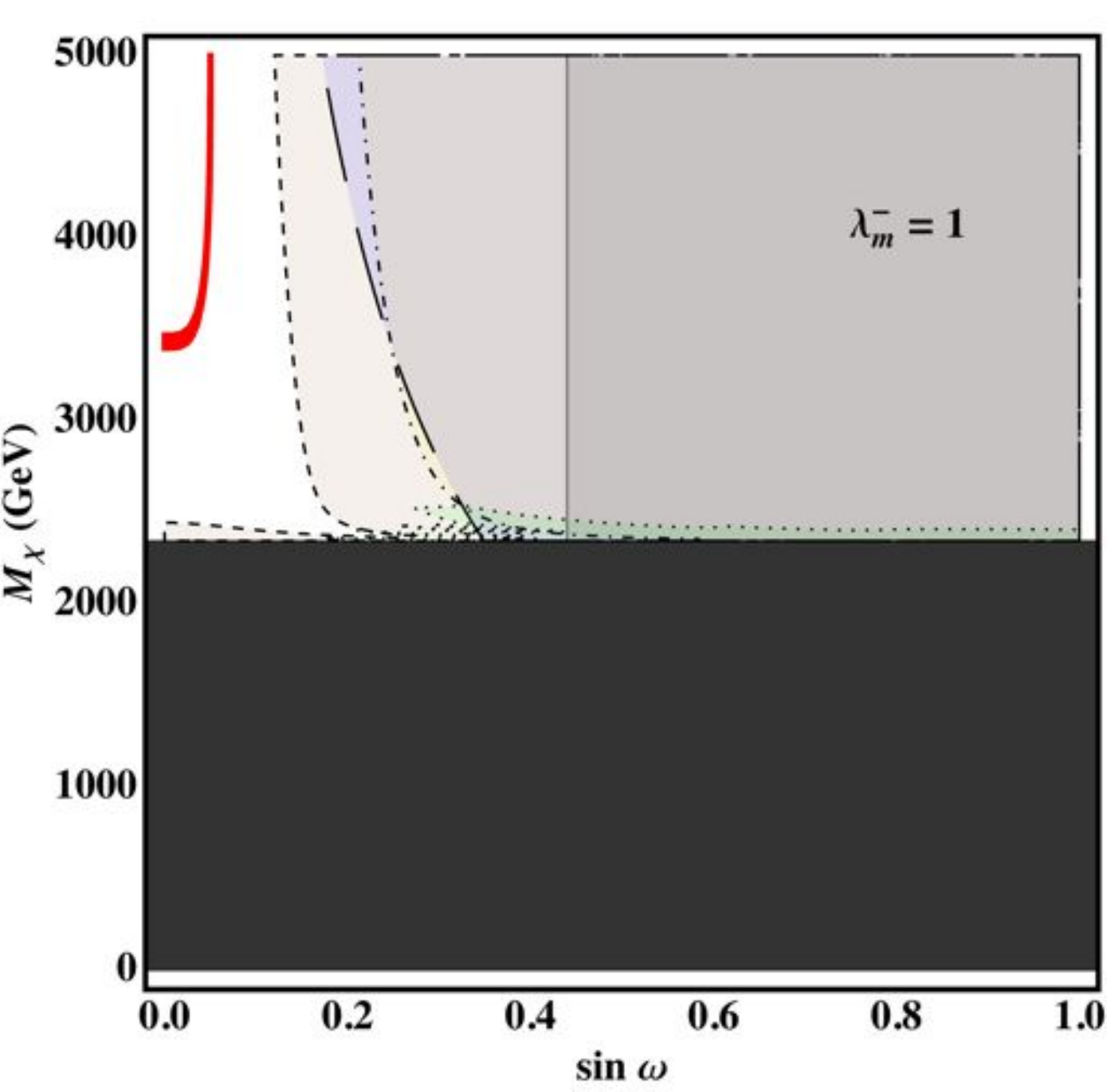}
\includegraphics[width=.329\textwidth]{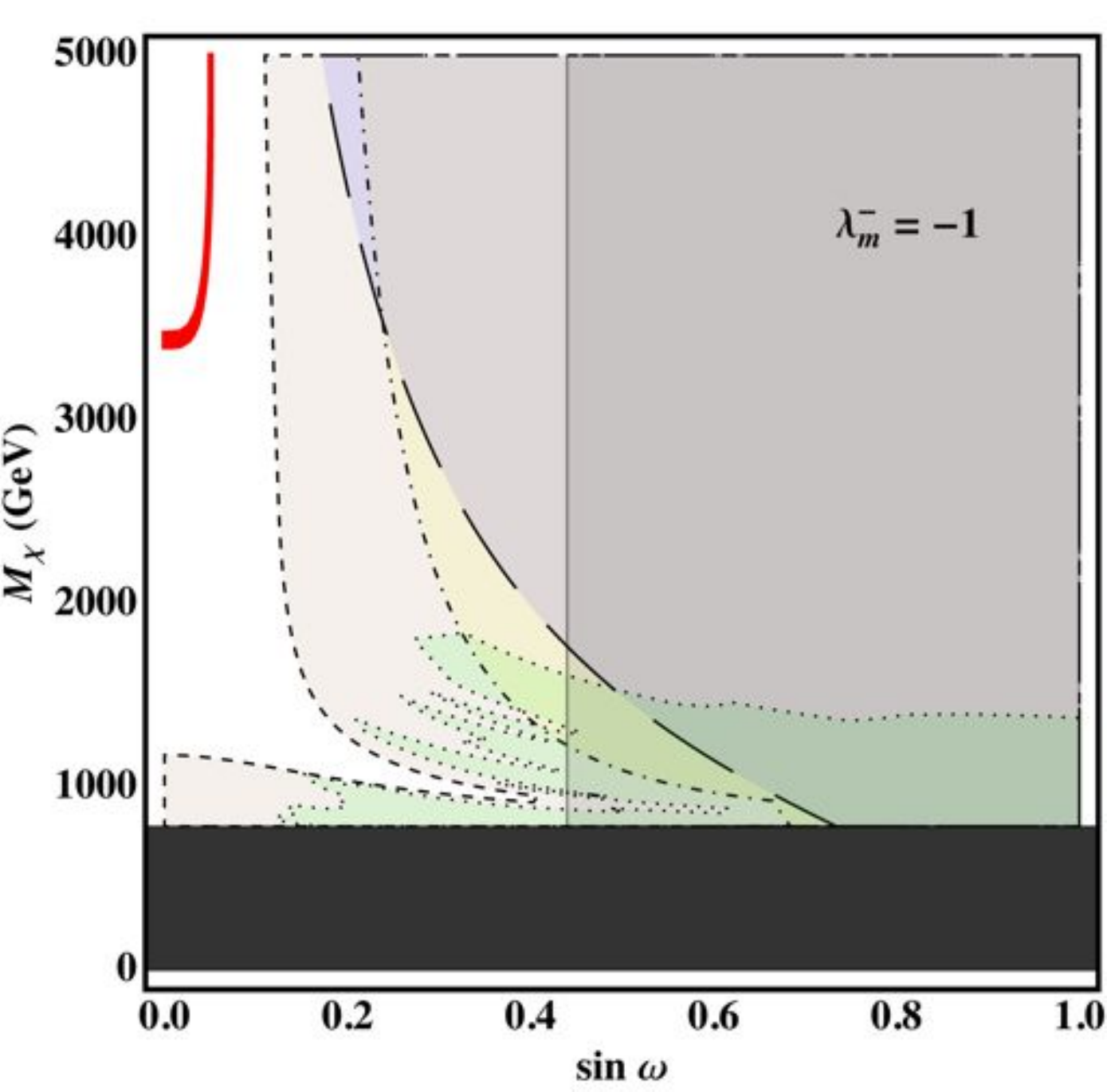}
\includegraphics[width=.329\textwidth]{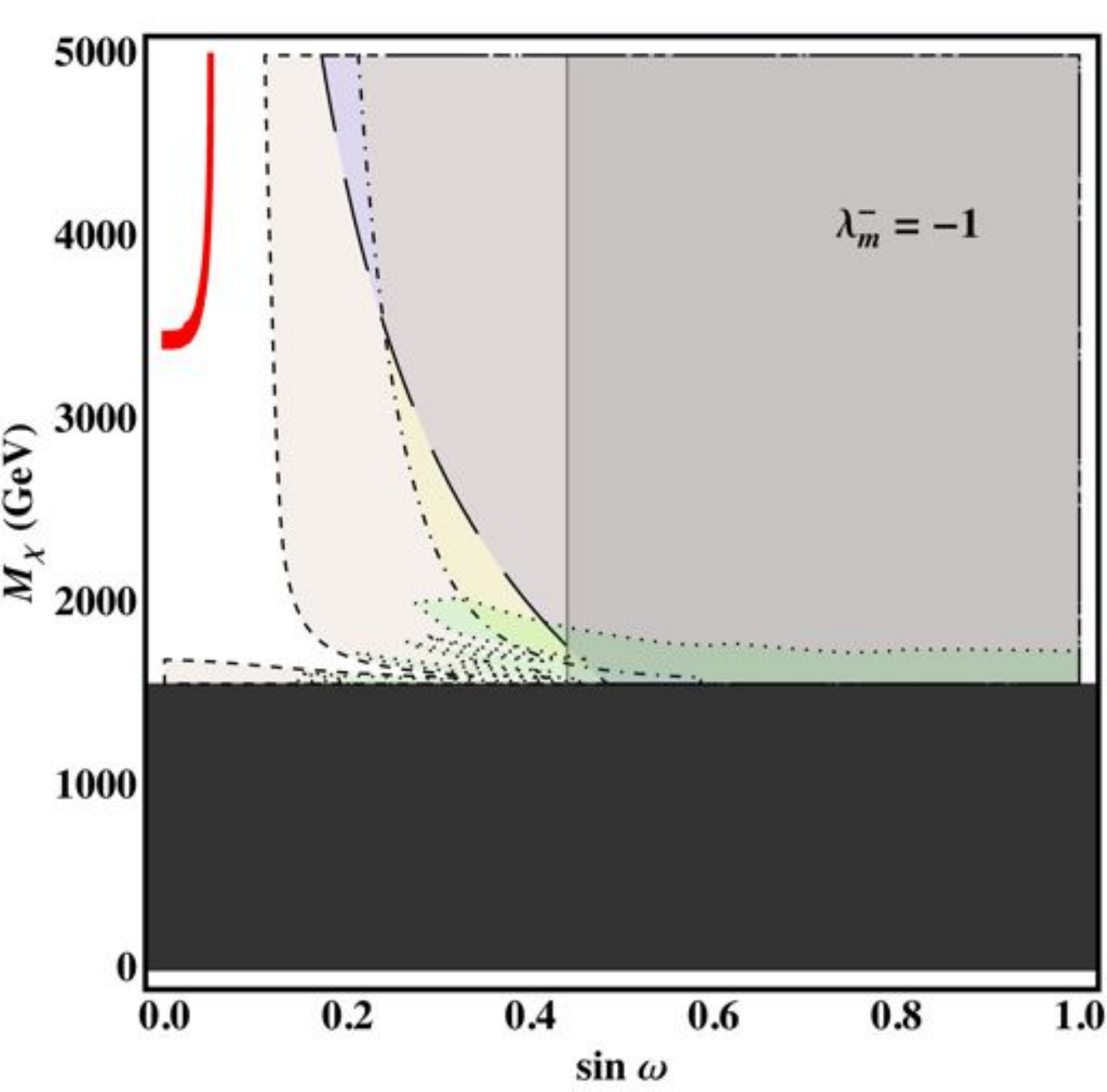}
\includegraphics[width=.329\textwidth]{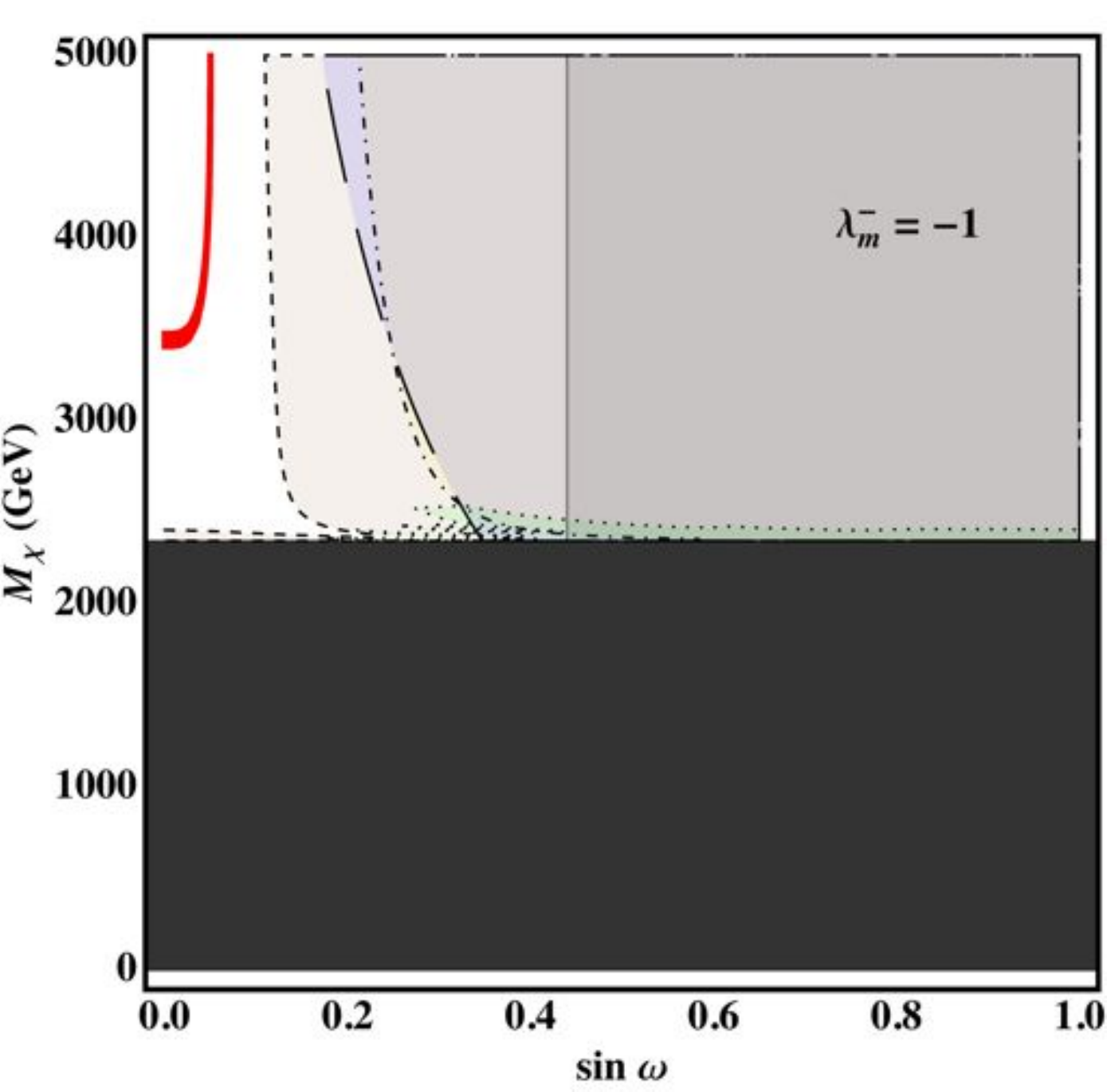}
\caption{Constraining the $\sin\omega - M_{\chi}$ plane for various choices of the parameters $\lambda_{m}^{-}$ and $M_{N}$. The solid black region, inferred from the stability condition of the one-loop potential \eqref{staboneloop}, determines the formal lower bound on the WIMP mass, $M_{\chi}$, for each selected value of $M_{N}$. (See the caption of Fig.~\ref{msom} for the details of the plots)}
\label{MXom}
\end{figure}

\begin{figure}
\includegraphics[width=.329\textwidth]{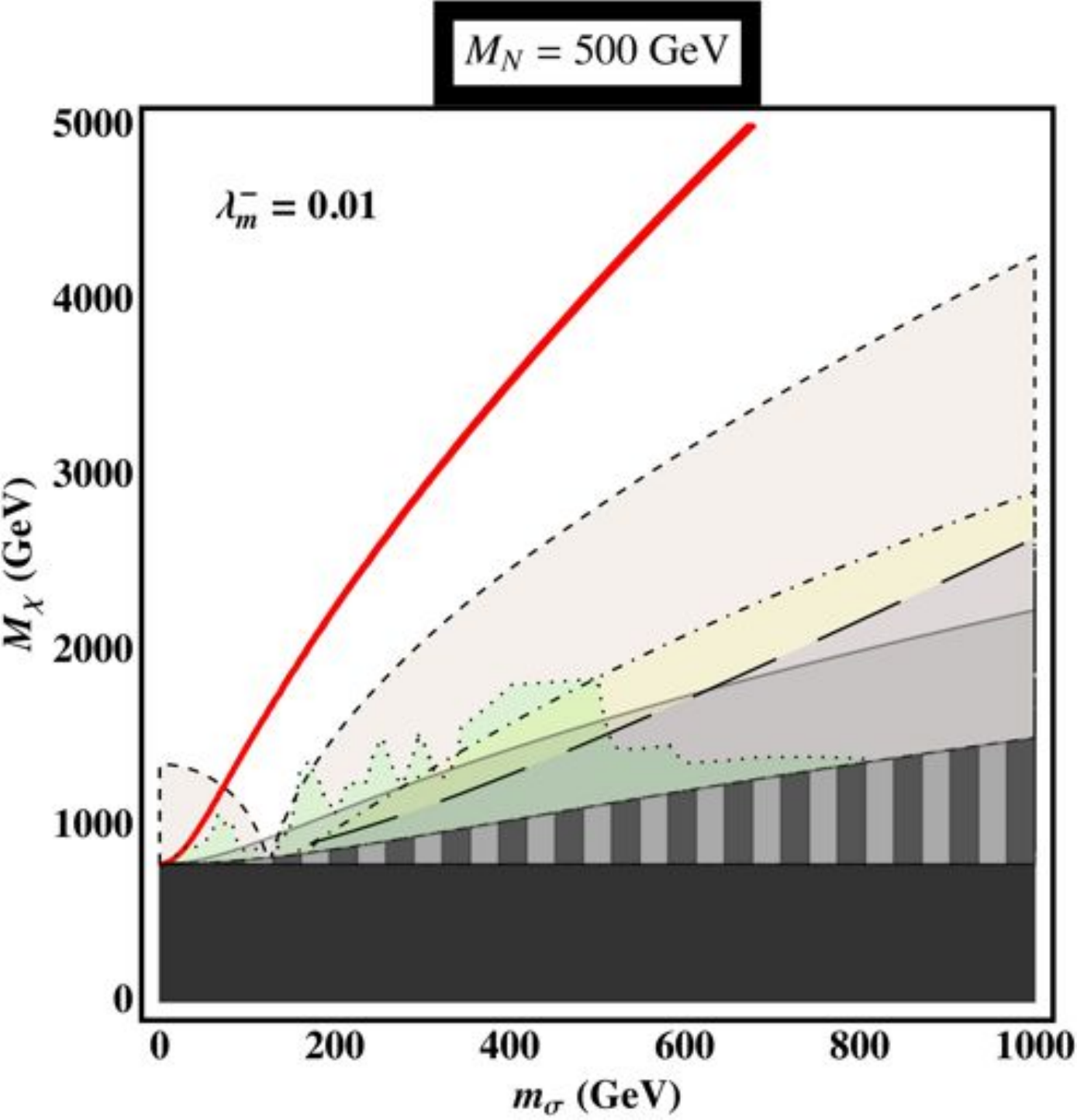}
\includegraphics[width=.329\textwidth]{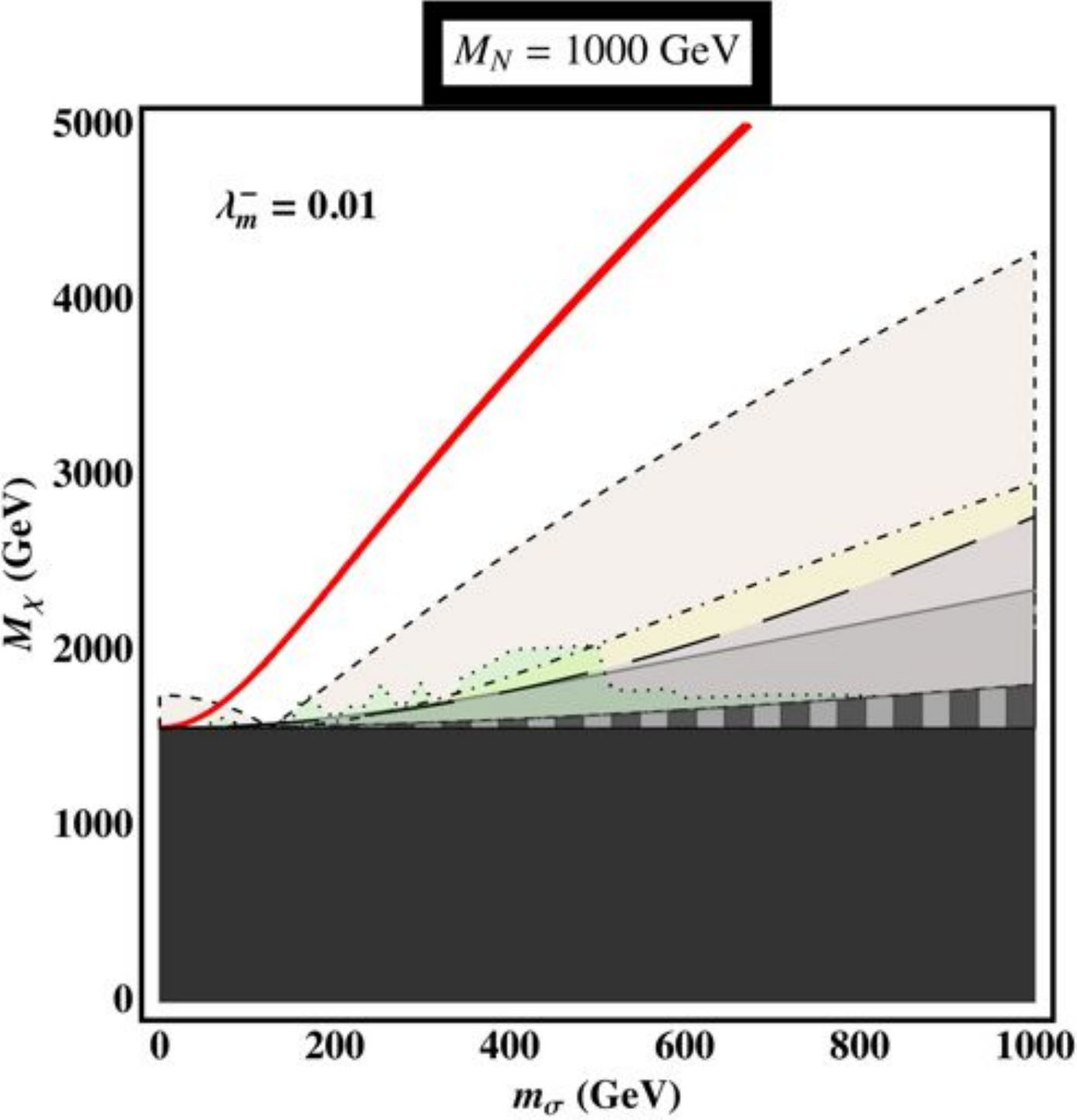}
\includegraphics[width=.329\textwidth]{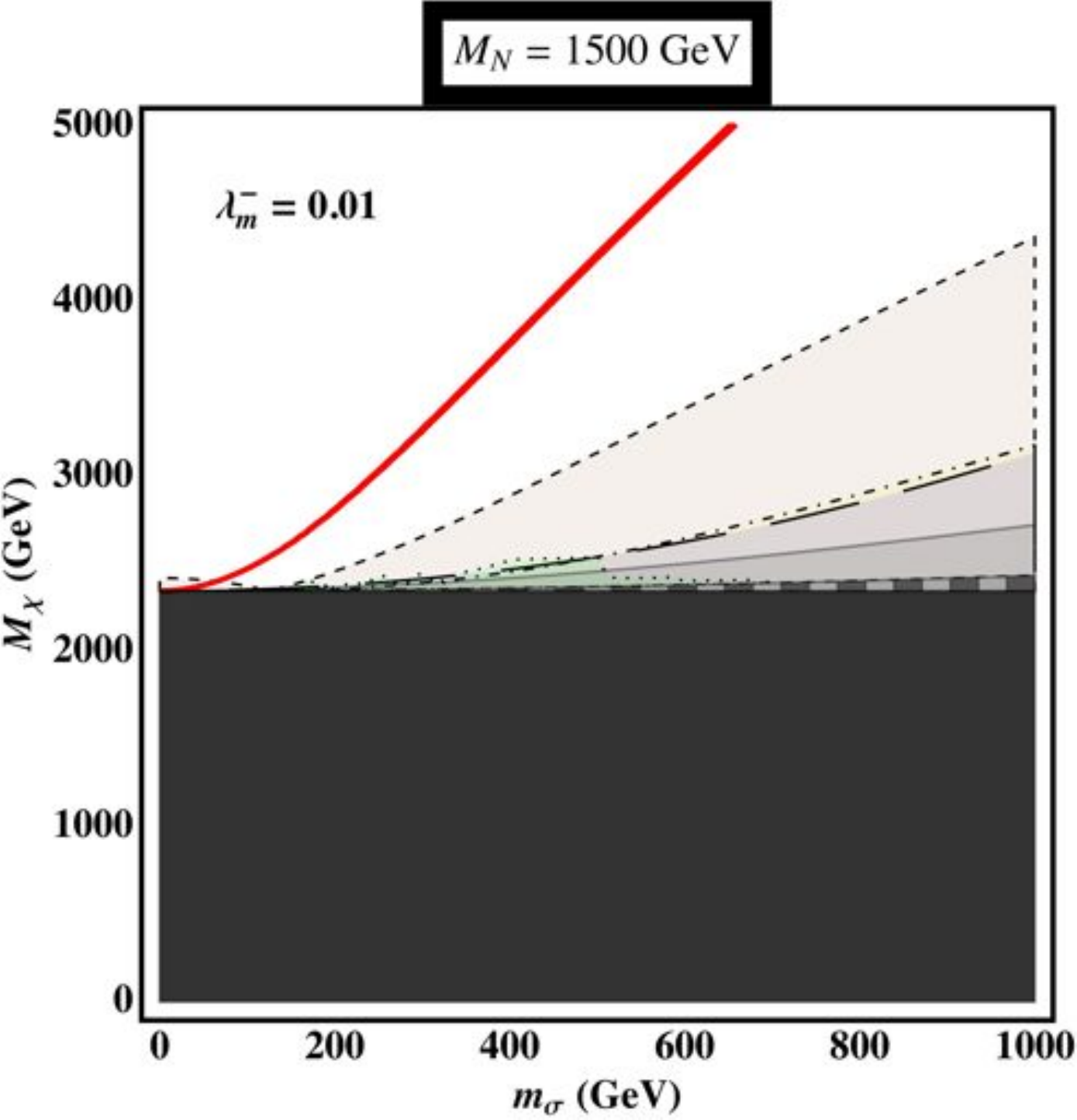}
\includegraphics[width=.329\textwidth]{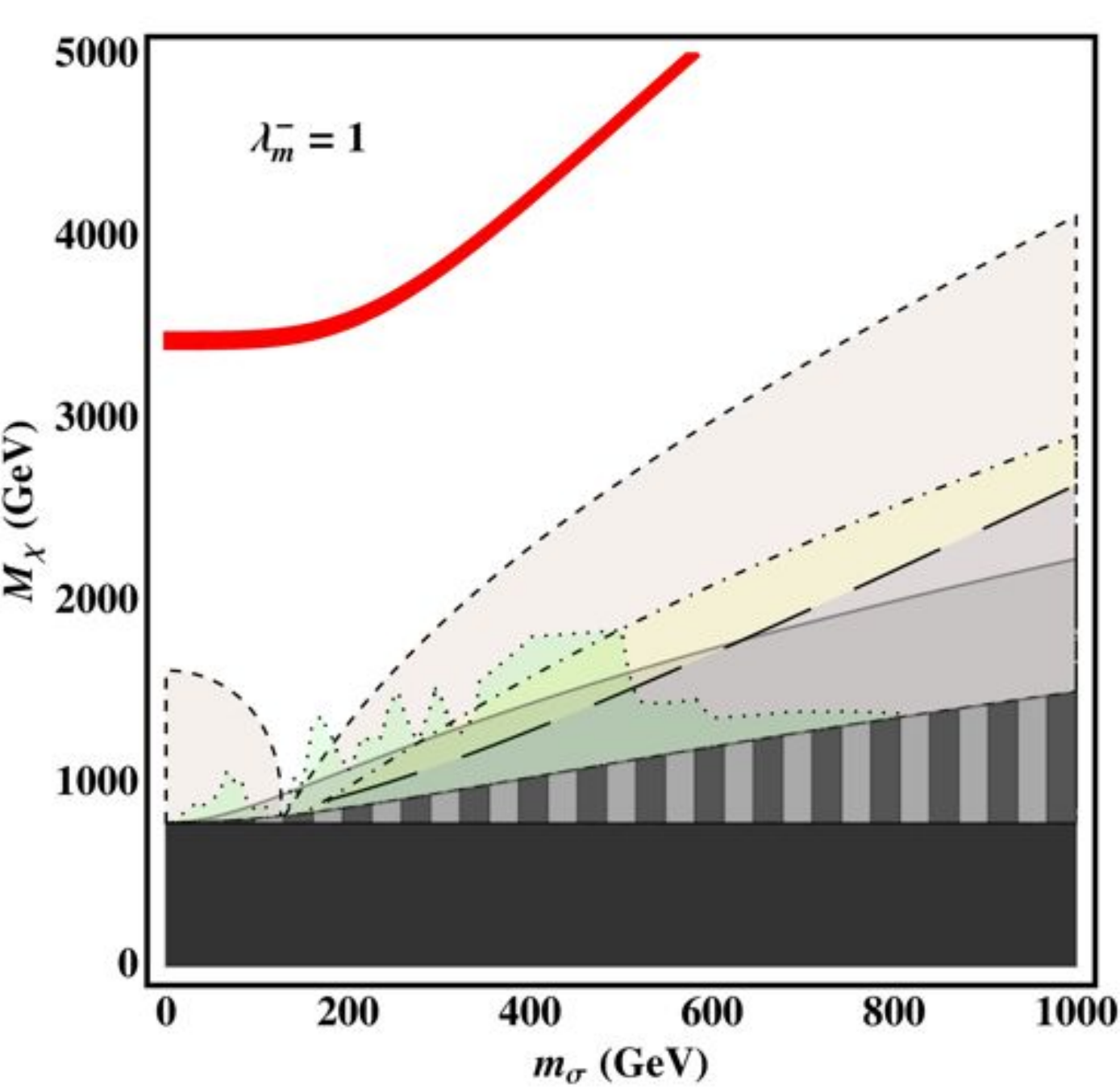}
\includegraphics[width=.329\textwidth]{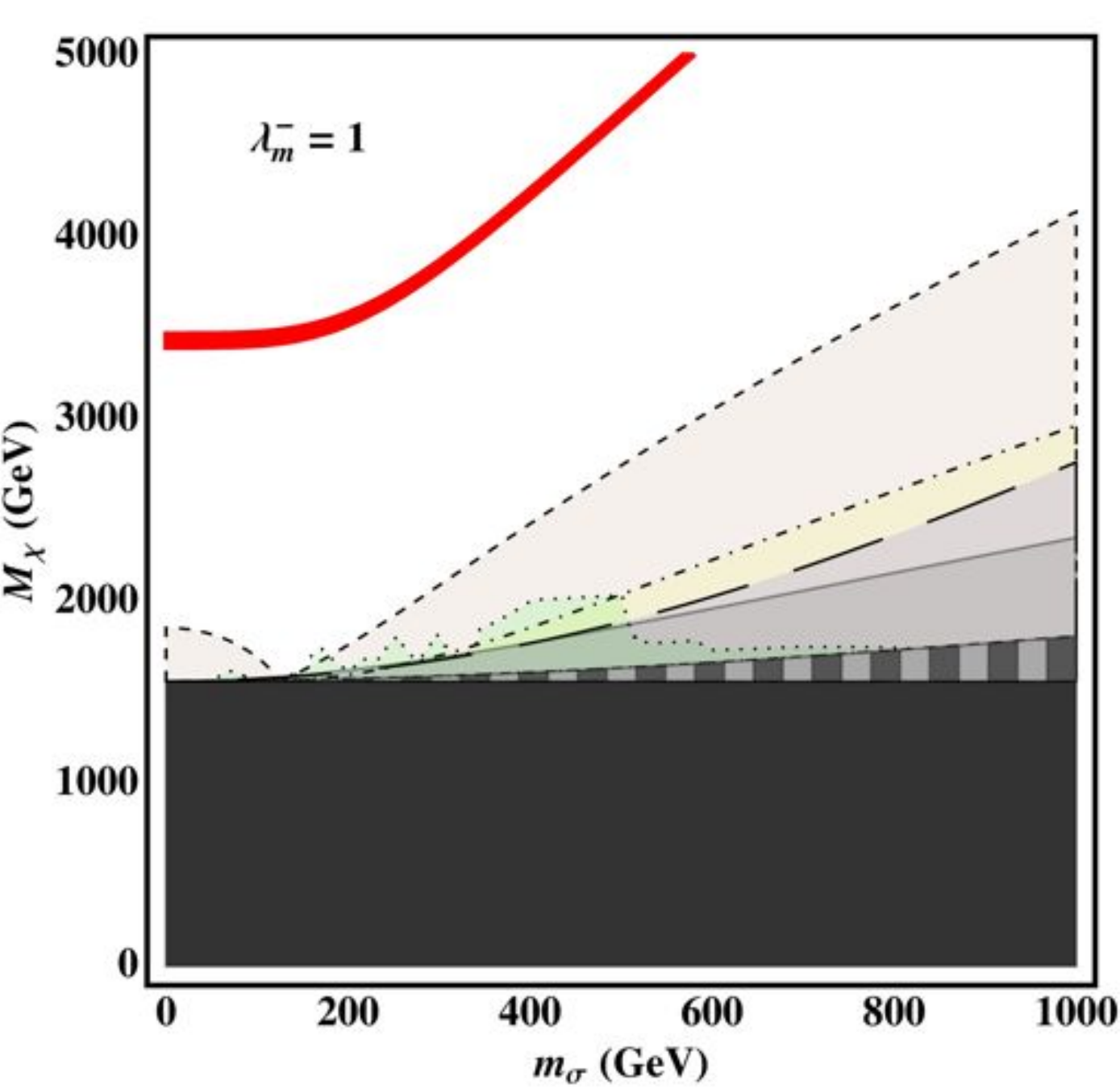}
\includegraphics[width=.329\textwidth]{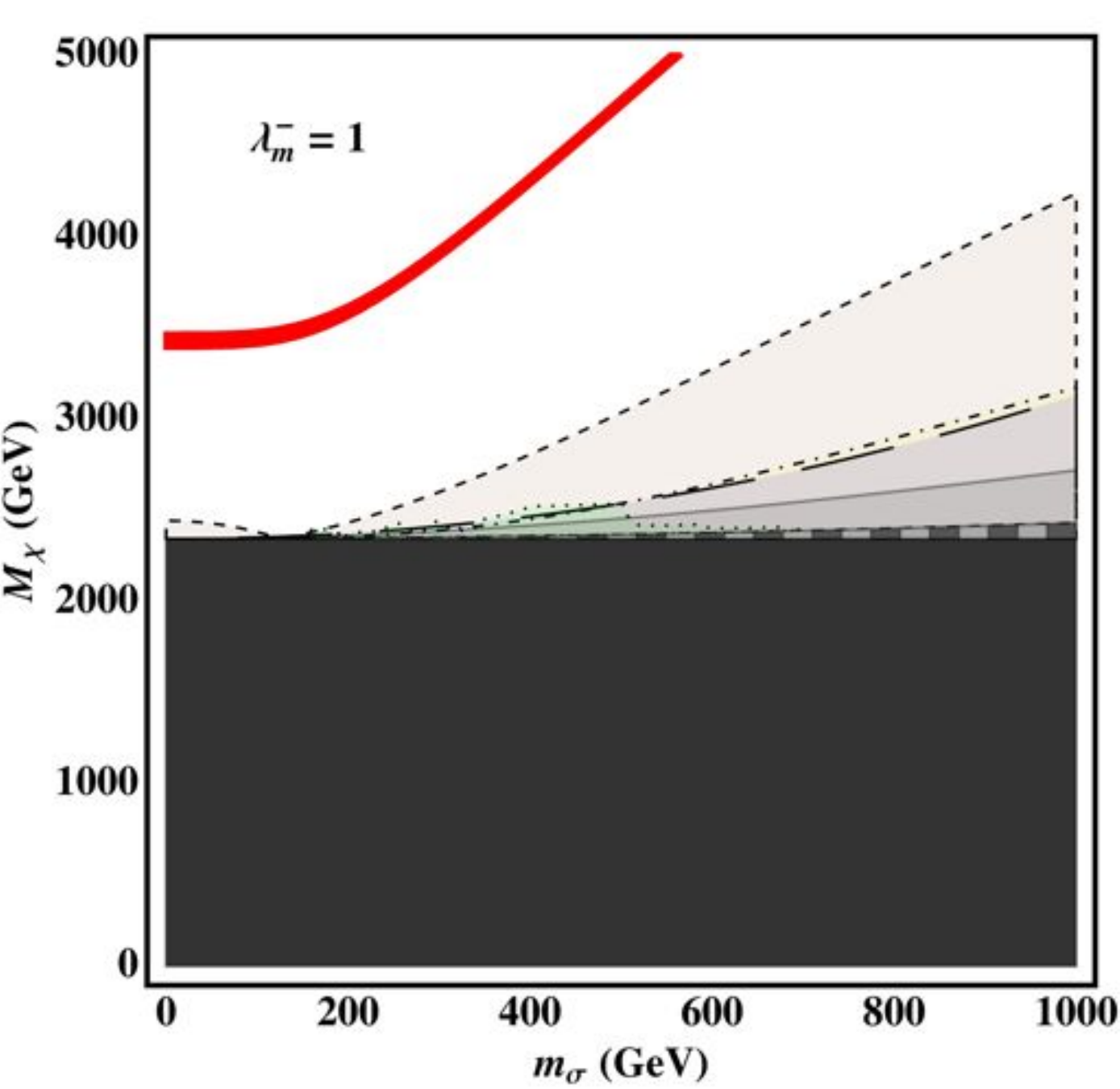}
\includegraphics[width=.329\textwidth]{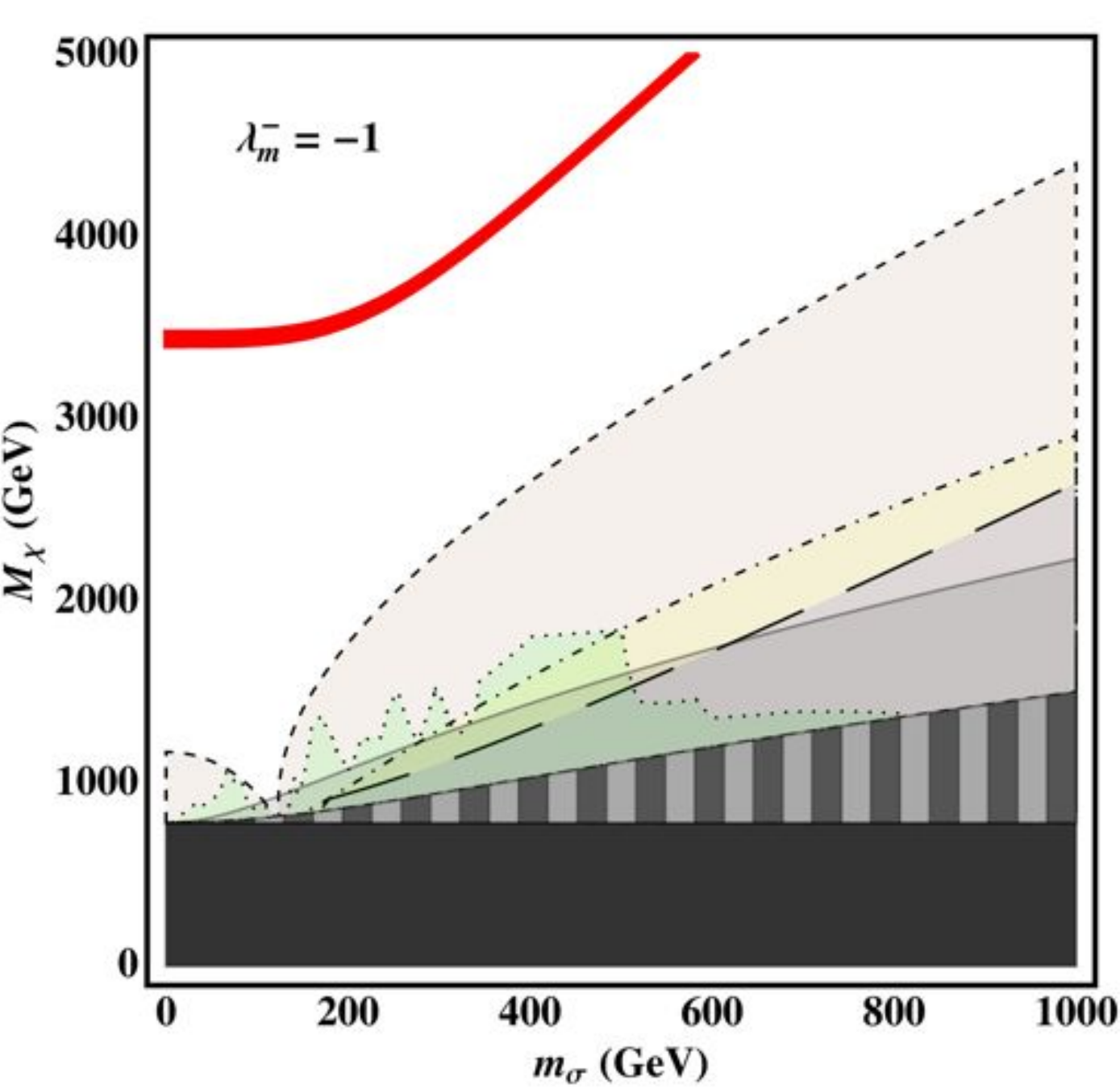}
\includegraphics[width=.329\textwidth]{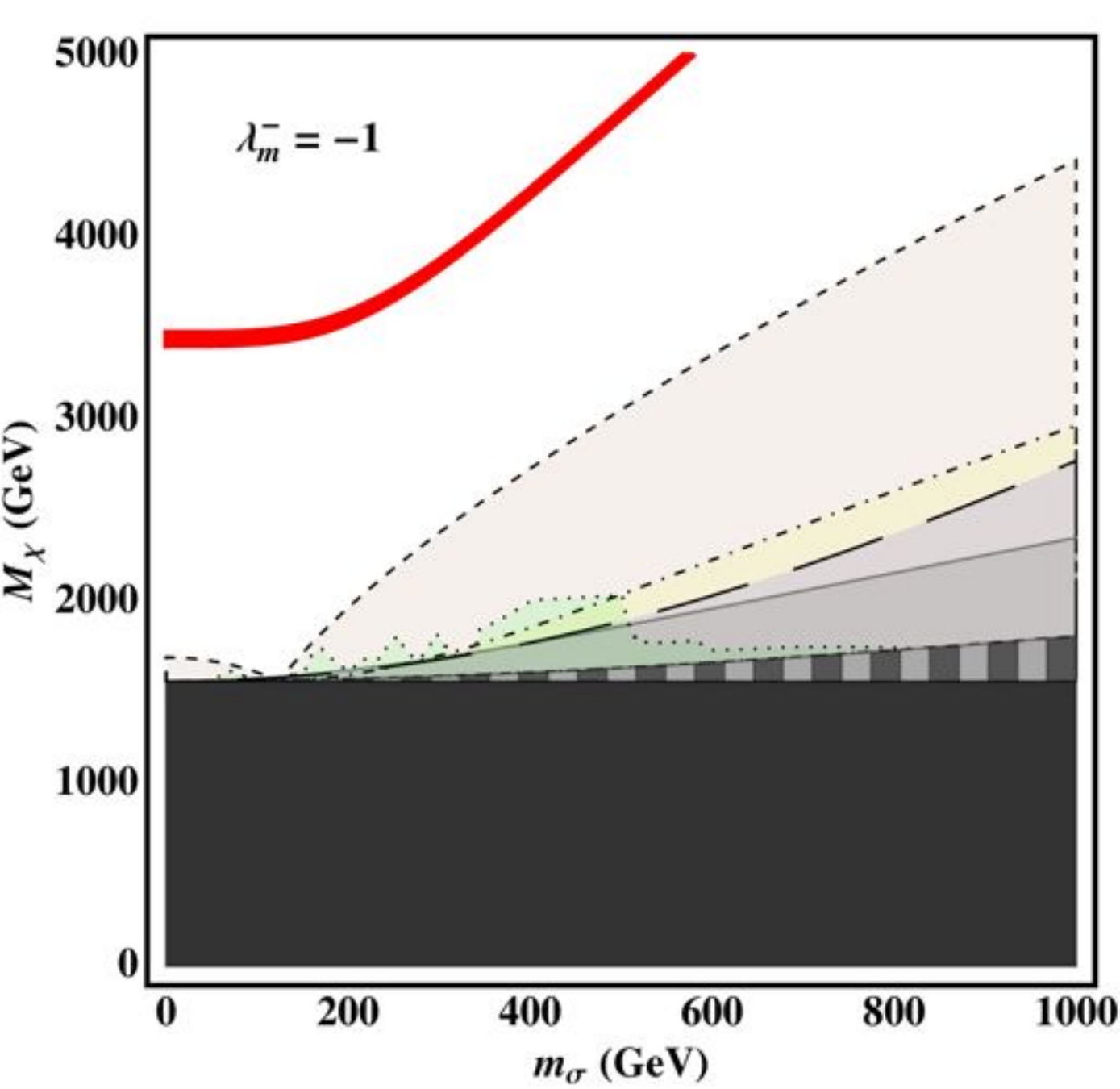}
\includegraphics[width=.329\textwidth]{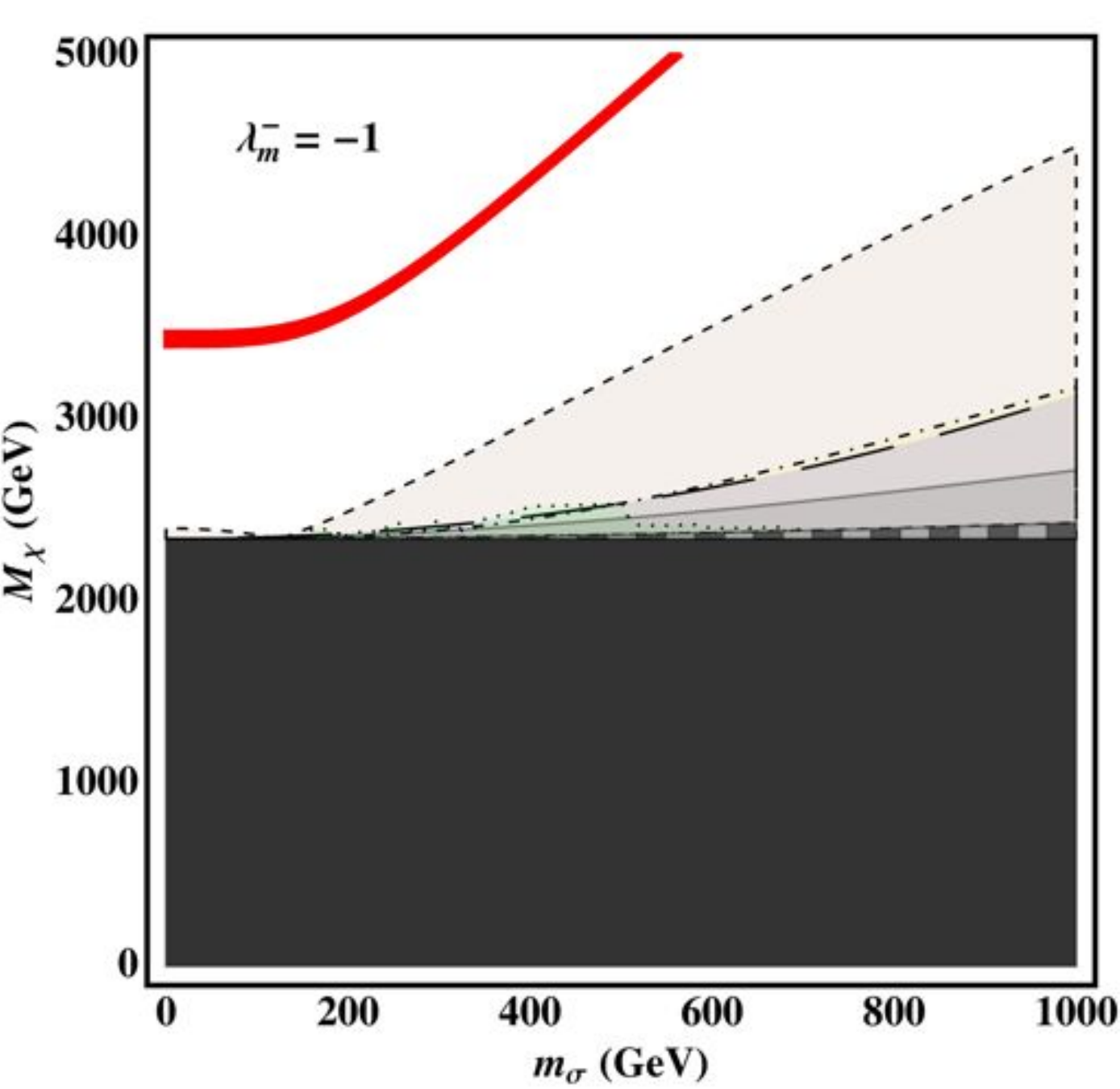}
\caption{Constraining the $m_{\sigma} - M_{\chi}$ plane for various choices of the parameters $\lambda_{m}^{-}$ and $M_{N}$. The solid black region, inferred from the stability condition of the one-loop potential \eqref{staboneloop}, determines the formal lower bound on the WIMP mass, $M_{\chi}$, for each selected value of $M_{N}$, while the vertically-shaded region is excluded by the $|\sin \omega| \leq 1$ condition. (See the caption of Fig.~\ref{msom} for the details of the plots)}
\label{MXms}
\end{figure}

In order to describe a comprehensive view of the parameter space, we illustrate the combined analysis in two-dimensional planes with the axes $M_{\chi}$, $m_{\sigma}$ or $\sin \omega$, while varying the remaining parameters. Since, neither the experimental investigations, nor the strongest unitarity condition (given by $\lambda_{\eta \chi}<8\pi$ \cite{Farzinnia:2013pga} for the parameters' range of interest) involve the input parameter $\lambda_{\chi}$, we fix $\lambda_\chi \sim \mathcal O (4\pi)$, and omit the latter henceforth.\footnote{As described below \eqref{eq:paraIV2}, the stability of the tree-level potential imposes the non-trivial inequality relation \eqref{treeineq} among $\lambda_\chi$, $\lambda_m^-$, and $\omega$. Given our primary interest in the experimental constraints, choosing $\lambda_m^- \gtrsim -1$ and $\lambda_\chi \sim \mathcal O (4\pi)$ results in a restriction on $\sin\omega$, which is contained within the experimental bounds. Accordingly, we ignore the tree-level stability condition in the exclusion plots (Figs.~\ref{msom}-\ref{MXms}), which will be automatically satisfied within the experimentally determined viable region of the parameter space.}

First, let us summarize the implications of our findings for the $\sigma$~boson. These are most clearly represented in the $\sin\omega-m_{\sigma}$~panels of Fig.~\ref{msom}, for $m_{\sigma} \leq 1$~TeV, and for various choices of the remaining two free parameters, $\lambda_{m}^{-}$ (row) and $M_{N}$ (column). A sensitivity to the right-handed Majorana neutrino mass mostly materializes in the perturbative unitarity bound, where heavier right-handed neutrinos are increasingly disfavored. In addition, the input parameter $\lambda_m^-$ is varied in magnitude from small to large and also in sign. The constraints from the LUX direct detection experiments \cite{LUX2013} are especially susceptible to this parameter, given its role in the cancellation among the two $h$-~and~$\sigma$-mediated competing channels in the spin-independent elastic scattering cross section \eqref{DMdirCS}. Larger mixing angle values satisfying the cancellation condition \eqref{dirrel} for a given $\lambda_m^-$ then become unconstrained by the observational data and open up a small window of compatibility. A larger positive $\lambda_{m}^{-}$ mildly mitigates the direct detection constraints. 

It is evident that incorporating the direct detection bounds into the limits already obtained from the other experimental considerations (c.f. the right panel of Fig.~\ref{mu}) results in more stringent constraints on the model's parameter space. It further narrows the viable values of the mixing angle to $\sin\omega \lesssim 0.2$ for most $\sigma$~boson masses, aside from the small window determined by the mentioned cancellation condition \eqref{dirrel}, as well as defining the lower limit of $m_{\sigma}$ compatible with the observational data. Moreover, requiring the pseudoscalar, $\chi$, to be a dark matter WIMP with the correct relic density \cite{Ade:2013zuv} places further restrictions on the interrelations of the parameters, and highly increases the predictability and testability of the current framework.

Fig.~\ref{MXom} illustrates, in an analogous manner, the aforementioned constraints on the $\sin\omega-M_{\chi}$~plane for similarly selected values of $\lambda_{m}^{-}$ (row) and $M_{N}$ (column) as in Fig.~\ref{msom}, where the WIMP mass $M_{\chi} \leq 5$~TeV. In this case, the stability of the potential at one-loop \eqref{staboneloop} imposes a formal lower bound on $M_{\chi}$ for each value of the right-handed neutrino mass. Once more, the dark matter direct detection data place the most stringent constraints on the parameter space, further narrowing the viable region as compared with the other experimental limits (c.f. Fig.~\ref{relpix}). The described observations in Fig.~\ref{msom} remain valid in the present case. A $\lambda_{m}^{-}$ parameter larger in magnitude, however, requires a heavier WIMP, in order to comply with the constraint from the thermal relic abundance, which may be fully accommodated by the viable parameter values of the current scenario in the $\sin \omega \lesssim 0.1$ region.

Furthermore, let us examine the interplay between the $\sigma$~boson and the dark matter $\chi$~pseudoscalar directly, by plotting the obtained experimental and theoretical constraints in the $m_{\sigma}-M_{\chi}$~plane. Similar choices of the input parameters $\lambda_{m}^{-}$ (row) and $M_{N}$ (column) as in the previous two figures are presented here in each panel as well. As before, the direct detection data from the LUX experiment offer the strongest constraints on the parameter space, necessitating a more massive dark matter particle for a heavier $\sigma$~scalar. However, for a larger magnitude $\lambda_{m}^{-}$, the WIMP mass, $M_{\chi}$, becomes independent of the $\sigma$~boson mass for $m_{\sigma} \lesssim 200$~GeV, assuming the $\chi$~pseudoscalar constitutes the sole or dominant component of the WIMP dark matter with the correct thermal relic density.

\begin{figure}
\includegraphics[width=.329\textwidth]{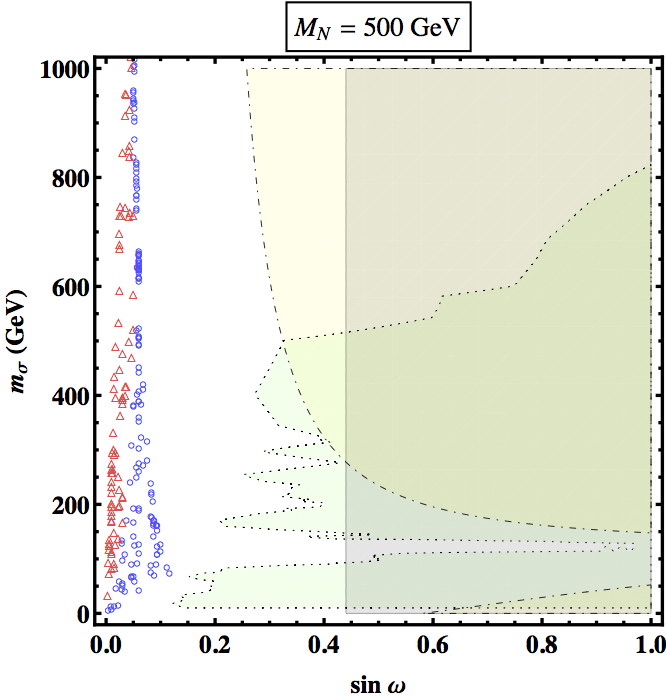}
\includegraphics[width=.329\textwidth]{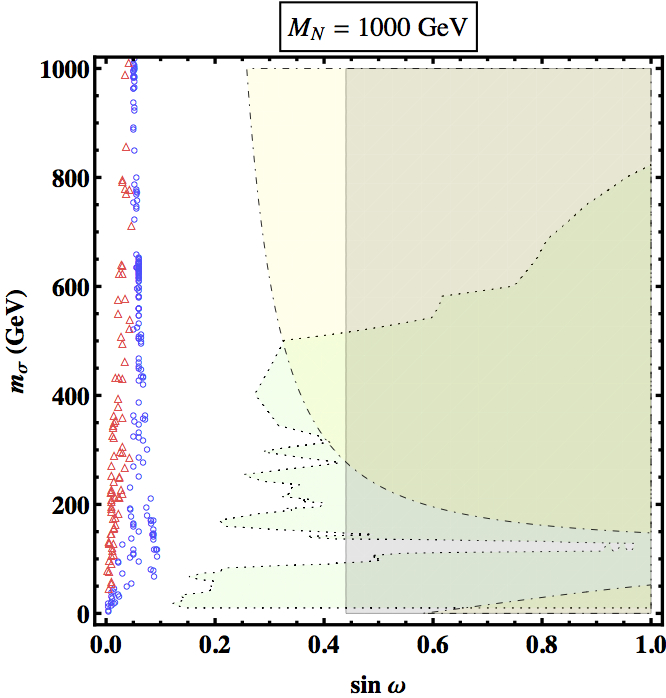}
\includegraphics[width=.329\textwidth]{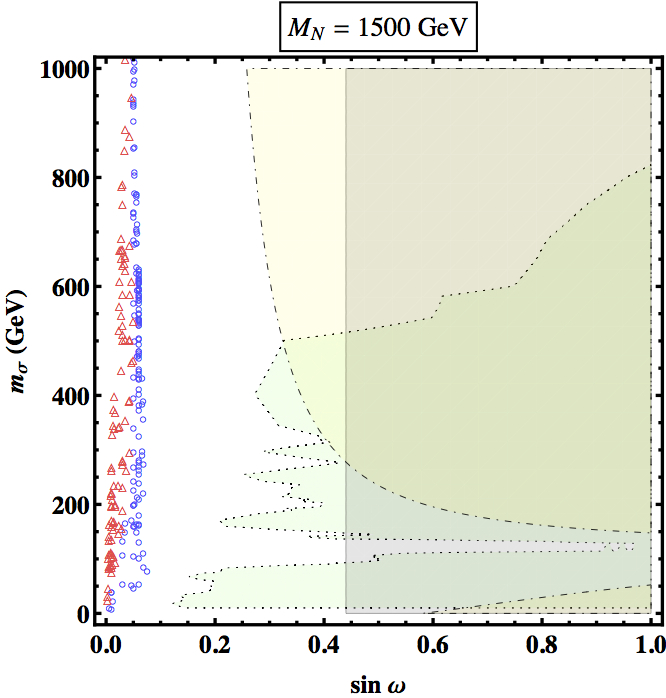}
\includegraphics[width=.329\textwidth]{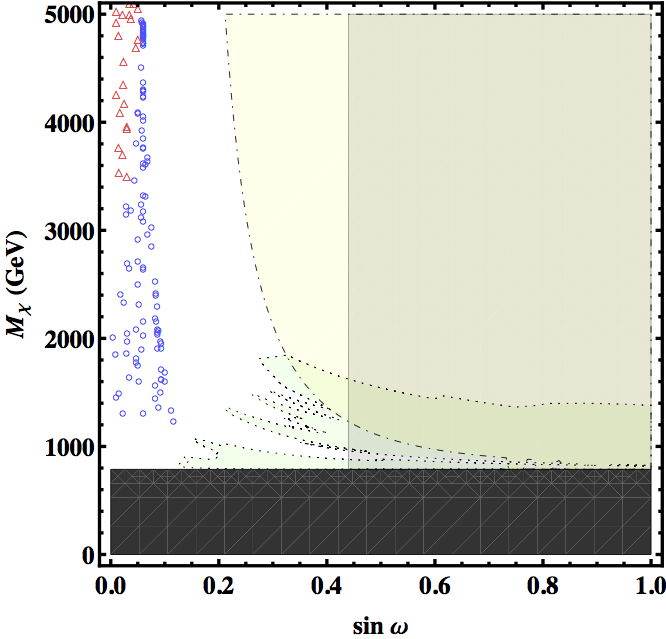}
\includegraphics[width=.329\textwidth]{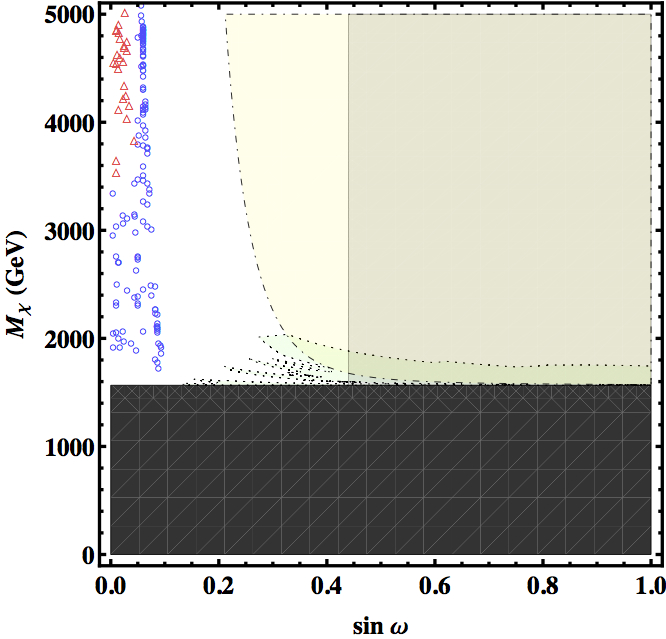}
\includegraphics[width=.329\textwidth]{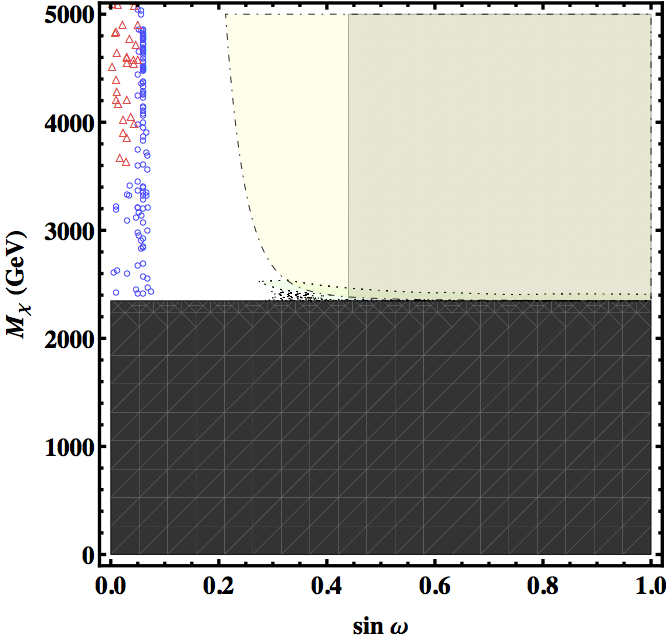}
\includegraphics[width=.329\textwidth]{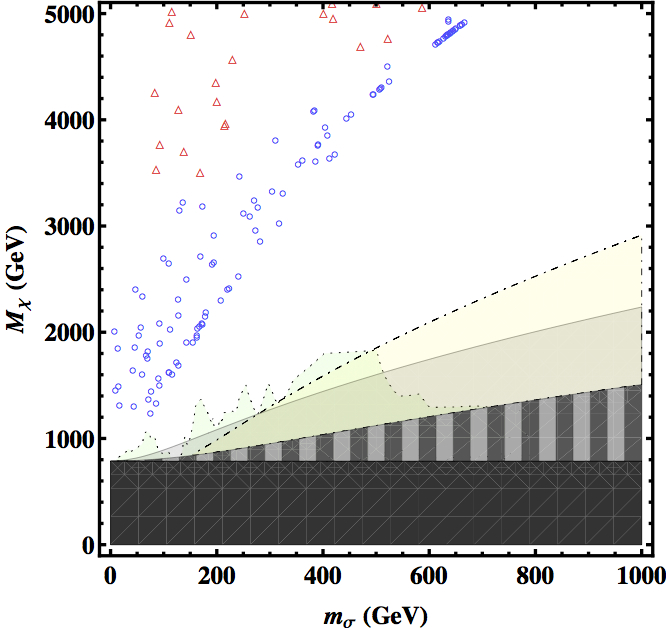}
\includegraphics[width=.329\textwidth]{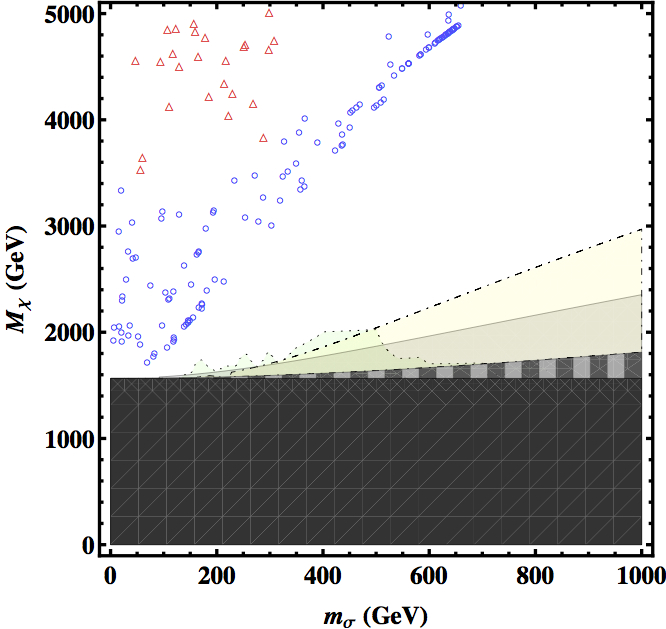}
\includegraphics[width=.329\textwidth]{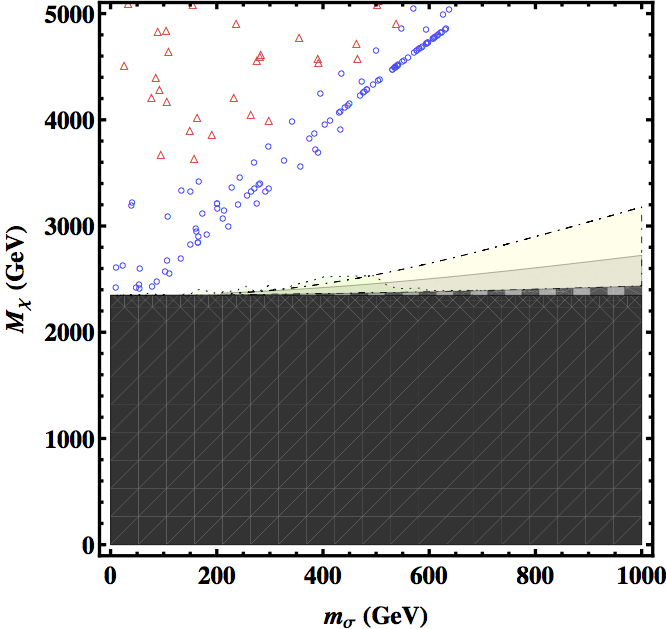}
\caption{Scatter plots displaying the viable region of the model's parameter space for three values of the right-handed Majorana neutrino masses (columns) in the $\sin\omega - m_{\sigma}$, $\sin\omega - M_{\chi}$, and $m_{\sigma} - M_{\chi}$ planes (rows). The scattered points pass perturbative unitarity, one-loop triviality and vacuum stability conditions (assuming a cutoff scale higher than $10^5$~GeV) \cite{Farzinnia:2013pga}, WIMP relic density within the $1\sigma$ uncertainty quoted by the Planck collaboration, and the LUX direct detection constraints at 90\%~C.L. The blue circles correspond to $|\lambda_m^-| \leq 1$, whereas the red triangles represent $|\lambda_m^-|>1$. All colored regions are excluded. (See the caption of Fig.~\ref{msom} for the details of the plots)}
\label{scat}
\end{figure}

Finally, for completeness, we also exhibit the scatter plots which determine the viable region of the parameter space, in Fig.~\ref{scat}. All free parameters of the model \eqref{inputs} are taken into account (including $\lambda_\chi$). The scattered points summarize the previously analyzed constraints in \cite{Farzinnia:2013pga}, arising from imposing perturbative unitarity, and one-loop triviality and vacuum stability where a cutoff scale higher than $10^5$~GeV was required, as well as the current dark matter analyses of the relic density \cite{Ade:2013zuv} and the direct detection bounds \cite{LUX2013}. The experimental constraints from the electroweak precision tests and the LHC measurements of the properties of the 125~GeV $h$~Higgs (also previously analyzed in \cite{Farzinnia:2013pga}), as well as the current collider study of the LEP \cite{Barate:2003sz} and LHC \cite{LHCHeavyH} Higgs searches are explicitly depicted. The overall analysis highly constrains the parameter space, demonstrating the predictive power of the present scenario.

\section{Conclusion}\label{concl}

In this treatment, we have further investigated some of the phenomenological aspects of the minimal viable scale invariant model introduced and studied previously in \cite{Farzinnia:2013pga}, by analyzing the available experimental and observational data from collider and dark matter searches, and their implications for the scenario's parameter space. In particular, we consistently applied the Higgs search data from LEP \cite{Barate:2003sz} and LHC \cite{LHCHeavyH} up to a mass of 1~TeV to the additional $CP$-even scalar predicted in this scenario, and presented the obtained constraints on the parameters, which are complementary to the analyses of the experimental data from electroweak precision tests and direct LHC measurements of the 125~GeV scalar state \cite{Farzinnia:2013pga}.

Furthermore, identifying the stable pseudoscalar---protected by the $CP$-symmetry of the theory---as a WIMP dark matter candidate, we calculated its thermal relic density at the present time. Assuming the pseudoscalar WIMP constitutes an $\mathcal O$(1) fraction of the dark matter in the universe, the compatibility of the predicted thermal relic abundance with the Planck satellite observations \cite{Ade:2013zuv} was demonstrated for a variety of the model's input parameter values. The latter analysis tightly constrained the parameter space. In addition, we studied the implications of the dark matter direct detection data from the LUX experiment \cite{LUX2013}, as applied to the heavy pseudoscalar WIMP candidate of the theory, and demonstrated that they impose further bounds on the viable parameter space. These constraints are more stringent than the ones obtained by the other experimental considerations.

Our results are summarized in extensive exclusion plots (Figs.~\ref{msom}-\ref{scat}), covering the relevant range of the model's parameters from a variety of representations, and demonstrating the interplay between the various input parameters. The combined analysis allows, in general, for a mixing between the SM Higgs and a $CP$-even singlet scalar restricted to $\sin \omega \lesssim 0.2$, pseudoscalar dark matter with a mass in the TeV range, and weak scale right-handed Majorana neutrinos. In particular, the thermal relic abundance consideration is accommodated within the $\sin \omega \lesssim 0.1$ region, and imposes tight bounds on the parameter space, rendering the scenario highly predictive.

\section*{Acknowledgments}

We would like to thank R. Sekhar Chivukula and Elizabeth H. Simmons for valuable comments on the manuscript, and Kristjan Kannike for bringing to our attention the improved stability conditions for the tree-level potential. J.R. thanks Hong-Jian He for early suggestions and discussions along the way. During the completion of this work, A.F. was in part supported by the Tsinghua Outstanding Postdoctoral Fellowship and by the NSF of China (under grants 11275101, 11135003). J.R. was supported by National NSF of China (under grants 11275101, 11135003) and National Basic Research Program (under grant 2010CB833000).

\appendix

\section{Feynman Rules}\label{FR}

\begin{figure}
\includegraphics[width=.57\textwidth]{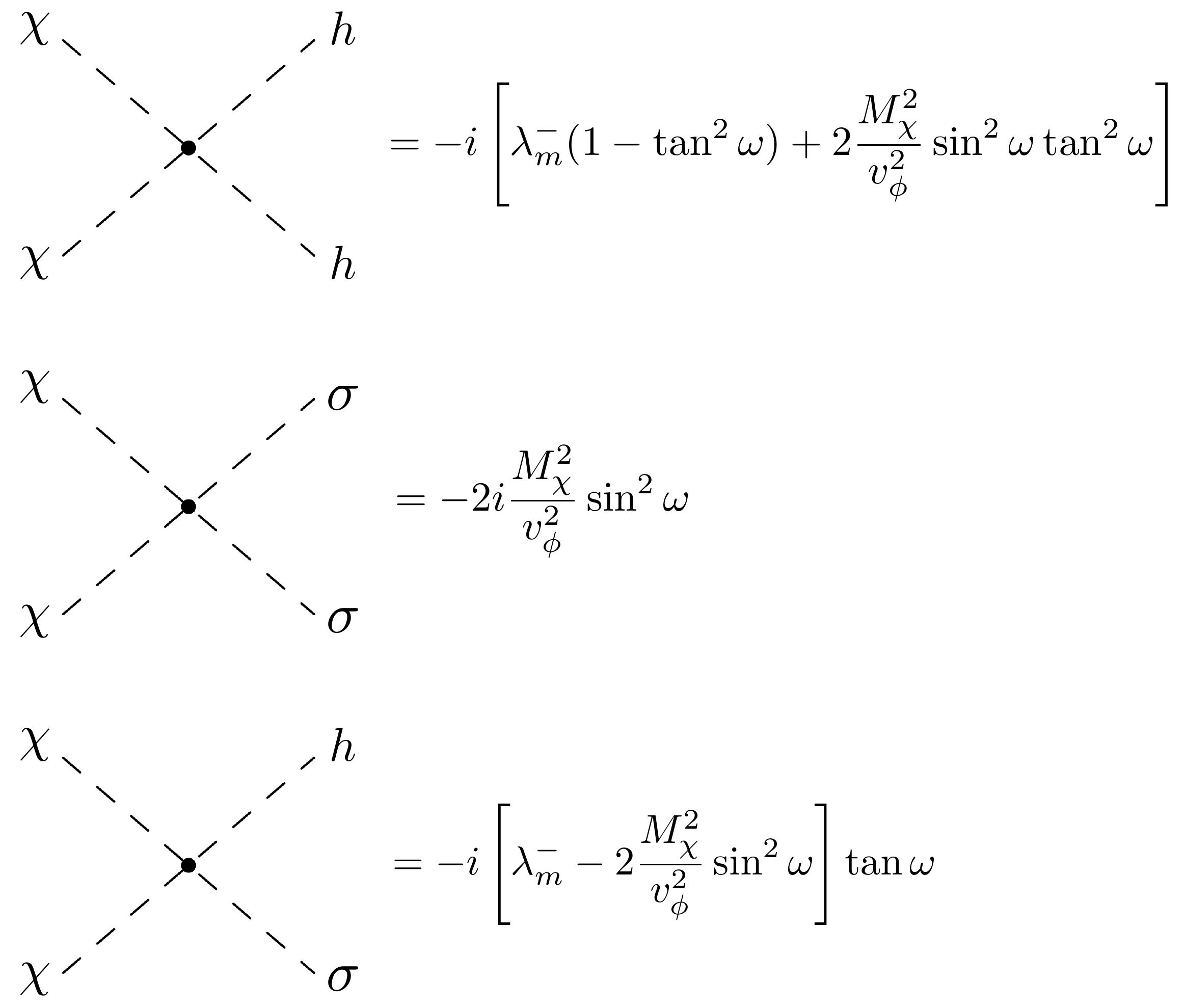}
\caption{Feynman rules for the relevant quartic couplings.}
\label{FR4}
\end{figure}

In this appendix, we exhibit the Feynman rules for the trilinear and quartic couplings, obtained from the tree-level Lagrangian, which are relevant for the dark matter pair-annihilation process and the corresponding calculation of the thermal relic density (see Appendix~\ref{CSexp}). Fig.~\ref{FR3} depicts the scalar trilinear couplings, as well as their Yukawa interactions with the right-handed Majorana neutrinos. The relevant scalar quartic couplings are shown in Fig.~\ref{FR4}.

\begin{figure}
\includegraphics[width=.45\textwidth]{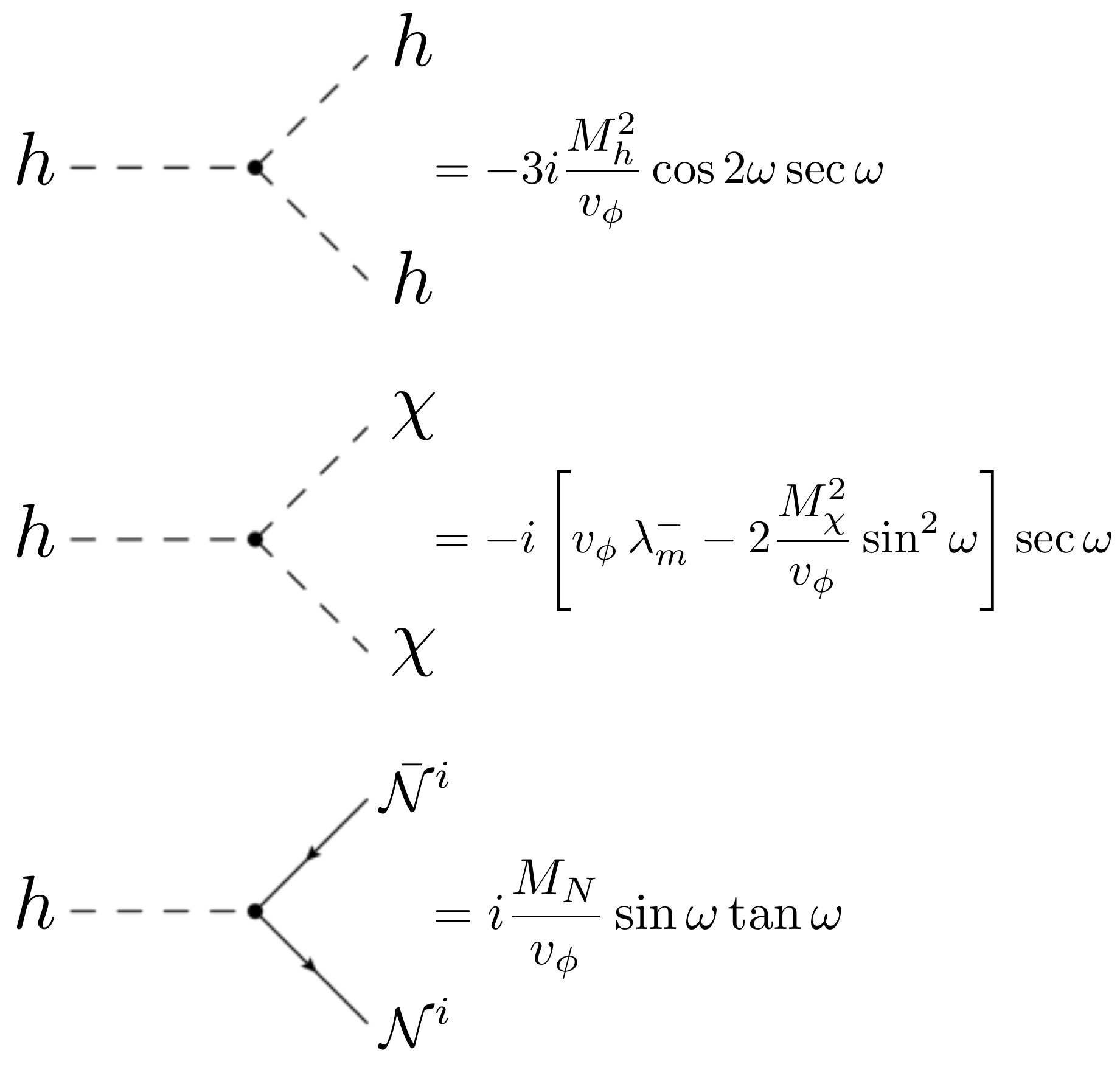}\qquad
\includegraphics[width=.315\textwidth]{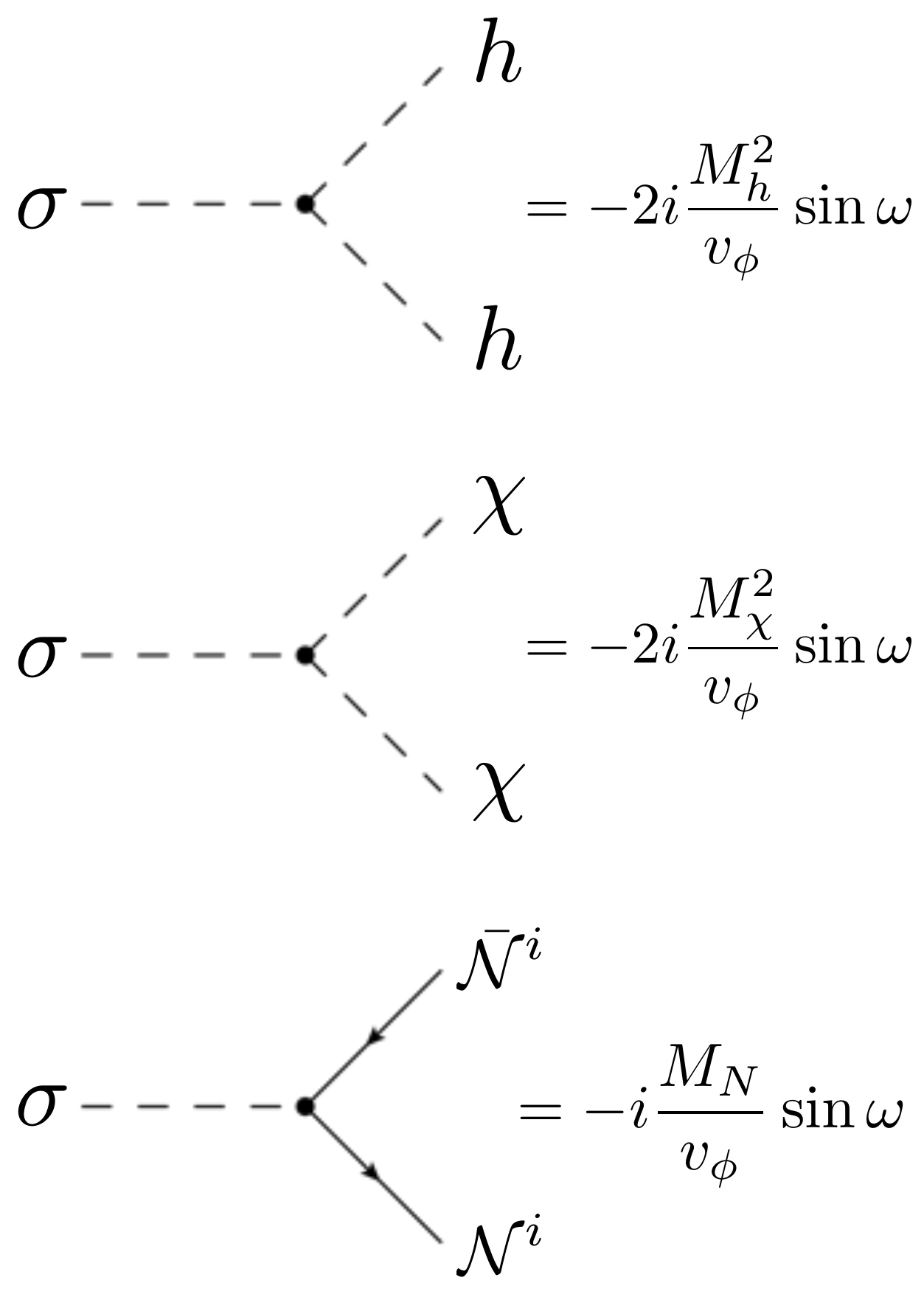}
\caption{Feynman rules for the relevant trilinear couplings.}
\label{FR3}
\end{figure}

\section{Thermally Averaged $\chi\chi \to j j'$ Cross Sections} \label{CSexp}

The leading-order term for the thermally averaged cross section of the WIMP pair-annihilation into $jj'$ final states is given in the right-hand side of \eqref{tacs}. In this appendix, we present the corresponding expressions for the dominant final state products, as depicted in Fig.~\ref{DMann}. Taking the trilinear ($i\lambda_{ijk}$) and the quartic ($i\lambda_{\chi\chi jj'}$) couplings from Appendix~\ref{FR}, one obtains
\begin{equation}\label{tacsj}
\begin{split}
\langle\sigma v \rangle_{\text{ann}}^{hh} =&\, \frac{1}{64\pi M_\chi^2}\left(1-\frac{M_h^2}{M_\chi^2}\right)^{1/2}\left[\lambda_{\chi\chi hh}-\frac{\lambda_{\chi\chi h}\lambda_{hhh}}{4M_\chi^2-M_h^2}-\frac{\lambda_{\chi\chi\sigma}\lambda_{hh\sigma}}{4M_\chi^2-M_\sigma^2}+\frac{2\lambda_{\chi\chi h}^2}{2M_\chi^2-M_h^2}\right]^2  \ , \\
\langle\sigma v \rangle_{\text{ann}}^{\sigma\sigma} =&\, \frac{1}{64\pi M_\chi^2}\left(1-\frac{M_\sigma^2}{M_\chi^2}\right)^{1/2}\left[\lambda_{\chi\chi \sigma\sigma}-\frac{\lambda_{\chi\chi\sigma}\lambda_{\sigma\sigma\sigma}}{4M_\chi^2-M_\sigma^2}+\frac{2\lambda_{\chi\chi \sigma}^2}{2M_\chi^2-M_\sigma^2}\right]^2 \ ,\\
\langle\sigma v \rangle_{\text{ann}}^{h\sigma} =&\, \frac{1}{32\pi M_\chi^2}\left(1-\frac{(M_h+M_\sigma)^2}{4M_\chi^2}\right)^{1/2}\left(1-\frac{(M_h-M_\sigma)^2}{4M_\chi^2}\right)^{1/2}\left[\lambda_{\chi\chi h\sigma}-\frac{\lambda_{\chi\chi h}\lambda_{hh\sigma}}{4M_\chi^2-M_h^2}+\frac{2\lambda_{\chi\chi h}\lambda_{\chi\chi\sigma}}{2M_\chi^2-M_h^2}\right]^2\ ,\\
\langle\sigma v \rangle_{\text{ann}}^{t\bar{t}} =&\, \frac{3M_t^2}{4\pi v_\phi^2}\left(1-\frac{M_t^2}{M_\chi^2}\right)^{3/2}\left[\frac{\lambda_{\chi\chi h}}{4M_\chi^2-M_h^2}\cos\omega+\frac{\lambda_{\chi\chi\sigma}}{4M_\chi^2-M_\sigma^2}\sin\omega\right]^2 \ ,\\
\langle\sigma v \rangle_{\text{ann}}^{\mathcal{N}\bar{\mathcal{N}}} =&\, \frac{M_N^2\tan^2\omega}{8\pi v_\phi^2}\left(1-\frac{M_N^2}{M_\chi^2}\right)^{3/2}\left[\frac{\lambda_{\chi\chi\sigma}}{4M_\chi^2-M_\sigma^2}\cos\omega-\frac{\lambda_{\chi\chi h}}{4M_\chi^2-M_h^2}\sin\omega \right]^2\ ,\\
\langle\sigma v \rangle_{\text{ann}}^{WW} =&\, \frac{M_W^4}{8\pi v_\phi^2M_\chi^2}\left(1-\frac{M_W^2}{M_\chi^2}\right)^{1/2}\left[2+\left(1-2\frac{M_\chi^2}{M_W^2}\right)^2\right]\left[\frac{\lambda_{\chi\chi h}}{4M_\chi^2-M_h^2}\cos\omega + \frac{\lambda_{\chi\chi\sigma}}{4M_\chi^2-M_\sigma^2}\sin\omega \right]^2\ ,\\
\langle\sigma v \rangle_{\text{ann}}^{ZZ} =&\, \frac{M_Z^4}{16\pi v_\phi^2M_\chi^2}\left(1-\frac{M_Z^2}{M_\chi^2}\right)^{1/2}\left[2+\left(1-2\frac{M_\chi^2}{M_Z^2}\right)^2\right]\left[\frac{\lambda_{\chi\chi h}}{4M_\chi^2-M_h^2}\cos\omega+\frac{\lambda_{\chi\chi\sigma}}{4M_\chi^2-M_\sigma^2} \sin\omega \right]^2 \ .
\end{split}
\end{equation}
The total thermally averaged WIMP pair-annihilation cross section is given by the sum of all channels in \eqref{tacsj},
\begin{equation}\label{tacstot}
\langle\sigma v \rangle_{\text{ann}}^{\text{total}} = \sum_{jj'} \langle\sigma v \rangle_{\text{ann}}^{jj'} \ .
\end{equation}

\end{document}